\documentclass[11pt,a4paper]{article}

\usepackage{jheppub}
\usepackage{amssymb,amsmath}

\usepackage{graphicx}
\usepackage{slashed}
\usepackage[normalem]{ulem}
\usepackage[caption=false]{subfig}

\usepackage{hyperref}

 \usepackage[final]{pdfpages}

\csname @addtoreset\endcsname{equation}{section}

\makeatletter
\def\timenow{\@tempcnta\time
  \@tempcntb\@tempcnta
  \divide\@tempcntb60
  \ifnum10>\@tempcntb0\fi\number\@tempcntb
  \multiply\@tempcntb60
  \advance\@tempcnta-\@tempcntb
  :\ifnum10>\@tempcnta0\fi\number\@tempcnta}
\makeatother

\def\oonoo#1#2#3{\vbox{\ialign{##\crcr
	\hfil\hfil\hfil{$#3{#1}$}\hfil\crcr\noalign{\kern1pt\nointerlineskip}
	$#3{#2}$\crcr}}}
\def\oon#1#2{\mathchoice{\oonoo{#1}{#2}{\displaystyle}}
	{\oonoo{#1}{#2}{\textstyle}}{\oonoo{#1}{#2}{\scriptstyle}}
	{\oonoo{#1}{#2}{\scriptscriptstyle}}}

\def\dt#1{\oon{\hbox{\bf .}}{#1}}  
\def\ddt#1{\oon{\hbox{\bf .\kern-1pt.}}#1}
\def\slap#1#2{\setbox0=\hbox{$#1{#2}$}
	#2\kern-\wd0{\hfuzz=1pt\hbox to\wd0{\hfil$#1{/}$\hfil}}}

\allowdisplaybreaks[3]

\makeatletter
\renewcommand\@fpheader{} 
\renewcommand\@journal{}
\makeatother

\title{
Amplitudes from anomalous superconformal symmetry
\bigskip
\bigskip}

\preprint{LAPTH-043/18, MPP-2018-259}

\author{D.\ Chicherin$^a$, J.\ M.\ Henn$^{a}$, E.\ Sokatchev$^{b}$}

\affiliation{
$^a$ Max-Planck-Institut f{\"u}r Physik, Werner-Heisenberg-Institut, 80805 M{\"u}nchen, Germany\\
$^b$ LAPTh, Universit\'e Savoie Mont Blanc, CNRS, B.P. 110, F-74941 Annecy-le-Vieux, France}

\emailAdd{chicheri@mpp.mpg.de}
\emailAdd{henn@mpp.mpg.de}
\emailAdd{emeri.sokatchev@cern.ch}

\abstract{
We initiate a systematic study of the consequences of (super)conformal symmetry of massless scattering amplitudes. The classical symmetry is potentially broken at the quantum level by infrared and ultraviolet effects. We study its manifestations on the finite hard part of the scattering process. The conformal  Ward identities  in momentum space are second-order differential equations,  difficult
to analyze. We prefer to study superconformal symmetry whose generators are first-order in the momenta. Working in a massless ${\cal N}=1$ supersymmetric Wess-Zumino model, we derive on-shell superconformal Ward identities. They contain an anomaly due to collinear regions of loop momenta. It is given by an integral with one loop less than the original graph, with an extra integral over a collinear splitting parameter. We discuss the relation to the holomorphic anomaly that was previously studied in tree-level amplitudes and at the level of unitarity cuts. We derive and solve Ward identities for various scattering processes in the model. We classify the on-shell superamplitudes according to their Grassmann degree, in close analogy with the helicity classification of gluon amplitudes. We focus on MHV-like and NMHV-like amplitudes with up to six external particles, at one and two loops. Interestingly, the superconformal generator acting on the bosonic part of the amplitudes is Witten's  twistor collinearity operator. We find that the first-order differential equations, together with physically motivated boundary conditions, uniquely fix the answer.  All the cases considered give rise to uniform weight functions.
Our most interesting example is a five-point non-planar hexa-box integral
with an off-shell leg.  It gives first indications on the function space needed for Higgs plus two jet production at next-to-next-to leading order.
 }

\begin{document}
\unitlength1cm

\maketitle



\newcommand{\norm}[1]{{\protect\normalsize{#1}}}
\newcommand{\p}[1]{(\ref{#1})}
\newcommand{\half}{\tfrac{1}{2}}
\newcommand \vev [1] {\langle{#1}\rangle}
\newcommand \ket [1] {|{#1}\rangle}
\newcommand \bra [1] {\langle {#1}|}
\newcommand \pd [1] {\frac{\pa}{\pa {#1}}}
\newcommand \ppd [2] {\frac{\pa^2}{\pa {#1} \pa{#2}}}

\newcommand{\cI}{{\cal I}}
\newcommand{\cM}{{\cal M}} 
\newcommand{\cR}{{\cal R}} 
\newcommand{\cS}{{\cal S}} 
\newcommand{\cK}{{\cal K}}
\newcommand{\cL}{{\cal L}} 
\newcommand{\cF}{{\cal F}}
\newcommand{\cN}{{\cal N}}
\newcommand{\cA}{{\cal A}}
\newcommand{\cB}{{\cal B}}
\newcommand{\cG}{{\cal G}}
\newcommand{\cO}{{\cal O}}
\newcommand{\cY}{{\cal Y}}
\newcommand{\cX}{{\cal X}}
\newcommand{\cT}{{\cal T}}
\newcommand{\cW}{{\cal W}}
\newcommand{\cP}{\cal P}
\newcommand{\mK}{{\mathbb K}}
\newcommand{\nt}{\notag\\} 
\newcommand{\pa}{\partial}
\newcommand{\ep}{\epsilon}
\newcommand{\om}{\omega}
\newcommand{\bom}{\bar\omega}
\newcommand{\etap}{\bar\epsilon}
\newcommand{\vep}{\varepsilon}
\newcommand{\vq}{\vartheta}
\renewcommand{\a}{\alpha}
\renewcommand{\b}{\beta}
\newcommand{\g}{\gamma}
\newcommand{\s}{\sigma}
\newcommand{\la}{\lambda}
\newcommand{\tl}{\tilde\lambda}
\newcommand{\tm}{\tilde\mu}
\newcommand{\tk}{\tilde k}
\newcommand{\tp}{\tilde p}
\newcommand{\tX}{\tilde \Xi}
\newcommand{\da}{{\dot\alpha}}
\newcommand{\db}{{\dot\beta}}
\newcommand{\dg}{{\dot\gamma}}
\newcommand{\dd}{{\dot\delta}}
\newcommand{\q}{\theta}
\newcommand{\bq}{{\bar\theta}}
\renewcommand{\r}{\rho}
\newcommand{\br}{\bar\rho}
\newcommand{\bp}{\bar\psi}
\newcommand{\be}{\bar\eta}
\newcommand{\bQ}{\bar Q}
\newcommand{\bS}{\bar S}
\newcommand{\bx}{\bar \xi}
\newcommand{\tx}{\tilde{x}}
\newcommand{\tr}{\mbox{tr}}
\newcommand{\+}{{\dt+}}
\renewcommand{\-}{{\dt-}}
\newcommand{\ti}{{\textup{i}}}
\newcommand{\tred}[1]{\textcolor{red}{\bfseries #1}}

\section{Introduction}

Conformal symmetry has played a central role in quantum field theory for many decades.
It has applications in condensed matter systems, statistical physics, and string theory, and is
also of interest to mathematicians. While the symmetry is most powerful in two dimensions,
where its algebra is infinite-dimensional, four-dimensional theories have
also been studied intensively. Current fields of interest include the AdS/CFT correspondence,
conformal bootstrap ideas, and high energy QCD \cite{Braun:2003rp}.

Most studies explore the consequences of the symmetry in position space,
with the goal of determining anomalous dimensions, or to try to obtain consistency conditions
for the space of conformal field theories via the operator product expansion.

On the other hand, a very important class of observables are scattering amplitudes. The
latter are essential ingredients for computing cross sections in particle collisions.
At high energies, often one may neglect the masses of the scattered particles, in which
case the Lagrangian of the Standard Model becomes classically conformal.
To give a concrete example, scattering amplitudes for three-jet production, or production
of a Higgs or a vector boson in association with two jets, neglecting quark masses, 
feature prominently on the current Les Houches wishlist \cite{Bendavid:2018nar}. 
To what extent does the powerful underlying conformal symmetry constrain the scattering
amplitudes of the theory? 

Besides this phenomenological motivation, there are also strong theoretical reasons
for studying this question, in particular in theories where the symmetry is exact at the quantum level. 
The prime example  is  the ${\cal N}=4$ super Yang-Mills (sYM)  theory. The last decade has
seen huge progress in understanding its scattering amplitudes, both perturbatively and at strong
coupling via the AdS/CFT correspondence. However, very little is known about how the 
conformal symmetry of the Lagrangian restricts its loop-level amplitudes.

There are several reasons why the above questions are very difficult to study, let alone to answer.
A major problem is that at the quantum level, typically divergences are present, which are either of
infrared (IR), or of ultraviolet (UV) origin. The regularization procedure necessarily introduces a 
mass scale, which obscures 
the underlying conformal symmetry of the Lagrangian. To make matters worse,
even finite integrals that are naively conformal 
(i.e., Feynman integrals that possess an exact conformal
invariance off-shell, e.g. in position space) 
need to be treated with care: after the amputation of an external leg the on-shell action of the
momentum-space conformal generators may produce  contact terms that lead to anomalies.
Finally, while the conformal generators are first order in position space, they become second order 
after Fourier transformation. This severely complicates even the study of the implications of the exact symmetry.

This sounds like a formidable set of problems that may seem too difficult to overcome. 
The skeptical reader has our full blessing to stop reading here (or to add further concerns to the list).
Understanding in detail how to cope with the difficulties will certainly require a considerable effort 
from many researchers. Here, we offer a way of thinking about these problems, and make the first steps 
toward solving some of them. This article is the extended  writeup of our Letter \cite{Chicherin:2018ubl}.
Here we present all  the detailed explanation and also new results.

Regarding the technical problem of dealing with second order generators and hence differential equations, 
we propose to circumvent this by first considering theories with superconformal symmetry.
The special conformal supersymmetry generators are first order in the bosonic variables.
This considerably simplifies the task of both finding particular solutions to the differential equations, and fixing the 
homogeneous freedom \cite{Chicherin:2018ubl}.

Next, let us comment on the issue of divergences. First of all, it is important to realize that ultraviolet and infrared divergences
are due to very different physical effects that occur at different scales. This leads to the idea of factorization, common in the QCD literature, which essentially
says that the two effects can be treated separately. Schematically, one may write an amplitude $\mathcal{A}$ in the factorized form \cite{Collins:1989gx}, $\mathcal{A} = Z_{\rm UV} Z_{\rm IR} \mathcal{A}_{f}$, where the $Z$-factors contain all UV and IR divergences (e.g., poles in dimensional regularization), respectively, while the `hard part'  $\mathcal{A}_{f}$ has a finite limit as the regulator is removed. The UV renormalization factor is of course well understood, and the $Z_{\rm IR}$ is known for any massless three-loop amplitude \cite{Catani:1998bh,Aybat:2006mz,Dixon:2008gr,Becher:2009cu,Almelid:2015jia}. In planar $\mathcal{N}=4$ sYM, the latter factor is known in closed form, up to certain anomalous dimensions. Note that the definition of the finite part is not unique, and depends, in particular, on the regularization and renormalization scheme.
Given that so much is known about the physics of the divergences, one may ask whether a suitable definition of a hard function, or finite remainder $\mathcal{A}_{f}$ exists that has simple transformation rules under conformal symmetry? In other words, is there a conformal symmetry friendly definition of this finite part?
This important question is beyond the scope of the present paper. Here, we bypass this difficulty by considering an infrared finite model. 

In practice, there are a number of ways to remove the unwanted ultraviolet divergences. We already mentioned
the $\mathcal{N}=4$ sYM model which has a zero $\beta$ function. Other ideas include going to a conformal fixed
point of a theory (e.g. as a function of the dimension, or some other parameter, such as the number of fermion flavors).
Another possibility is to restrict the analysis to terms without ultraviolet divergences, as it is done in the 
quenched approximation of QED. In this paper, we will choose a variant of the latter approach.

In order to study this question in more concrete terms, it is useful to have some simple models at hand.
We already mentioned that it is advantageous to consider a superconformal model. 
There are several reasons for choosing a model with as little supersymmetry as possible.
One reason is to stay as close as possible to reality. The other reason is to
have manifestly supersymmetric Feynman rules. Such an off-shell formulation is not known,
for example, for $\mathcal{N}=4$ sYM. 

We choose the massless Wess-Zumino model \cite{Wess:1973kz}, which describes $\mathcal{N}=1$ supersymmetric matter
and possesses a superconformal symmetry at the classical level. In components, it describes fermions and scalars that interact
via Yukawa and quartic vertices. The UV renormalization of this model is well understood, and in particular
it only involves wave function renormalization \cite{Ferrara:1974fv}. Therefore we can exclude UV divergences by restricting
ourselves to Feynman graphs without propagator subdiagrams. At the same time, as already mentioned, the amplitudes in
this model are IR finite.

One might think that the finite part of the amplitudes in the Wess-Zumino model defined as above will be trivially (super)conformally invariant,
and that this model is somewhat (over)simplistic. This is not the case. We will see that, although finite, the amplitudes
obey the (super)conformal symmetry in a very interesting way.
Analyzing carefully the action of the special conformal supersymmetry generator, we find that it produces subtle contact terms in
the loop integrals. The mechanism, to be explained in detail in the paper,  can be summarized as follows. When the momenta of an external on-shell leg and of two adjacent internal lines are aligned, the product of propagators turns into a singular distribution. The action of the superconformal generator results in an ambiguity $0\times\infty$, whose careful resolution yields finite anomaly terms of the collinear contact type. The latter have the effect of removing one loop integration, making the anomaly term of the Ward identity 
relatively simple. 
In the conformal case, previously studied in \cite{Chicherin:2017bxc}, the anomalous Ward identity  takes the schematic form 
\begin{align}\label{equation_ward_identity_intro}
K^\mu  I^{(L,n)} =  \sum_{i} p^\mu_i\int_{0}^{1} d\xi\ I_{i}^{(L-1,n+1)}(\xi)\,.
\end{align}
The special conformal variation of an $L$-loop, $n$-particle integral is expressed in terms of $(L-1)$-loop $(n+1)$-particle integrals. 
Two particles of the latter are in the collinear limit, with the collinear splitting parameter $\xi$ being integrated over. Each anomaly term is proportional to the on-shell momentum $p^\mu_i$ of the $i$-th leg.
This means that the integrals on the right-hand side of the equation are much easer to compute than the integral on its left-hand side.
Thus the equation is a very useful tool with predictive power.

In the superconformal case, we will find a similar equation for the generators of   special conformal supersymmetry,
where again the most important feature is that the anomaly term is given by an integral with one loop less. Schematically, the anomalous Ward identity for the chiral superconformal generator $S^\a$ reads
\begin{align}\label{equation_superconformal_ward_identity_intro}
S^\a  I^{(L,n)} =  \sum_{i} \la^\a_i\int_{0}^{1}d\xi\int d\eta\  I_{i}^{(L-1,n+1)}(\xi,\eta)\,.
\end{align}
Here the anomaly is determined by the collinear limit of an $(n+1)$-particle integral, and the splitting (super)parameters $(\xi, \eta)$ are integrated over. Each anomaly term is proportional to the chiral helicity spinor $\la^\a_i$ of the $i$-th leg, if that leg corresponds to an antichiral on-shell state. A similar Ward identity exists for the antichiral superconformal generator $\bar S^\da$.
One crucial difference from the conformal case, as already mentioned, is that the generators are first-order in the bosonic kinematic variables.
This feature makes the Ward identity much easier to analyze and to solve.

We would like to mention that the breakdown of superconformal symmetry of amplitudes because of collinear singularities is not an entirely new observation. In \cite{Bargheer:2009qu,Korchemsky:2009hm,Beisert:2010gn} an  $\bar S$ anomaly has been revealed at the level of the unitarity cuts of the $\cN=4$ sYM amplitude.  It is due to singularities occurring when two on-shell legs, one external and one cut, become  collinear. This phenomenon is known as a `holomorphic anomaly' \cite{Cachazo:2004by} {(see also Ref.~\cite{Bidder:2004tx} for amplitudes with $\cN=1$ supersymmetry)}. In our case, the anomaly concerns the integrated quantity rather than its cuts. {It takes the form of an inhomogeneous differential equation for the integral with high predictive power. } 

Readers knowledgeable in $\mathcal{N}=4$ sYM will have noticed that the anomaly equation \p{equation_superconformal_ward_identity_intro} is reminiscent in form to the descent equations formulated for null polygonal Wilson loops in that theory \cite{CaronHuot:2011ky,Bullimore:2011kg}. The Wilson loops are dual to planar scattering amplitudes in $\mathcal{N}=4$ sYM, and hence the descent equations also apply to the latter.   These equations have been very useful in the analysis of $\mathcal{N}=4$ sYM scattering amplitudes, more precisely in restricting the possible form of the remainder function, see e.g. \cite{CaronHuot:2011ky,Dixon:2011pw,Caron-Huot:2016owq}.
We wish to point out that although these equations are similar in form to our Eq.~(\ref{equation_superconformal_ward_identity_intro}), they are a different manifestation of the same phenomenon. 
In the case of \cite{CaronHuot:2011ky,Bullimore:2011kg}, the equation applies to the dual Wilson loops, while in our case they apply directly to amplitudes. The main difference is that the anomalous symmetry of the Wilson loop is the Poincar\'e supersymmetry $\bar Q$, which is dual to the superconformal symmetry $\bar S$ of the amplitude. Also, since the duality Wilson loops/amplitudes only works in the planar limit, nothing could be predicted about the nonplanar sector of the amplitude. 

In this paper, we explain in detail how the anomalous superconformal Ward identity \p{equation_superconformal_ward_identity_intro} is derived,
and study its consequences. We emphasize that our approach does not rely on planarity or integrability.
Moreover, since we only use $\mathcal{N}=1$ supersymmetry, the techniques developed here can be expected to apply to much larger classes of superconformal theories compared to just $\mathcal{N}=4$ sYM.

We organize the component amplitudes into superamplitudes depending on Grassmann parameters, and classify them according to their R charge, i.e. their Grassmann degree.
In $\cN=4$ sYM the latter is related to the helicity distribution, but here it is not.
In a slight abuse of language, we refer to superamplitudes as `maximally-helicity-violating-like' (MHV-like),
`next-to-MHV-like' amplitudes (NMHV-like), and so on. 
Readers familiar with $\mathcal{N}=4$ sYM will find many similarities to the structure of superamplitudes in that theory, but also certain differences.
For example, the structure of super-invariants is remarkably similar to that known from $\mathcal{N}=4$ sYM \cite{Drummond:2008vq}.
One new aspect is that the N${}^{k}$MHV-like amplitudes in our model of $\mathcal{N}=1$ supersymmetric matter do not exist for arbitrary number of particles.
For example, MHV-like amplitudes exist for $N=3,4,5,6$ external particles only.

We find that the MHV-like and the equivalent $\overline{\rm MHV}$-like superamplitudes  are characterized by a single bosonic function. This means that all of their component amplitudes are related. {Such amplitudes are particularly easy to study and we provide many examples.
On the other hand, non-MHV-like superamplitudes contain several independent component amplitudes, which makes their structure richer.} 

In our Letter \cite{Chicherin:2018ubl}, we used the superconformal Ward identity (\ref{equation_superconformal_ward_identity_intro}) 
to compute a previously unknown non-planar two-loop five-particle integral. 
In the present paper, we illustrate the usefulness of the method by a number of further non-trivial examples. We show how the inhomogeneous first-order partial differential equations that follow from  \p{equation_superconformal_ward_identity_intro} can be solved for various planar and nonplanar one- and two-loop Feynman integrals. These are finite integrals constructed from Yukawa and $\phi^4$ vertices.
Our method does not rely on any prior knowledge of special properties of the functions, such as the symbol alphabet \cite{Goncharov:2010jf,Duhr:2011zq}, which we derive by integrating the differential equations.

We wish  to mention an important property of the special conformal supersymmetry generators $S^\a$ and $\bar S^\da$. 
When they pass through the Grassmann structure of the superamplitude and reach the bosonic functions, they are converted to `twistor collinearity operators' \cite{Witten:2003nn}.
In some cases, the fact that the latter operator only acts on half of the variables (i.e., the $\lambda$, but not the $\tilde{\lambda}$ variables), turns out to be very useful when fixing the boundary conditions of the differential equations.

Interestingly, we find that all the integrals considered have uniform transcendental weight, and that they are closely related to the `local integrals' from Ref.~\cite{ArkaniHamed:2010gh}. 

As a particular highlight, we work out the superconformal Ward identities for a generalization of the integral considered in \cite{Chicherin:2018ubl} to the case where one of the external legs is off-shell. 
This case is very interesting, as the kinematics corresponds to that of the process where a Higgs and two jets are produced.
Computing next-to-next-to-leading order five-particle Feynman integrals \cite{Chicherin:2017dob,Gehrmann:2018yef,Abreu:2018rcw,Chicherin:2018mue},
integration-by-parts reductions \cite{Chawdhry:2018awn,Boehm:2018fpv}
and the corresponding amplitudes \cite{Badger:2017jhb,Abreu:2017hqn,Abreu:2018jgq} is a very active area of research. 
Only some results for certain planar integrals \cite{Papadopoulos:2015jft} are available,
and our paper is the first to provide insights into the non-planar case.
We derive and solve the Ward identities in the form of a two-parameter integral of known functions.

We begin  in Sect.~\ref{sec1} by recalling the description of the massless Wess-Zumino model in $\cN=1$ chiral superspace and defining (anti)chiral  on-shell states.   
In Sect.~\ref{sec2} we discuss the general structure and symmetries of the $\cN=1$ matter superamplitudes and explain the N${}^{k}$MHV-like classification according to their Grassmann degree. We trace the origin of the Ward identity (\ref{equation_superconformal_ward_identity_intro}) to the superconformal anomaly of the elementary three-point vertex functions.
The following Sect.~\ref{56leg} is devoted to the detailed study of a number of nontrivial examples of superamplitudes of the $\overline{\rm MHV}$ type with five legs. The single bosonic function that defines them is obtained by solving the superconformal Ward identities. We illustrate the method of integrating the differential equations and finding the relevant boundary conditions. In Sect.~\ref{6leg} we study a previously unknown non-planar two-loop integral of hexa-box topology. This result gives first insights into the function space of Higgs plus two jets amplitudes at next-to-next-to-leading order. In Sect.~\ref{6nmhv} we present an example of a six-leg one-loop superamplitude of the NMHV type. It is described by two bosonic functions, but thanks to the  additional symmetry (cyclic and dual conformal) we are again able to solve the differential equations. 

We provide five appendices for   the reader's convenience. Appendices \ref{Appendix_Spinor_Conventions} and \ref{Appendix_Susy_Algebra} contain details on our two-component spinor conventions and the supersymmetry algebra, respectively. Appendix~\ref{Appendix_Derivation_Anomaly} presents two different derivations of the superconformal anomaly formula of the three-point chiral vertex function. 
Appendix \ref{Appendix1LoopIntegrals} collects analytic formulas for one-loop integrals  used in the calculation of the anomaly terms at higher loop orders.
In Appendix \ref{appE} we explain how  the previously known holomorphic anomaly of the unitarity cuts of an integral fits in our more general picture of the anomaly of the integral itself.

\section{$\cN=1$ superfields and Wess-Zumino model}\label{sec1}

We consider the massless Wess-Zumino model with $\cN=1$ supersymmetry in four dimensional space-time.  It is described by a chiral and an antichiral off-shell superfield
$\Phi(x_L,\q)$ and $\bar\Phi(x_R,\bq)$, respectively, with component expansion 
\begin{align}\label{e11}
& \Phi(x_L,\q) = \phi(x) + \q^\a   \psi_\a(x) + \q^2 F(x)\,,\nt
&\bar\Phi(x_R,\bq) = \bar\phi(x) + \bq_\da  \bar\psi^\da(x) + \bq^2 \bar F(x)\,.
\end{align}
Here $\phi$ is a complex scalar, $\psi,\bp$ form a Majorana spinor  and $F$ is a complex  auxiliary (non-propagating) field. The (anti)chiral superfields are defined in the appropriate superspace bases, 
\begin{align}\label{e2.2}
\mbox{Chiral basis:} & \qquad x^\mu_L = x^\mu + \frac{i}{2}\q\sigma^\mu\bq\,, \quad \q^\a\,; \nt
\mbox{Antichiral basis:} & \qquad x^\mu_R = x^\mu - \frac{i}{2}\q\sigma^\mu\bq\,, \quad \bq^\da \,,
\end{align}
which are closed under the supersymmetry transformations (see Appendix \ref{Appendix_Susy_Algebra}). 

The massless Wess-Zumino action 
\begin{align}\label{1.2}
S_{\rm WZ} &= \int d^4 x d^2\q d^2\bq\, \bar\Phi(x_R,\bq)\, \Phi(x_L,\q) \nt
&+ \frac{g}{3} \int d^4x_L d^2\q\, \Phi^3(x_L,\q) + \frac{g}{3} \int d^4x_R d^2\bq\, \bar\Phi^3(x_R,\bq)
\end{align}
involves a kinetic term (note that the two superfields have been written in the common real basis $(x,\q,\bq)$) and cubic self-interaction terms.   
Substituting the superfield expansions \p{e11} in \p{1.2} we obtain the component field form of the Lagrangian
\begin{align}\label{}
L_{\rm WZ} (x) = \bar\phi\Box\phi-i\bar\psi\pa\psi+ \bar FF + g\left[\phi^2 F -\frac1{2}\psi^2\phi + {\rm c.c.}  \right]\,.
\end{align}
Eliminating the auxiliary fields $F,\bar F$ via their algebraic field equations, we obtain the familiar  $g\psi^2\phi$ (Yukawa) and $g^2\phi^2\bar\phi^2$ couplings. 

The action \p{1.2} has a manifest $U(1)$ symmetry. Ascribing to the odd variables $\q,\bq$ and to the superfields $\Phi,\bar\Phi$ $U(1)$ charges (called R-charges) according to Table \ref{e16}, we see that the action is invariant. The R-charge counting will play an essential role in what follows. 

This model is superconformal at the classical level. At the quantum level, the symmetry is broken, but only
by propagator corrections, and the beta function is proportional to the anomalous dimension of the superfield \cite{Ferrara:1974fv}.
This property allows us to study individual UV finite supergraphs, with the only requirement that they do not contain propagator correction subgraphs. Further, we are interested in supergraphs contributing to the scattering of massless supermultiplets. Such supergraphs are potentially IR/collinear divergent, but  their component graphs are composed of Yukawa and $\phi^2\bar\phi^2$ vertices, so they are finite in this sense as well.

The on-shell massless states in the $\cN=1$ matter scattering theory, corresponding to  the (anti)chiral matter superfields $\Phi(x_L,\q)$ and $\bar\Phi(x_R,\bq)$,
are obtained by replacing the fields in \p{e11} by the solutions of the classical free equations of motion. For the spinor fields with a lightlike momentum $p^2 = 0$, i.e. $p_{\a\da} = \la_\a \tl_\da$,  these are $\psi_\a(p) = \la_\a \psi_+(p)$ and $\bar\psi_\da=\tl_\da \psi_-(p)$. The subscript $\pm$ indicates the helicity of the fermion state. The auxiliary fields vanish on the free shell. In this way we obtain the superstates  
\begin{align}\label{72}
\Phi(p,\be) = \phi(p) + \psi_+(p)\be  \,, \qquad \bar\Phi(p,\eta) = \bar\phi(p) + \eta \psi_-(p)\,.
\end{align}
In the chiral superstate  the odd variable $\be=\vev{\la\q} \equiv \la^\a \q_\a$  has helicity $(-1/2)$; in the antichiral superstate the variable $\eta =[\tl\bq] \equiv \tl_\da \bq^\da$ has  helicity $(+1/2)$. These definitions follow the standard $\cN=4$ sYM conventions for ascribing helicity to the odd on-shell variables $\eta^A$ (with $A=1\ldots4$).
The on-shell states $ \psi_+,\, \phi,\, \bar\phi ,\,  \psi_-$ carry helicities $+\tfrac{1}{2}, 0 , 0, -\tfrac{1}{2}$, respectively. The odd variables $\be,\eta,\q,\bq$ also carry R-charges listed in Table \ref{e16}.  

We prefer to use only one type of on-shell Grassmann variable. To this end we 
perform a  Grassmann Fourier transform of $\Phi(p,\be)$,  
\begin{align} \label{FTgr}
\Psi(p,\eta) = -\int d \be \, e^{\eta\be} \Phi(p,\be) = \psi_+(p) + \eta \phi(p)\,.
\end{align}
We are allowed to use the same symbol $\eta$ for the Fourier image of $\be$ and for the Grassmann variable of $\bar\Phi$ since they have the same helicity $(+1/2)$ and R-charge $(+1)$.

The  assignments of dimension $D$ (in mass units), R-charge $R$  and helicity $H$ are summarized in Table \ref{e16}.\footnote{The dimension of the  off-shell superfields $\Phi\,, \bar\Phi$ is  $(-3)$ in momentum space (in mass units). After the amputation the on-shell states have dimension $(-1)$. The Fourier transform \p{FTgr} does not change the dimension. The R-charge of $\q_\a$ and the helicity of $\la_\a$ are conventional. The rest follows from the definitions and the interaction terms in \p{1.2}. We find it convenient to have integer R-charges of the odd variables. } 
\begin{table}[htp]
\begin{center}
\begin{tabular}{|c|c|c|c|c|c|c|c|c|c|}
\hline
 & $\q$ & $\bq$ & $\eta$ & $\be$ & $\Phi(\eta)$ & $\bar\Phi(\be)$ & $\Psi(\eta)$ & $\la$ & $\tl$   \\
\hline
$D$ & -1/2 & -1/2 & 0 & 0 & -1 & -1 & -1 & 1/2 & 1/2 \\
\hline
$R$ &-1 &1 &1 &-1 &-2/3 &2/3 &1/3 &0 &0 \\
\hline
$H$ &0 &0 &1/2 &-1/2 &0 &0 &1/2 &-1/2 &1/2 \\
    \hline
\end{tabular}
\end{center}
\caption{Quantum numbers of the variables and superstates}
\label{e16}
\end{table}%

In the quantized theory, we define the superpropagator (vacuum expectation value)\footnote{Here and in what follows we tacitly assume time ordering of the fields (Feynman prescription) $1/(q^2+i0)$.} \footnote{For the two-component spinor notation see App.~\ref{Appendix_Spinor_Conventions}.}
\begin{align}\label{e19}
\vev{\bar\Phi(-q,\bq_1) \Phi(q,\q_2)} =  \frac{e^{\bq_1 \tilde q \q_2 }}{q^2}\,,
\end{align}
where we omit the momentum conservation delta function. 
The Grassmann dependence in \p{e19} is  fully fixed by the invariance under the off-shell generators $Q$ and $\bar Q$ in  \p{e111}.\footnote{We use the notation $q$ for the off-shell momenta ($q^2\neq0$) and $p=\la\tl$ for the on-shell ones. } 

Amputating one end of the propagator \p{e19} we obtain the super wave-functions (or rather their Fourier transform to momentum space)  
\begin{align}
&\vev{\Phi(-p,\q)\bar\Phi(p,\eta)} = e^{\vev{\la \q}\eta} \,, \qquad      \vev{\bar\Phi(-p,\bq) \Psi(p,\eta)} = \eta+ [\tl\bq] 
 \,.
\label{534}
\end{align}
Alternatively, they can be uniquely fixed as $Q$- and $\bar Q$-invariants, see \p{b3} and \p{e111}. 

We would like to mention that our Feynman rules differ   from the familiar ones in the literature (see, e.g., Ref.~\cite{Wess:1992cp}), in  that we do not use spinor derivatives to achieve chirality. This is possible for massless fields because the momentum space propagator $1/q^2$ is Fourier transformed to $1/x_{12}^2$, which has a manifestly supersymmetric extension (see \p{533}).  This saves us a lot of algebraic manipulations (known as `D algebra') in the supergraphs.

\section{$\cN=1$ matter superamplitudes}\label{sec2}

In the following we  consider $\cN = 1$ supergraphs with on-shell external legs. They contribute to a certain scattering superamplitude. Our model does not involve gauge fields, therefore we are not obliged to sum up all the supergraphs for a given scattering process to achieve gauge invariance. We are studying the superconformal properties of individual supergraphs involving {\it finite Feynman integrals}. Nevertheless, by abuse of language we will call them {\it superamplitudes}.

These objects are (naively) superconformal since the Lagrangian of the theory has the symmetry and it is not spoiled by UV  or IR divergences. Still, as we show below, the superconformal symmetry can become anomalous because of collinear singularities in certain Feynman integrals, in the case when  external and loop momenta are aligned. Our main task is to derive anomalous superconformal Ward identities for such integrals and to learn how to solve them.

\subsection{$\cN=1$ on-shell supergraphs}
\label{sectgraphs}

\begin{figure}
\begin{align}
\begin{array}{c}\includegraphics[width = 6cm]{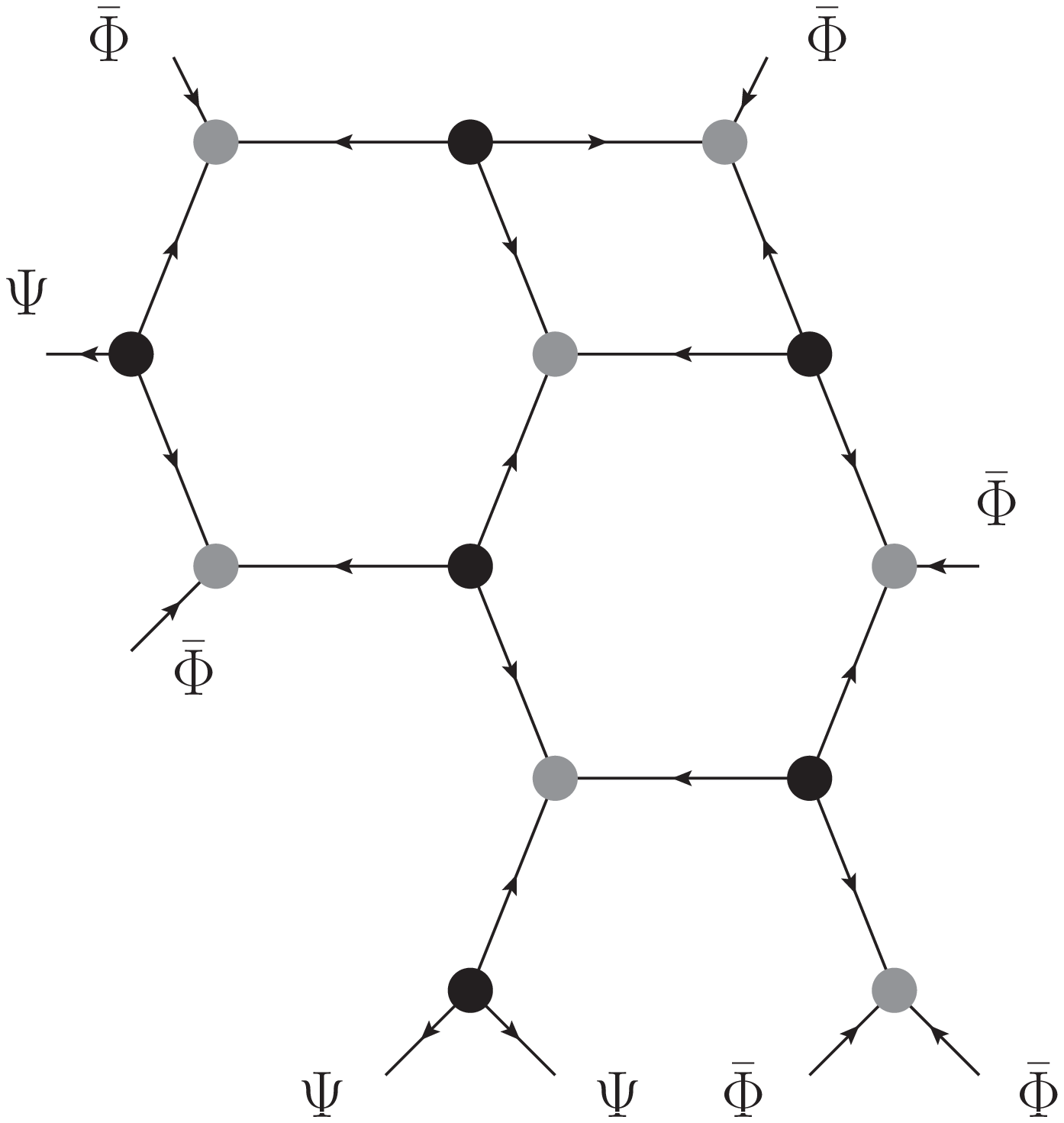}\end{array} \qquad\Longrightarrow\qquad 
\begin{array}{c}\includegraphics[width = 6cm]{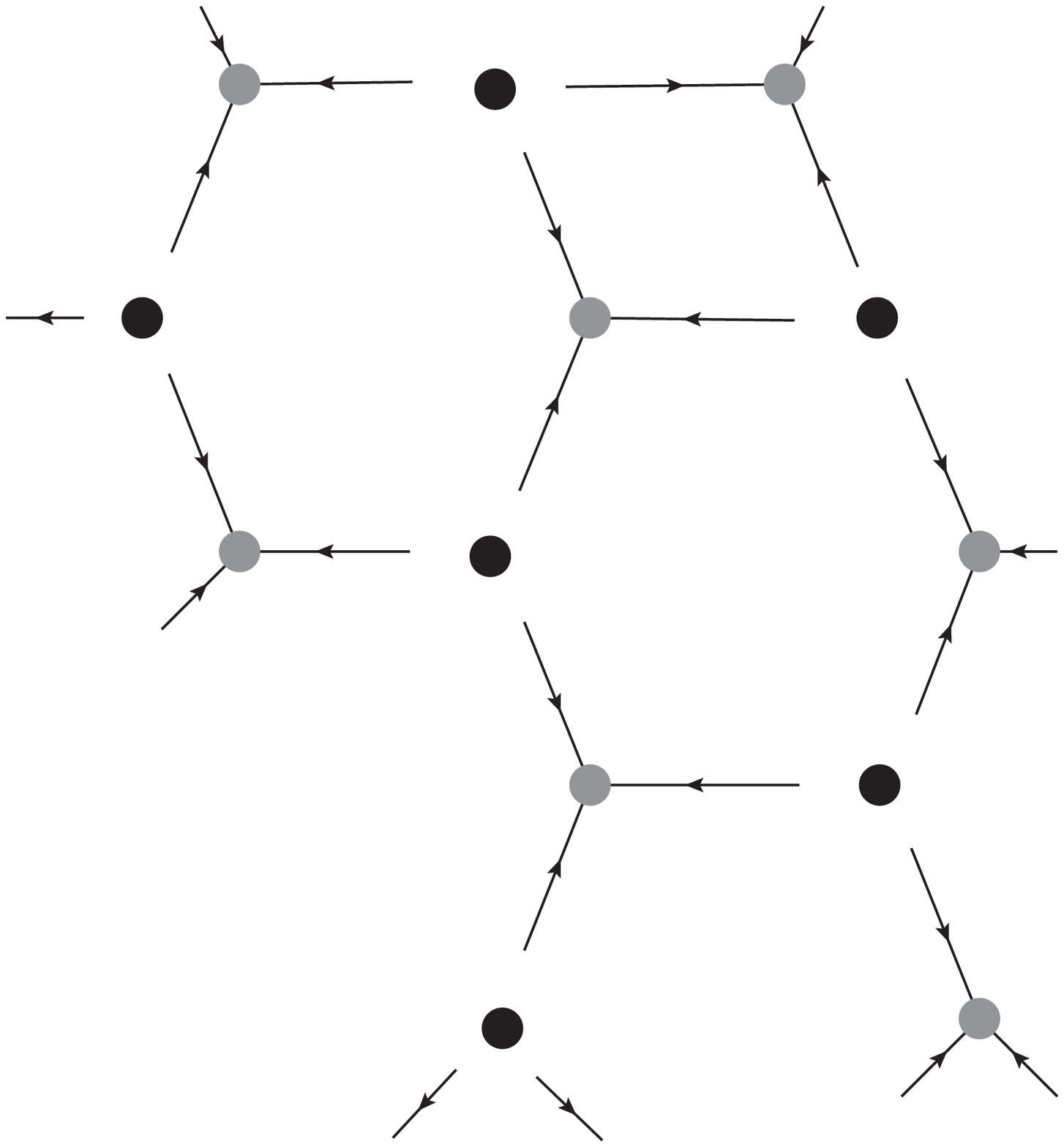} \notag
\end{array}
\end{align} 
\caption{A generic $\cN = 1$ supergraph is split into vertex functions. Gray blobs denote chiral vertices 
$\Phi^3$ and black blobs -- antichiral vertices $\bar\Phi^3$. Arrows represent the chirality flow.} \label{gengraph}
\end{figure}

As we have already stated, we study $\cN = 1$ supergraphs of the scattering amplitude type, based on the Wess-Zumino action \p{1.2}.
Let us make some comments on their structure. A generic graph is depicted on the lhs of Fig.~\ref{gengraph}. 
It is a bipartite graph, i.e. it is comprised of chiral  and antichiral vertices. Chiral (or antichiral) vertices are denoted by gray (or black) blobs with   $\int d^2\theta$ (or $\int d^2\bar\theta$) assigned to them. The vertices of opposite chirality are connected by propagators \p{e19}. The arrows on the figure denote the `chirality flow' from antichiral to chiral. The external on-shell states $\bar{\Phi}(p,\eta)$ and $\Psi(p,\eta)$ are described by the wave functions \p{534}. 

Using these Feynman rules and doing the Grassmann integrations, we can reduce the supergraphs to familiar  Feynman graphs for bosons and fermions. However, we can further simplify the task. We implement at first all chiral integrations, see the rhs of Fig.~\ref{gengraph}. In this way we form  antichiral tree-level {\it vertex functions} $\vev{\bar\Phi \bar\Phi \bar\Phi}$, where each $\bar\Phi$ can either be an on-shell state $\bar\Phi(\eta)$ or an off-shell superfield $\bar\Phi(\bar\theta)$. 
In the fully off-shell case we have
\begin{align}
\begin{array}{c}\includegraphics[width = 3.5cm]{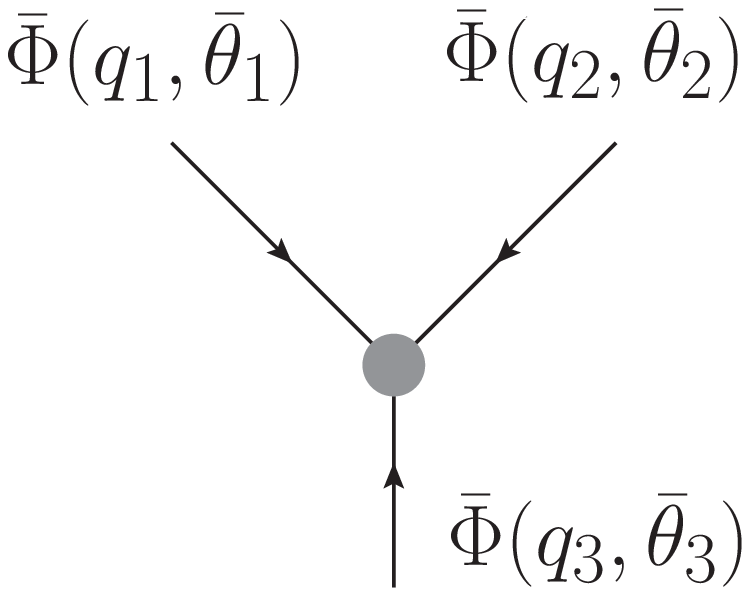}
\end{array} \qquad = 
\frac{\delta^{(4)}(P)\delta^{(2)}(Q)}{q_1^2 q_2^2 q_3^2}\,,
\label{e3.16}
\end{align} 
where $P=\sum_{i=1}^3 q_i$ is the total momentum  and $Q=\sum_{i=1}^3 \bq_i \tilde q_i$ is the total supercharge. We can replace one or more of the off-shell superfields $\bar\Phi(\theta)$ by an on-shell state $\bar{\Phi}(\eta)$ by just dropping the corresponding propagator $1/{q^2}$ in \p{e3.16}, replacing $q \to p =\la \tl$ in the total momentum and $\bq\tilde q \to \la\eta$ in the total supercharge $Q$.  For example, putting a single leg on shell gives (for more detail see App.~\ref{Appendix_Derivation_Anomaly})
\begin{align}
\begin{array}{c}\includegraphics[width = 3.5cm]{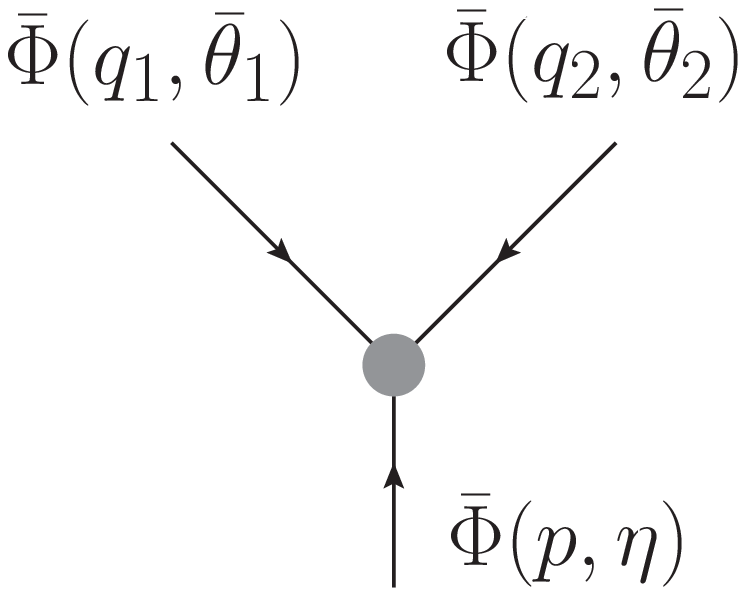}
\end{array} \qquad = 
\frac{\delta^{(4)}(P)\delta^{(2)}(Q)}{q_1^2 q_2^2}\,,
\label{e3.17}
\end{align}
where now $Q=\bq_1 \tilde q_1 +\bq_2 \tilde q_2+\la\eta$. With
two legs on shell we get
\begin{align}
\begin{array}{c}\includegraphics[width = 3.5cm]{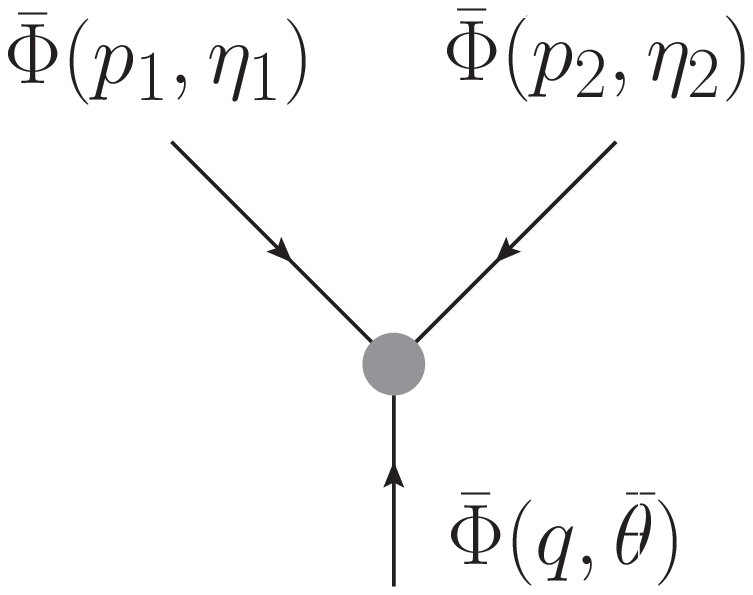}
\end{array} \qquad = 
\frac{\delta^{(4)}(P)\delta^{(2)}(Q)}{q^2 }\,,
\label{e3.17'}
\end{align}
where $Q=\bq\tilde q+\la_1\eta_1+\la_2\eta_2$.

After integrating at the  chiral vertices in this way, the resulting graph on the rhs of Fig.~\ref{gengraph} consists of antichiral vertex functions and wave functions $\vev{\Psi(\eta) \bar\Phi(\bar\theta)}$ \p{534}, which are glued together at the black blobs by antichiral integrations $\int d^2\bar\theta$. 
 
We also note that the bipartite graphs can only have $2n$-gon one-loop subdiagrams, e.g. box and hexagon one-loop sub-diagrams in Fig.~\ref{gengraph}. As mentioned earlier, we exclude graphs with propagator corrections (`bubbles') because they are UV divergent.

\subsection{Grassmann structure and symmetries} 

\subsubsection{Corrolaries of Poincar\'e supersymmetry}\label{s311}

The general  superamplitude with $N=(m+n)\geq3$ particles,  made from $m$ antichiral $\bar\Phi$ and $n$ chiral $\Psi$ matter superstates has the following manifestly $Q$-supersymmetric form\footnote{The three-point chiral amplitude $A_{0,3}$ is an exception, see \p{49}.}
\begin{align}\label{43}
A_{m,n} &= \vev{\bar\Phi(p_1,\eta_1) \cdots \bar\Phi(p_m,\eta_m) \, \Psi(p_{m+1},\eta_{m+1}) \cdots \Psi(p_{m+n},\eta_{m+n})}\nt
&  = \delta^{(4)}(P) \, \delta^{(2)}(Q)\, \cA_{m,n}(p,\eta)\,,
\end{align}
where $P= \sum_i p_i$ and $Q = \sum_i \la_i \eta_i$ are the total momentum and supercharge, respectively. We call the function $\cA_{m,n}$ the {\it reduced amplitude}.

Let us now act with the generator of  $\bQ$ supersymmetry. Using the relations $[P, \bQ]=0$ and $\{Q,\bQ\}=P$ (see \p{b1}), we obtain that $[\bQ,  \delta^{(4)}(P) \, \delta^{(2)}(Q)]=0$. Then the $\bQ$ invariance of $A_{m,n}$ implies the invariance  of the reduced amplitude,
\begin{align}\label{e42}
\bar Q^\da \cA =0\,.
\end{align}

Next we determine the R-charges of the amplitude \p{43} and of the reduced amplitude. According to  Table \ref{e16},   $R_A=(2m+n)/3$, which has to be integer, since  only the odd variables $\eta$ carry R-charge $(+ 1)$.  The delta function $\delta^{(2)}(Q)$ has R-charge $(+2)$, therefore $R_{\cA}= (2m+n)/3-2$. This implies that the reduced amplitude $\cA_{m,n}$ is of Grassmann degree $R_{\cA} \geq 0$, hence $2m+n \geq 6$.  We find the following relations
\begin{align}\label{e3.3}
m=3R_{\cA} - N+6 \geq 0\,, \qquad n = 2N-6 - 3R_{\cA} \geq0\,,
\end{align}
hence $R_{\cA}\geq 0$ is an integer in the interval
\begin{align}\label{e35}
  \frac1{3}(N-6) \leq R_{\cA} \leq \frac2{3}(N-3)\,.
\end{align}

Each amplitude $A_{m,n}$ has a conjugate $\bar A_{n,m}$ obtained by complex conjugation followed by a Grassmann Fourier transform \p{FTgr} of all the points. This establishes the equivalence relation $\bar A_{m,n}(\bar\Phi,\Psi) = A_{n,m}(\Psi,\bar\Phi)$.  For even $N$ and for $m=n=N/2$  the amplitude with $R_\cA=N/2-2$  is self-conjugate.  

\begin{figure}
\begin{center}
\includegraphics[width = 14cm]{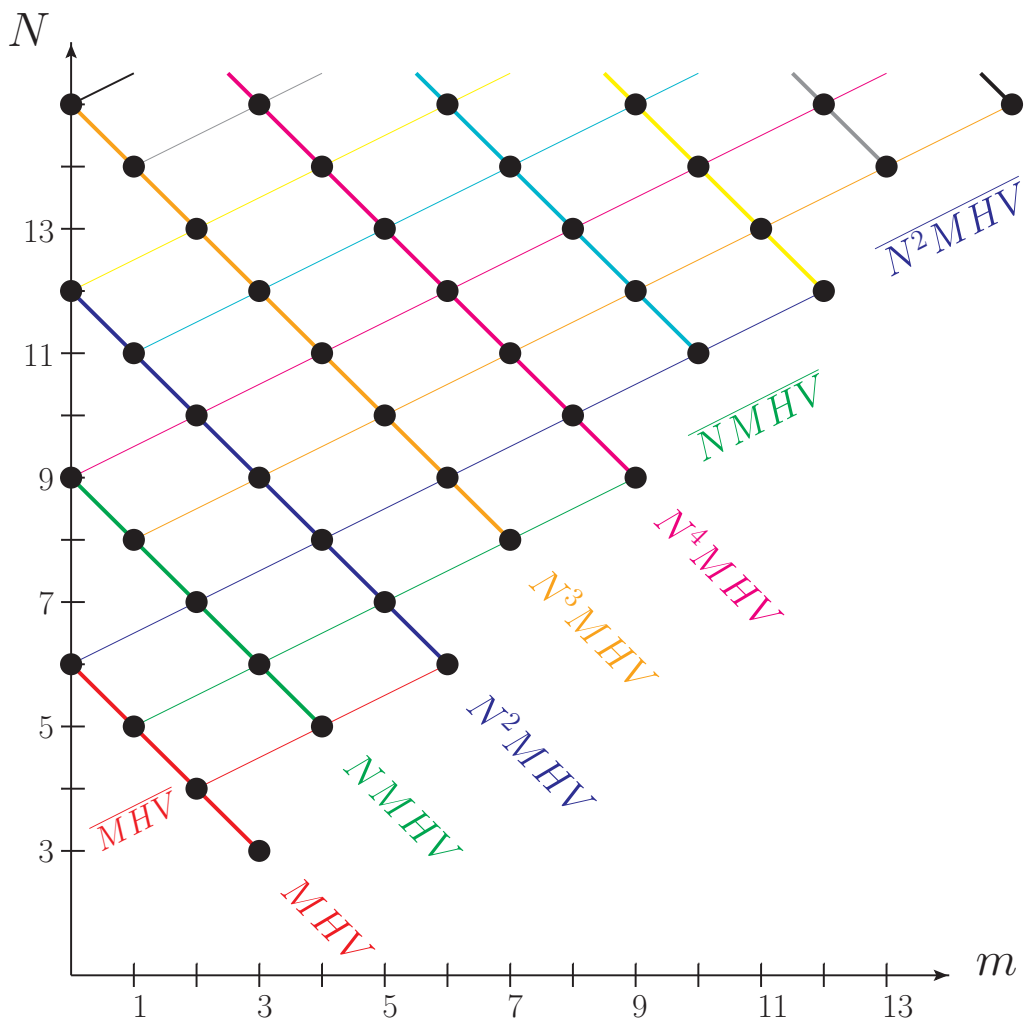}
\end{center}
\caption{$N$-point superamplitudes  with $m$ states $\bar\Phi$ and $(N-m)$ states $\Psi$ allowed by supersymmetry. Points on the anti-diagonals represent the $\rm N^{k}MHV$ amplitudes with $R_{\cA} = k$; points on the diagonals represent the conjugate amplitudes $\overline{\rm N^{k}MHV}$. Amplitudes lying on the intersection of lines with the same $k$ are self-conjugate. The MHV point $(N,m)=(3,3)$ is special, since its $\overline{\rm MHV}$ conjugate has an exceptional description.}
\label{figgrid}
\end{figure}

For small $N$, we have listed the  allowed cases in Table~\ref{smallN}. Figure~\ref{figgrid} illustrates the structure of the families of $\cN=1$ matter superamplitudes. 


 \begin{table}[htp]
\begin{center}
\begin{tabular}{|c|c|c|c c| c c c |c c|ccc|cccc|}
\hline
$N$ & 3 & 4 & 5 &  &6 & & &7 & & 8 & & & 9  &&& \\
\hline
$R_\cA$ & 0 & 0 & 0 & 1 &0&1&2 &1 &2 &1&2&3 &1&2&3&4\\
\hline
$m$ & 3 & 2 &1 & 4 &0&3&6 &2&5 &1&4&7 &0&3&6&9 \\
\hline
\end{tabular}
\end{center}
\caption{Allowed amplitudes for small total number of particles $N=m+n$}
\label{smallN}
\end{table}%

We find it useful to make an analogy with the classification of  $\cN=4$ sYM superamplitudes according to their Grassmann degree. Schematically, the $n$-particle $\cN=4$ superamplitude has the form \cite{Drummond:2008vq}
\begin{align}\label{}
A^{\cN=4\, {\rm sYM}}_n = \delta^{(4)}(P) \, \delta^{(8)}(Q)\, \cA^{\cN=4\, {\rm sYM}}_n\,,
\end{align}
where the reduced amplitude is a polynomial in the odd variables $\eta^A$ (with $A=1\ldots4$) expanded as follows
\begin{align}\label{N4}
\cA^{\cN=4\, {\rm sYM}}_n(\eta) = \cA^{\rm MHV}_n + \cA^{\rm NMHV}_n +\ldots+ \cA^{\rm N^k MHV}_n +\ldots +\cA^{\rm \overline{MHV}}_n \,.
\end{align}
Here the ${\rm N^k MHV}$ term is a homogeneous polynomial of  Grassmann degree ${4k}$. The last term  ${\rm \overline{MHV}}= {\rm N^{n-4} MHV}$ has the maximal degree  $4(n-4)$ allowed by $\cN=4$ supersymmetry. Since the $\cN=4$ on-shell multiplet is self-conjugate under PCT, the terms in the expansion \p{N4} are pairwise equivalent via complex conjugation and Grassmann Fourier transform, $\overline{\rm N^k MHV}_n = {\rm N^{n-k-4} MHV}_n$.   In particular, the  top and bottom terms in the expansion \p{N4} form the pair $\overline{\rm MHV}_n = {\rm MHV}_n$.

Coming back to our $\cN=1$ matter amplitudes \p{43}, the Grassmann degree of the reduced amplitude equals its  R-charge. Then, by analogy with $\cA^{\cN=4\, {\rm sYM}}_n$, we call $\cA_{m,n}$ MHV-like  if $R_{\cA}=0$, NMHV-like if $R_{\cA}=1$, etc.  

From \p{e35} we see that MHV-like amplitudes exist only for $3\leq N\leq 6$, with $m=6-N, \ n=2N-6$. This  case is very special, since the reduced amplitude has Grassmann degree zero and is thus given by a single bosonic function of the particle momenta,
\begin{align}\label{MHV}
\cA^{\rm MHV}=\cI(p)  \,.
\end{align}
The conjugate $\cA^{\rm \overline{MHV}}$ amplitude has $m$ and $n$ exchanged, and, as follows from \p{e3.3}, its R-charge  $R_\cA=N-4$ takes the maximal allowed value. Consequently, $\cA^{\rm \overline{MHV}}$ is also described by the single   function $\bar \cI(p)$.  As in the MHV case,  ${\rm \overline{MHV}}$-like amplitudes exist only if $N\leq 6$ (see footnote \ref{MHVbar}).

For all the other cases $\cA$ is spanned by a basis of nilpotent $\bQ$ invariants, each coming with its own bosonic coefficient function. 
To determine their form and number we can use  supersymmetry to make the following prediction.  $Q$ supersymmetry eliminates two of the odd variables, e.g. $\eta_{N-1}$ and $\eta_N$,   and  $\bar Q$ supersymmetry another two, e.g. $\eta_1$ and $\eta_2$. The remaining variables $\eta_i$  give  rise to $N-4$ independent combinations\footnote{There exist various equivalent choices of bases of $\Xi$-invariants, see Sect.~\ref{sect5NMHV}.}
\begin{align}\label{e3.7}
&\Xi_{12i} = [12] \eta_i + [2i]\eta_1 + [i1]\eta_2  \qquad  {\rm for} \ i=3,\ldots,N-2\,.
\end{align}
They are $\bQ$-invariants,  $\bQ \Xi_{12i} = 0$ with $\bQ$ from \p{42}, as follows from the Schouten identity.

So, the NMHV-like reduced amplitude has the following generic form
\begin{align}\label{NMHV}
\cA^{\rm NMHV} = \sum_{i=3}^{N-2} \Xi_{12i}\, \cI_i(p)\,,
\end{align}
with $N-4$ independent bosonic coefficient functions $\cI_i(p)$. From \p{e35} we deduce that NMHV-like amplitudes exist only for $5\leq N\leq 9$. The case $N=5$ is in fact an ${\rm \overline{MHV}}$ amplitude, since $\overline{\rm MHV}_5 = {\rm NMHV}_5$. So, genuine NMHV-like amplitudes exist  for $6\leq N\leq 9$, with the number of functions $\cI_i(p)$ ranging from 2 to 5. Further $\cA^{\rm N^kMHV}$ are constructed by multiplying together various subsets of $k$  from the total number of $N-4$ $\Xi$-invariants.\footnote{One might think that the maximal allowed value of $k$  is always $N-4$. In reality, the R-charge of the reduced amplitude is limited from above by the inequality \p{e35}. If $N>6$ this upper bound is lower than the maximal number $N-4$ of $\Xi$ invariants. This explains  why $\cA^{\rm \overline{MHV}}$, which involves the product of all the available $\Xi$ invariants, exists only for $N \leq 6$. \label{MHVbar}}

We remark that the Grassmann structure of the $\cN=1$ amplitude resembles, to some extent, that of the $\cN=4$ sYM amplitude \p{N4}. There one constructs a basis of the so-called R-invariants, first introduced for NMHV amplitudes in \cite{Drummond:2008vq} and later generalized to all N${}^k$MHV amplitudes in \cite{Drummond:2008cr}. Our $\Xi$ invariants in \p{e3.7} are the $\cN=1$ analogs of the $\cN=4$ R-invariants. The main difference is that in the $\cN=4$ case there is no R-charge, so the various N${}^k$MHV amplitudes can appear together in the expansion of the full $n$-particle amplitude. To put it differently, the $\cN=1$ on-shell matter multiplets \p{72} are not self-conjugate under PCT. This is why the odd expansion of an $\cN=1$ amplitude with  $N$ particles also depends on  how these particles are distributed between chiral and antichiral multiplets, $N=m+n$. 

In this paper we consider the two simplest cases of MHV (or equivalently $\overline{\rm MHV}$) and NMHV-like amplitudes. We present several examples of MHV and ${\rm \overline{MHV}}$-like amplitudes in Sects.~\ref{4legs}, \ref{56leg} and \ref{6leg} and one example of an NMHV-like amplitude in Sect.~\ref{6nmhv}. 

\subsubsection{Special conformal supersymmetry}\label{s312}

Next we turn to the generators of special conformal supersymmetry, $S$ and $\bar S$. They commute with the (super)momentum deltas (see App.~\ref{appeC}), so we have
\begin{align}\label{e33}
S_\a A = \delta^{(4)}(P)\, \delta^{(2)}(Q)\, S_\a \cA\,, \qquad 
\bar S_\da  A = \delta^4(P) \delta^{(2)}(Q)\, \bar S_\da \cA\,.
\end{align}

Let us see how the symmetry is realized in the two simplest cases of MHV and NMHV-like amplitudes. We start with $S$ supersymmetry. According to \p{MHV} and \p{42}, the fermionic derivatives in $S$ annihilate the bosonic function $\cI(p)$, so the Ward identity 
\begin{align}\label{e3.11}
S_\a\, \cA^{\rm MHV} = 0
\end{align}
is trivially satisfied. 

For NMHV-like amplitudes \p{NMHV} we need the anticommutator
\begin{align}\label{twco}
\{S^\a, \Xi_{12i}\} =  [12]\frac{\pa}{\pa \la_i^\a} +[2i]\frac{\pa}{\pa \la_1^\a} + [i1]\frac{\pa}{\pa \la_2^\a} \equiv F^\a_{12i}\,.
\end{align}
Then the expected Ward identity takes the form
\begin{align}\label{e3.13}
S^\a\, \cA^{\rm NMHV} = \sum_{i=3}^{N-2} F^\a_{12i}\, \cI_i(p) =0\,.
\end{align}
In it  we see the 1st-order chiral spinor operators $F^\a_{12i}$. Interestingly, these are the so-called collinearity operator in twistor space   (cf. Eq.~(3.37) in Ref.~\cite{Witten:2003nn}). This fact is not entirely new. In Refs.~\cite{Bargheer:2009qu,Korchemsky:2009hm} it was shown that the dual $\bQ$  supersymmetry of $\cN=4$ sYM NMHV amplitudes (or equivalently, the ordinary $\bar S$ supersymmetry) amounts to a condition for twistor collinearity, if applied to the unitarity cut of the amplitude. Here we see the same relationship between $S$ supersymmetry and twistor collinearity, but now at the level of the amplitude itself and not its cuts. 

Let us now examine $\bar S$ supersymmetry for MHV-like amplitudes. The generator in \p{42} is 1st-order in the bosonic variables and involves a linear combination of the odd variables $\eta$. These are not all independent, as follows from supercharge conservation. We can solve this condition for, e.g.,
\begin{align}\label{}
\eta_1 = \frac1{\vev{12}} \sum_{i=3}^N \vev{2i}\eta_i\,, \qquad \eta_2 = \frac1{\vev{12}} \sum_{i=3}^N \vev{i1}\eta_i\, . 
\end{align}
Substituting this in the expression for $\bar S$ and equating the coefficients of the $N-2$ independent $\eta_i$ to zero, we obtain the following $\bar S$ supersymmetry Ward identities for the single function $\cI(p)$ in \p{MHV}:
\begin{align}\label{e3.15}
\bar S^\da\, \cA^{\rm MHV} \ \Rightarrow\ \tilde F_{12i}^\da\, \cI(p) = 0 \qquad {\rm for} \ i=3,\ldots,N\,.
\end{align}
Here $\tilde F$ is the antichiral conjugate of the twistor collinearity operator \p{twco}.

For a tree amplitude we expect $S_\a \cA=\bar S_\da \cA=0$, and consequently, the Ward identities \p{e3.13} and \p{e3.15} to hold exactly. However,  at loop level we encounter anomalies. The main point of this paper is to explain their origin and how to use the corresponding anomalous Ward identities. This will be done in Sect.~\ref{se3.3}. Let us first look at some examples of tree-level superamplitudes which illustrate the features seen so far.

\subsection{Three- and four-particle tree-level examples}\label{s6}

In this subsection we present very simple first examples which illustrate how a graph is composed from elementary vertices, and allows us to see the Grassmann structure due to $Q$ supersymmetry and how the superconformal Ward identities work.

The cubic vertex $\bar\Phi^3$ gives rise to the elementary three-leg MHV-like amplitude\footnote{For the three-point examples \p{46} and \p{49} we may assume space-time signature $(2,2)$, in which three-particle massless scattering is possible. }
\begin{align}\label{46}
A_{3,0}=\vev{\bar\Phi(p_1,\eta_1)  \bar\Phi(p_2,\eta_2)\bar\Phi(p_3,\eta_3) }_{\rm tree} =  \delta^{(4)}(P) \, \delta^{(2)}(Q)\,.
\end{align}
Its total R-charge $(+2)$ is carried by $\delta^{(2)}(Q)$. The reduced amplitude is trivial, $\cA_{3,0} = 1$, as follows from comparing the $\eta$ expansion with, e.g., the component tree amplitude
\begin{align}\label{}
\eta_1^0 \eta_2 \eta_3\, : \qquad \vev{\bar\phi \psi_- \psi_-}=  \vev{23}\,.
\end{align}

The cubic vertex $\Phi^3$ gives rise to the elementary three-leg $\overline{\rm MHV}$-like amplitude
\begin{align}\label{49}
A_{0,3}=\vev{\Psi(p_1,\eta_1)  \Psi(p_2,\eta_2)\Psi(p_2,\eta_2) }_{\rm tree} =  \delta^{(4)}(P) \, \Xi_{123}\,,
\end{align}
with the $\bQ$ invariant $\Xi_{123}$ defined in \p{e3.7}.
It has the right overall R-charge $3\times(1/3)=1$, as well as the expected helicity $(+1/2)$ at each point. Notice the absence of the supercharge conservation delta function $\delta^{(2)}(Q)$, whose R-charge $(+2)$ is too high. The invariance under $Q$ supersymmetry, $Q\,\Xi_{123} = 0$,  follows from the solution of the three-point kinematic condition on the momenta. As a check of \p{49}, we can consider the component 
\begin{align}\label{}
\eta_1 \eta^0_2 \eta^0_3\, : \qquad  \vev{\phi \psi_+ \psi_+}= [23]\,.
\end{align}

As explained earlier, the $\overline{\rm MHV}$-like amplitude \p{49} is equivalent to the MHV-like \p{46}. To see this, we first complex conjugate $\bar\Phi(\eta) \to \Phi(\be)$ and then we Fourier transform $\be \to \eta$ using the identity 
\begin{align}\label{}
 \Xi_{123}=\frac1{2}\int \prod_{i=1}^3 d\be_i\, e^{\sum_{j=1}^3 \eta_j\be_j} \, \delta^{(2)}\left(\sum_{k=1}^3 \tl_k \be_k  \right) \,.
\end{align}

According to Table~\ref{smallN}, for the four-leg amplitude the allowed values are $(m,n)=(2,2)$ and $R_{\cA}=0$, hence this amplitude is of the MHV type:\footnote{This amplitude is self-conjugate, ${\rm MHV}_{2,2}=\overline{\rm MHV}_{2,2}$.}
\begin{align}\label{6.6}
\vev{\bar\Phi(p_1,\eta_1)  \Psi(p_2,\eta_2) \bar\Phi(p_3,\eta_3)  \Psi(p_4,\eta_4) } = \delta^{(4)}(P) \, \delta^{(2)}(Q)\, \frac1{\vev{24}}\, \cA_{2,2}(p)\,.
\end{align}
The bosonic factor $1/\vev{24}$ gives the amplitude the required helicities at points 2 and 4.  It can be determined by comparing with the four-scalar component amplitude $\eta_1^0 \eta_2 \eta_3^0 \eta_4 \, \vev{\bar\phi\phi\bar\phi\phi}$. At tree level we find
\begin{align}\label{}
\vev{\bar\phi\phi\bar\phi\phi}^{\rm tree} = 1 \quad \Rightarrow \quad \cA_{2,2}^{\rm tree} = 1\,.
\end{align}

It is easy to see that the above tree-level amplitudes satisfy the expected superconformal Ward identities.

\subsection{Superconformal anomaly of vertex functions and loop amplitudes} 
\label{se3.3}

Let us now analyze the action of the superconformal symmetry on loop-level amplitudes. In order to do this, we need to study 
how the three types of vertex functions \p{e3.16}, \p{e3.17} and \p{e3.17'} behave under superconformal transformations, paying close attention to possible distributional effects. 

To begin with, the antichiral vertex functions considered here do not have an $\bar S$ anomaly. Such an anomaly may appear in their chiral conjugates. 

The argument in App.~\ref{appeC} why the generator $S$ commutes with the (super)momentum conservation delta functions applies equally well to correlation functions with some or all of the legs off shell. Then, counting the dimensions and the R-charges of the three off-shell superfields in  \p{e3.16}, one can check that this off-shell vertex is $S$-invariant.
 
Next, the vertex function with two on-shell legs does not present any anomaly. Indeed, the single propagator in  \p{e3.17'} does not develop a singularity that may break the (super)conformal symmetry. This is also clear at the component level. Expanding in $\eta$ we find the component Yukawa vertex 
\begin{align}\label{}
\eta_1\eta_2\, \vev{\psi_- \psi_- \bar\phi} \sim \eta_1\eta_2\, \frac{\vev{12}}{(p_1+p_2)^2}\,.
\end{align}
It is conformal and hence the super-vertex is superconformal as well. 

The situation changes radically in the case where one leg is on shell. There exists a region in the momentum space where the two off-shell momenta become collinear with the on-shell one, $q_1\sim q_2\sim p$. Then the product of propagators in \p{e3.17} becomes a singular distribution (see App.~\ref{Appendix_Derivation_Anomaly} for the detailed explanation). Consequently, the antichiral vertex function \p{e3.17} is invariant only up to a contact term (see \p{2.19}), 
\begin{align}\label{e411}
&S_\a \, \vev{\bar\Phi(q_1,\bq_1) \bar\Phi(q_2,\bq_2)|\bar\Phi(p,\eta)}_{\rm tree}\nt
  & = \frac{i \pi^2}{2} \la_{\a}\,  \int_0^1 d\xi\, \Big(\eta + [\tl\bar\q_1]\xi + [\tl\bar\q_2] \bar\xi \Big) \, 
\delta^{(4)}(q_1+\xi p)\, \delta^{(4)}(q_2+\bx p) \,,
\end{align}
where $\bx=1-\xi$. This contact term becomes relevant when $q_i$ are loop integration momenta. 
This mechanism is at the heart of our anomalous superconformal Ward identities.

\begin{figure}[t]
\begin{center}
\includegraphics[height = 3.0cm]{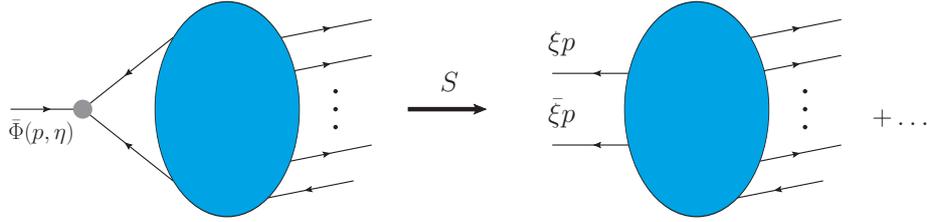}
\end{center}
\caption{The $S$-variation of a multi-loop supergraph is governed by the anomaly of the antichiral vertex function \p{e411}. We show explicitly only the contribution of one vertex, and also we omit the integration over $\xi$. }\label{S-anom-pict}
\end{figure}

In order to derive the $S$-supersymmetry Ward identity for a supergraph, we need to evaluate explicitly its $S$-variation.
Consider a generic supergraph, such as the one shown in Fig.~\ref{gengraph}. 
As explained in Sect.~\ref{sectgraphs}, the graph is built from antichiral vertex functions \p{e3.17}, which have an $S$-anomaly in the form of a contact term, see \p{e411}.   It becomes relevant due to the loop integrations over $q_i$. So, we act with $S_\a$ at each of the antichiral external legs and replace the corresponding vertex functions by their $S$-anomaly \p{e411}, see Fig.~\ref{S-anom-pict}. 

Since the anomaly of the elementary vertex is $Q$-invariant (see \p{c11}), the anomaly of the whole amplitude is $Q$-invariant as well. This allows us to write the anomaly term with the standard (super)momentum conservation prefactor $\delta^{(4)}(P) \, \delta^{(2)}(Q)$. The generator $S_\a$ lowers the Grassmann degree by one unit. The Lorentz index $\a$ of the anomaly is carried by a spinor factor $\la_{i\,\a}$ for each antichiral leg. Putting these facts together, we find the {\it anomalous Ward identity}  
\begin{align}\label{SAgeneric}
S^\a  A = \delta^{(4)}(P) \, \delta^{(2)}(Q)\,  \sum_{i} \la^\a_{i}\cA_{i}(p,\eta) \,,  
\end{align}
where the sum runs over all antichiral external legs of the graphs. The anomaly terms $\cA_{i}(p)$ are determined by Feynman integrals that have one loop less than the original ones in the graph $A$, thanks to the delta functions in \p{e411}. So, when calculating the anomaly, we gain one loop order. To be more precise, we still have to carry out the parameter integral in \p{e411}, but this is easier than doing another loop integral.

We can perform a similar analysis for $\bar{S}^\da$, exchanging the roles of the chiral and anti-chiral vertices. In this way, we obtain a Ward identity of the form
\begin{align}\label{SAbargeneric}
\bar{S}^\da  A = \delta^{(4)}(P) \, \delta^{(2)}(Q)\,  \sum_{i} \tilde{\la}^\da_{i} \tilde{\cA}_{i}(p,\eta) \,.
\end{align}
Equations (\ref{e411}),  (\ref{SAgeneric}) and (\ref{SAbargeneric}) are the main conceptual results of this paper. They are Ward identities showing how superconformal symmetry is realized on loop-level amplitudes. In the following sections we will illustrate this by many examples, and will use the symmetry to compute previously unknown amplitudes.

\begin{figure}[t]
\begin{center}
\includegraphics[height = 4.0cm]{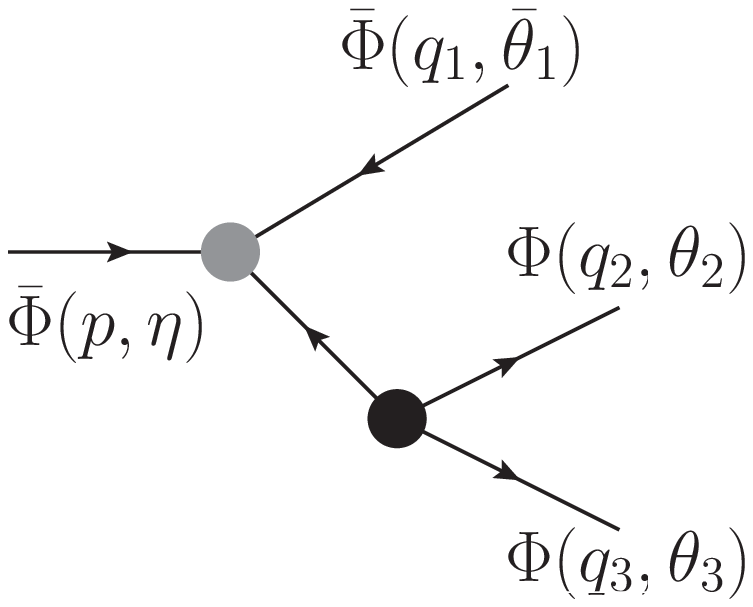} \qquad\qquad
\includegraphics[height = 4.0cm]{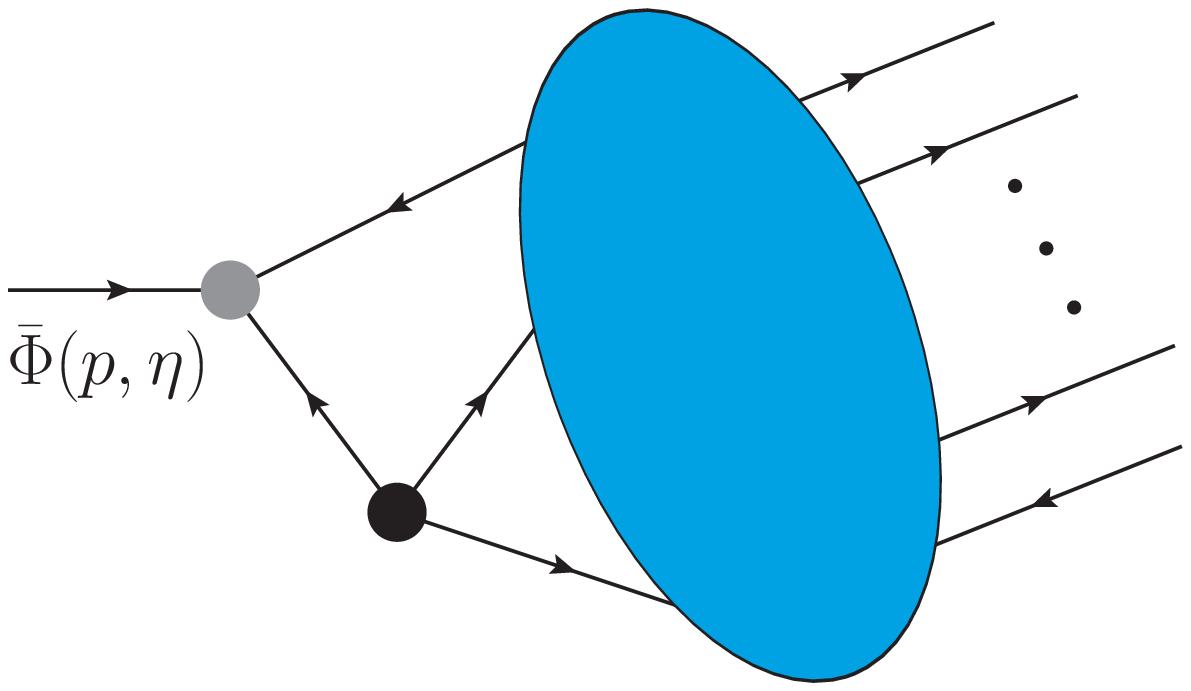}
\end{center}
\caption{Left: Tree level supergraph formed by chiral and antichiral vertices with one on-shell and three off-shell legs \p{c18}. Right: As a subdiagram of a bigger multi-loop supergraph it does not produce a new collinear $S$-anomaly with $p\sim q_1 \sim q_2 \sim q_3$.}\label{2vert}
\end{figure}

We wish to comment on a subtle point. In the derivation of the anomaly formula (\ref{SAgeneric}), we took into account the anomaly of the three-point vertex function. In a multi-loop graph, it could happen that more complicated anomaly terms exist that have support, for example, on multiple collinear configurations of several loop momenta and one external momentum. For the $\cN=1$ matter superamplitudes under consideration, we find that such `long-range' contact terms are absent. In order to see this, let us consider the four-leg diagram on the lhs of Fig.~\ref{2vert}, 
\begin{align}\label{c18}
\vev{\bar\Phi(q_1,\bq_1)\,\Phi(q_2, \q_2)\, \Phi(q_3, \q_3)|\bar\Phi(p,\eta)}\,.
\end{align}
Its three off-shell legs could depend on two independent loop momenta, when this diagram is thought of as part of a higher-loop superamplitude.
The diagram has an $S$ anomaly inherited from the antichiral vertex (the gray blob in the lhs of Fig.~\ref{2vert}), but $\bar S$ is not anomalous. In principle, we could expect a collective effect due to the collinear regime $q_1 \sim q_2 \sim q_3 \sim p$. The detailed calculation (not shown here) shows that this collective singularity does not produce an additional $S$ anomaly, owing to the Grassmann integrations.

\subsection{Four-particle MHV-like amplitudes}\label{4legs}

Four-particle scattering amplitudes are the simplest examples that 
can exist for generic configurations of loop momenta.
As we will see presently, they are a bit too simple from the point
of view of superconformal symmetry. In fact, despite the anomaly
mechanism described above, these amplitudes turn out to be exactly 
invariant. According to \p{e3.11}, this is automatically so for  the generator $S_\a$. 
In addition, also $\bar{S}_\da$ annihilates it, and this can be seen in two ways.

We already considered a four-leg example (at tree-level),
and saw that the general form of the amplitude is given by Eq.~(\ref{6.6}).
By dimensional analysis one can see that $\cA_{2,2}(p)$ is scale invariant.
As a consequence, it can only be a function of the unique scale-invariant 
variable in the problem, namely $s_{12}/s_{23}$.
Now, it turns out that any such function is annihilated by the collinearity
operators $\tilde F_{ijk}^\da$, e.g.
\begin{align}\label{}
\tilde F_{123}^\da \frac{s_{12}}{s_{23}} = \frac{\vev{12}}{\vev{23}}  \tilde F_{123}^\da \frac{[21]}{[32]}  = 0\,,
\end{align}
as a corollary of momentum conservation.

The automatic $\bar S$ invariance also follows from the fact that the four-leg MHV-like amplitude $A_{2,2}$ is equivalent to the $\overline{\rm MHV}$ one. For the latter $\bar S$ is a trivial symmetry.

At the technical level, this means that when using the anomaly formula for $\bar{S}$ (i.e. the conjugate of Eq.~(\ref{SAgeneric})), the anomaly terms evaluate to zero. One may verify that this is indeed the case.

The reader might think that this is rather disappointing: the superconformal symmetry does not seem to restrict the single-variable function $\cA_{2,2}(s_{12}/s_{23})$. We will see, however, that starting from five particles, the symmetry is very powerful. Having obtained the five-particle answer in this way, one may recover the four-particle answer as a simple corollary. For this reason we find it useful to include a four-particle example here.

\begin{figure}
\begin{center}
\includegraphics[width = 6cm]{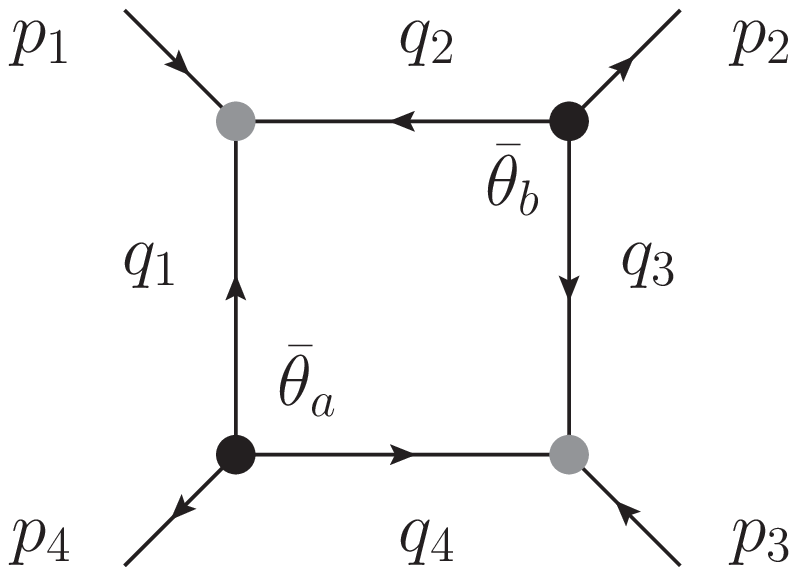}\qquad
\includegraphics[width = 6cm]{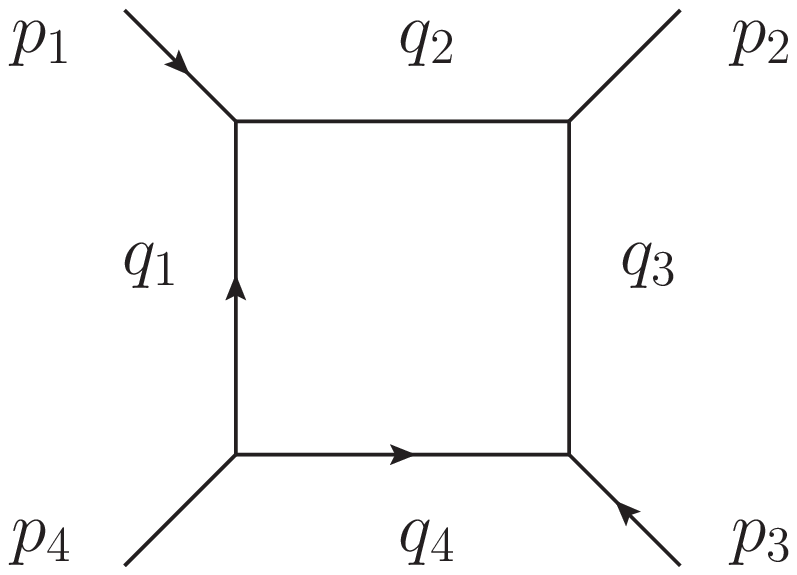}
\end{center}
\caption{Left:  One-loop superamplitude $\vev{\bar\Phi(1)  \Psi(2) \bar\Phi(3)  \Psi(4)}$. The loop propagator momenta  are denoted by $q_1,\ldots,q_4$ and their direction is chosen to coincide with chirality. All external momenta $p_1,\ldots,p_4$ are inflowing. Right: The corresponding bosonic integral in Eq.~\p{e318} is a zero-mass box with magic numerator. Arrows denote fermionic propagators and fermionic external states.} \label{bobox}
\end{figure}

At one loop we consider the Feynman diagram shown in Fig.~\ref{bobox}.
According to Sect.~\ref{sectgraphs} it is composed of two vertex functions \p{e3.17} (see also \p{219}), glued together with two wave functions $\vev{\bar\Phi(\bq) \Psi(\eta)}$ (see \p{534}), by Grassmann integrations $\int d^2\bq$ at the black points:
\begin{align}\label{e317}
&A_{2,2}^{\rm one-loop} = \delta^{(4)}(P) \, \int \frac{d^4 \ell}{i \pi^2}\, 
\frac{d^2\bq_a d^2\bq_b}{q^2_1 q^2_2 q^2_3 q^2_4}\, \, (\eta_4+[4\bq_a]) \, (\eta_2+[2\bq_b]) \, 
 \nt
&\qquad \qquad \times \, 
\delta^{(2)}( q_1 \bq_a  +  q_2 \bq_b + \ket{1}\eta_1) \, \delta^{(2)}( q_4 \bq_a  +  q_3 \bq_b + \ket{3}\eta_3)\,,
\end{align}
where the loop momentum $\ell$ is identified with one of the $q_i$.
This time it is more convenient to choose the component $\eta_1 \eta_2^0 \eta_3^0 \eta_4$  corresponding to the amplitude$\vev{\psi_-\psi_+\bar\phi\phi}$. Setting $\eta_2=\eta_3=0$ in \p{e317} and in \p{6.6}, and doing the Grassmann integrations  in \p{e317} with the help of the delta functions, we find 
\begin{align}\label{e318}
\cA_{2,2}^{\rm one-loop} = \frac{s_{24}}{\vev{13}} \int \frac{d^4 \ell}{i\pi^2}\, \frac{\bra{1} q_1 \tilde q_4 \ket{3}}{q^2_1 q^2_2 q^2_3 q^2_4}=  \frac{s_{24}}{2 s_{13}}\left[ \log^2\left(\frac{s_{12}}{s_{23}}\right) + \pi^2 \right]\,. 
\end{align}
This one-loop integral is the 4D zero-mass box with `magic' numerator, and it coincides with the 6D zero-mass box, see e.g. \cite{Dixon:2011ng}.

\section{$\overline{\rm MHV}$-like five-leg amplitudes}\label{56leg}

In this and the following sections we give examples of one- and two-loop amplitudes of the $\overline{\rm MHV}$ type. The reduced amplitude contains  a single bosonic function $\cI(p)$ of the momenta, given by some Feynman integral. We derive simple anomalous Ward identities involving this single function. These are first-order partial differential equations whose rhs is determined by the anomaly. We explain how such equations can be solved.

\subsection{General properties of five-leg amplitudes} \label{sect5NMHV}

According to Table~\ref{smallN}, the allowed values are $(m,n)=(1,4)$ or $(4,1)$, corresponding to $R_{\cA}$ charges $(+1)$ and $0$, or to NMHV and MHV types, respectively. As explained in Sect.~\ref{s311}, the two cases are equivalent, $\overline{\rm MHV}_5 = {\rm NMHV}_5$. Let us consider the NMHV-like amplitude,
\begin{align}\label{5leg}
A_{4,1} = \vev{\bar\Phi(p_1,\eta_1)  \bar\Phi(p_2,\eta_2) \bar\Phi(p_3,\eta_3)\bar\Phi(p_4,\eta_4) \Psi(p_5,\eta_5) } = \delta^{(4)}(P) \, \delta^{(2)}(Q) \, \cA_{4,1}(p,\eta)\,.
\end{align}
It has helicity $(+1/2)$ at point 5 and zero helicity at the remaining points. 

According to \p{NMHV}, the  reduced amplitude $\cA_{4,1}(\la,\tl,\eta)$  takes the general form\footnote{The bosonic function $\cI(\la,\tl)$ has helicity, see e.g. \p{422}, hence it is a function of the helicity spinors  $\la,\tl$ rather than the momenta $p=\la\tl$. However, for the sake of brevity  we use the notations $\cI(p)$ and $\cA(p)$.}
\begin{align}\label{414}
\cA_{4,1}(\la,\tl,\eta) =  \Xi_{123}\, \cI(\la,\tl)
\end{align}
with a bosonic coefficient function $\cI$ that can be determined by comparing with some component amplitude. For example,   the components $\eta_1^0 \eta_2^0 \eta_3 \eta_4 \eta_5$ and $\eta_1^0 \eta_2 \eta_3^0 \eta_4 \eta_5 $ 
 yield equivalent expressions for $\cI$:
\begin{align}\label{Campl}
\cI= \frac{\vev{\bar\phi\bar\phi \psi_- \psi_- \phi}}{\vev{45}[12]} 
= \frac{\vev{\bar\phi\psi_- 
\bar\phi \psi_- \phi}}{\vev{45}[31]} 
 \,.
\end{align}
This is an example of a supersymmetry Ward identity implying relations among the Feynman integrals that represent the various component amplitudes.  

We remark that any three $\eta$'s define a $\bar Q$ superinvariant of the type \p{e3.7},
\begin{align} \label{XIijk}
\Xi_{ijk} = \eta_i [jk] + \eta_j [ki] + \eta_k [ij] \,,
\end{align}
where any cyclic permutation of the indices is allowed, i.e. $\Xi_{ijk}= \Xi_{(ijk)}$, and swapping two indices results in a sign change, $\Xi_{ijk} = -\Xi_{ikj}$. Thus, for $N=5$ particles we have $10$ invariants which are not related by permutation of their indices.
However, all of them are equivalent due to $Q$ supercharge conservation. Indeed, from the condition $Q=0$ we can eliminate any two $\eta$'s in terms of the remaining three. 
We find the equivalence relations
\begin{align}\label{XiEquiv}
N=5: \qquad \frac{(-1)^{\sigma(i_1,\ldots,i_5)}}{\vev{i_1 i_2}} \Xi_{i_3 i_4 i_5} = \frac{(-1)^{\sigma(j_1,\ldots,j_5)}}{\vev{j_1 j_2}} \Xi_{j_3 j_4 j_5}\,,
\end{align}
where $\{i_1,i_2,i_3,i_4,i_5\}$ and $\{j_1,j_2,j_3,j_4,j_5\}$ are permutations of the set $\{1,2,3,4,5\}$ and $(-1)^\sigma$ denotes the permutation sign.
Thus, choosing another superinvariant in \p{414}  will only slightly modify the definition of the bosonic function $\cI$.    

\subsection{Anomalous superconformal Ward identity}\label{s4.2}

Here we establish the generic form of the superconformal Ward identity for the five-leg amplitude \p{5leg}. We start with the $S_{\a}$ transformations of $A_{4,1}$. 
In view of \p{e33} we push the generator $S_{\a}$ through the total momentum and supercharge conservation delta functions. Then we apply the identity \p{e3.13} and deduce  the lhs of the Ward identity 
\begin{align}\label{515}
S^{\a} A_{4,1} = \delta^4(P) \, \delta^{(2)}(Q)\, F^{\a}_{123} \cI\,.
\end{align}
The freedom in choosing the superinvariant $\Xi_{ijk}$ (see \p{XiEquiv}) also affects the collinearity operator $F_{ijk}$ in \p{515} and modifies the bosonic function $\cI$ according to \p{XiEquiv}. We have
\begin{align}\label{KC}
(-1)^{\sigma(i_1,\ldots,i_5)} \, F_{i_1 i_2 i_3} \frac{\cI}{\vev{i_4 i_5}} = 
(-1)^{\sigma(j_1,\ldots,j_5)}\, F_{j_1 j_2 j_3} \frac{\cI}{\vev{j_4 j_5}}\,,
\end{align}
where we impose momentum conservation after the differentiation.

In order to determine the rhs of the Ward identity we need to evaluate explicitly the $S$-variation of the supergraph $A_{4,1}$. 
We do this following the steps outlined in section \ref{se3.3}.
In the present case, we get contributions from the anomalies corresponding to external legs $i=1,2,3,4$, so that we have
\begin{align}\label{SA}
S^\a  A_{4,1} = \delta^{(4)}(P) \, \delta^{(2)}(Q)\,  \sum_{i=1,2,3,4} \la^\a_{i}\cA_{i}(p) \,.
\end{align}
If in general the anomaly terms may have a Grassmann dependence, in this case 
taking into account the total Grassmann degree of $A_{4,1}$ one sees that the
$\cA_{i}(p)$ are bosonic functions. The latter are determined by Feynman integrals that have one loop less than the original ones in the graph $ A_{4,1}$, 
thanks to the delta functions in \p{e411}. 

Comparing the rhs of \p{SA}  and \p{515}, we arrive at the generic anomalous superconformal Ward identity 
\begin{align}\label{e513}
F_{123}^\a\, \cI = \sum_{i=1,2,3,4} \la^\a_{i}\cA_{i} \,,
\end{align}
where we have the liberty of choosing the lhs according to \p{KC}. This freedom implies additional consistency relation for the anomaly on the rhs of \p{e513}. Indeed, since any pair of collinearity operators commute, 
$[F_{123}^\a, F_{j_1 j_2 j_3}^\b] =0$, we conclude that the following spinorial equation should hold
\begin{align} \label{consist5}
F_{j_1 j_2 j_3}^\b \sum_{i=1,2,3,4} \la^\a_{i}\cA_{i}  = 
\frac{(-1)^{\sigma(j_1\ldots j_5)}}{\vev{45}}F_{123}^\a \, \vev{j_4 j_5}\sum_{i=1,2,3,4} \la^\b_{i}\cA_{i}  \,,
\end{align}
where the notation is explained around Eq.~\p{XiEquiv}.
We evaluate the anomaly functions $\cA_i$ using Feynman supergraphs and the consistency relation \p{consist5} serves as a strong cross-check of the calculation.

Despite the fact that the supergraph $A_{4,1}$ \p{5leg} contains one chiral leg, there is no $\bar{S}$-anomaly.
We act on the amplitude \p{5leg} with $\bar{S}_{\da}$ taking into account \p{SbarA}, and 
we want to show that 
\begin{align}\label{}
\bar S^\da \, A_{4,1} = \delta^4(P)\,  \delta^{(2)}(Q)\, \bar S^\da \, [\Xi_{123}\, \cI]=0 \,.
\end{align}
Counting dimension, helicity and R-charge as in \p{e47}, one sees that $\{\bar Q, \bar S\}\, [\Xi_{123}\, \cI]=0$. This allows us to set $\eta_2=\eta_3=0$ and reduce $\Xi_{123} \sim \eta_1$. Then, with the help of $\delta^{(2)}(Q)$ we can eliminate $\eta_4, \eta_5$ from $\bar S$, reducing it also to $ \sim \eta_1$. Thus $\bar S^\da \, [\Xi_{123}\, \cI] \sim (\eta_1)^2=0$. 

The absence of an $\bar S$-anomaly for NMHV${}_5 = \overline{\rm MHV}_5$ is equivalent to the absence of an $S$-anomaly for MHV${}_5$. Table~\ref{smallN} determines the form  of the latter
\begin{align}\label{4.29}
A_{1,4}=\vev{\bar\Phi(p_1,\eta_1)\Psi(p_2,\eta_2)  \Psi(p_3,\eta_3) \Psi(p_4,\eta_4)\Psi(p_5,\eta_5)  } = \delta^{(4)}(P) \, \delta^{(2)}(Q)\, \bar \cI\,.
\end{align}
According to \p{e33}, the generator $S$ \p{42} goes through the delta functions and hits the reduced amplitude $\bar \cI$, which has no Grassmann dependence.

We now show two explicit examples of five-leg integrals of the above type. In Sect.~\ref{sec_1loop} we consider a one-loop box example,
and in Sect.~\ref{sec_dblbox}  a two-loop example of the double-box topology. In Ref.~\cite{Chicherin:2018ubl} we presented a more complicated two-loop example with nonplanar hexa-box topology.

\subsection{Five-leg one-loop box 
} \label{sec_1loop}

\begin{figure}
\begin{center}
\includegraphics[height = 4cm]{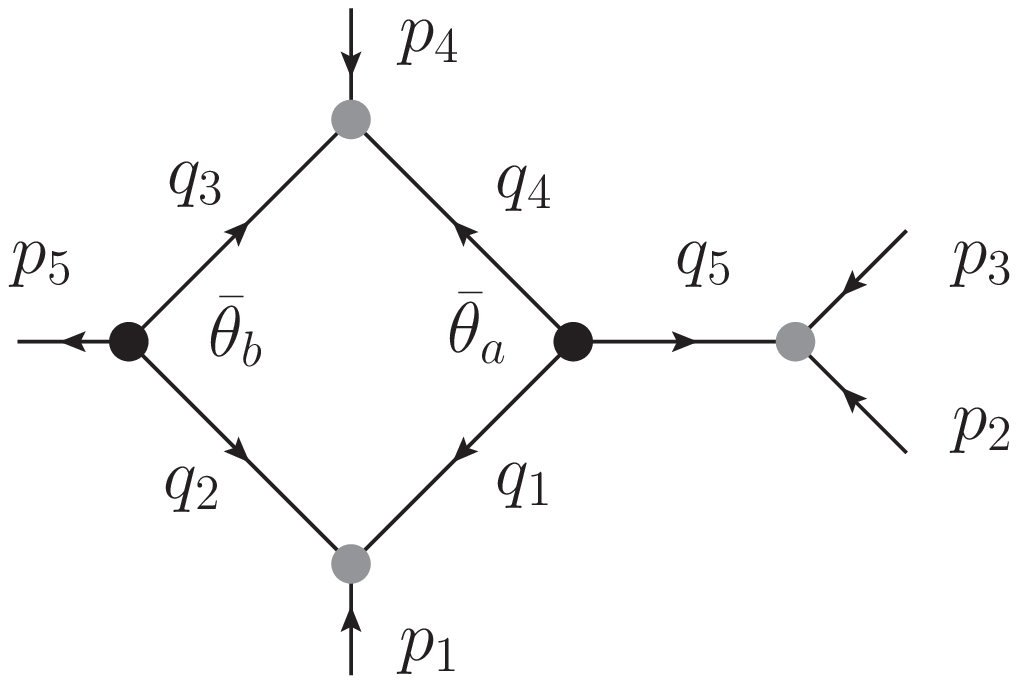}\qquad
\includegraphics[height = 4cm]{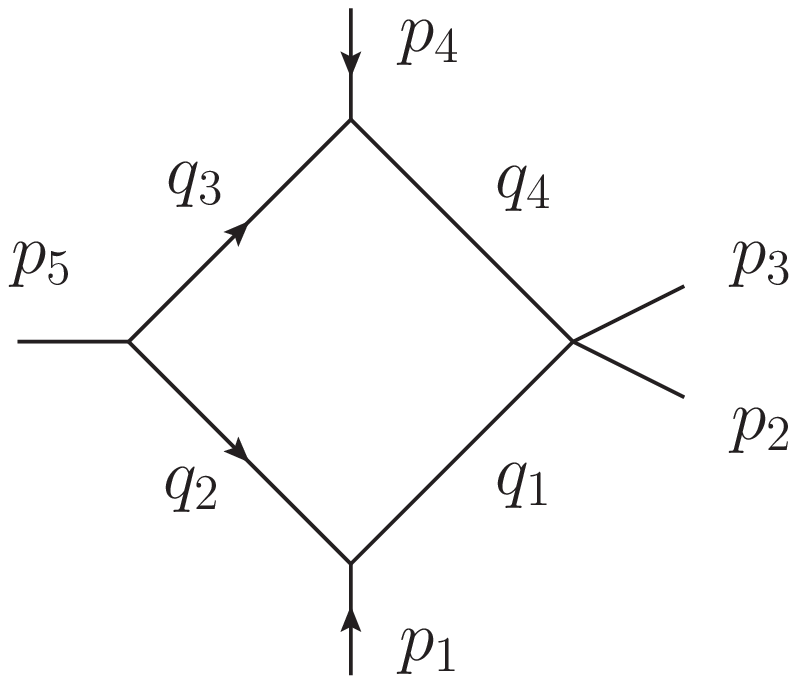}
\end{center}
\caption{Left: One-loop NMHV$_5$ on-shell supegraph $\vev{\bar\Phi(1)\bar\Phi(2)\bar\Phi(3)\bar\Phi(4)\Psi(5)}$. 
Right: The corresponding bosonic integral $\cal I$, Eq.~\p{422}, is a one-mass box with magic numerator.} \label{fig1}
\end{figure}

Let us consider the one-loop supergraph in Fig.~\ref{fig1} contributing to the amplitude \p{5leg}. This example was briefly outlined in Ref.~\cite{Chicherin:2018ubl}, and here we present the detailed calculation. The corresponding expression for the supergraph can easily be worked out using the Feynman rules from Section~\ref{sectgraphs}. We put together the 3-leg antichiral vertices \p{e3.17}, \p{e3.17'} with one or two on-shell legs and the wave function \p{534} at  leg 5. This accounts for the Grassmann integration at the gray points  in  Fig.~\ref{fig1}. What is left to do are the integrals over the antichiral odd variables at the black points, 
\begin{align}\label{420}
&A_{4,1} = \delta^{(4)}(P) \, \int \frac{d^4 \ell}{i\pi^2}
\frac{d^2\bq_a d^2\bq_b}{q_1^2 q_2^2 q_3^2 q_4^2 q_5^2}\, \, (\eta_5+[5\bq_b]) \, 
\delta^{(2)}( \bq_a \tilde q_5 + \ket{2}\eta_2 + \ket{3}\eta_3) \nt
&\qquad \qquad \times\delta^{(2)}(\bq_a \tilde q_1 + \bq_b \tilde q_2 + \ket{1}\eta_1)\, \delta^{(2)}(\bq_a \tilde q_4 + \bq_b \tilde q_3 + \ket{4}\eta_4)\,.
\end{align}

Like in the calculation in Sect.~\ref{4legs}, we take into account the form of the odd part of the NMHV$_5$ supergraph, Eqs.~\p{5leg} and \p{414}, and extract the component $\eta_1 \eta_2^0 \eta_3^0 \eta_4\eta_5$ in \p{420}, i.e. $\vev{\psi_- \bar\phi \bar\phi \psi_- \phi}$. In this way we find that \p{420} is  equal to 
\begin{align}\label{e416}
A_{4,1} = \delta^{(4)}(P)\,\delta^{(2)}(Q)\, \Xi_{123}\, \cI\,,
\end{align}
where the nontrivial bosonic part of the supergraph is given by the following 4D Feynman integral of the box topology with `magic' numerator and with one massive corner, 
\begin{align}\label{422}
&\cI = \frac{1}{\vev{45} [23]} 
\int \frac{d^4 \ell}{i\pi^2}\frac{\bra{1}q_2 \tilde q_3\ket{4}}{q^2_{1} q^2_{2} q^2_{3} q^2_{4}}\,.
\end{align}
The expression for this integral is well known (see Eq.~\p{e428} below). Below we will show how to rederive it using the superconformal symmetry of our model.

\begin{figure}
\begin{center}
\includegraphics[height = 3cm]{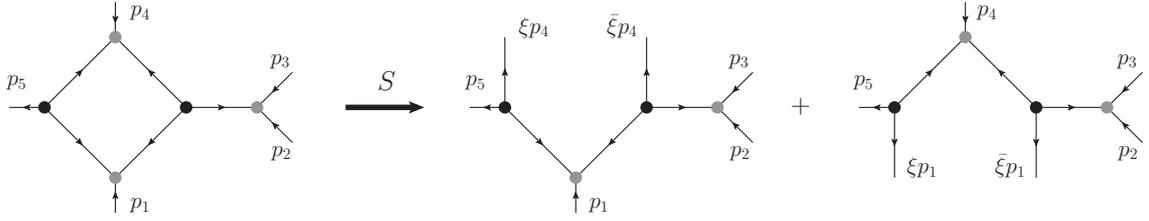}
\end{center}
\caption{The $S$-variation of the supergraph \p{420} contains two contributions: $\cA_1$ and $\cA_4$, see Eqs. \p{WIboxtail} and \p{5.1}. They originate from the $S$-anomaly of the antichiral vertex \p{e411}. We omit the $\xi$-integration in the picture.} \label{fig7}
\end{figure}

In order to write down the anomalous Ward identity \p{SA} explicitly, we need to evaluate the bosonic anomaly functions $\cA_{i}(p)$.
To this end we replace, in turn, the antichiral vertices adjacent to the external legs  1 and 4 in Fig.~\ref{fig1} by their $S$-anomaly according to \p{e411} (legs 2 and 3 do not lead to
an anomaly, see Sect.~\ref{se3.3}). Then we implement the loop integration with the help of the delta functions in \p{e411}. We find  (see Fig.~\ref{fig7})
\begin{align}\label{WIboxtail}
S^\a & A_{4,1} = \delta^{(4)}(P) \delta^{(2)}(Q)\left( \la^\a_1 \cA_1(p) + \la^\a_4 \cA_4(p) \right)\,,
\end{align}
where {
\begin{align}\label{5.1}
& \delta^{(2)}(Q) \cA_1  =  \int_0^1 d\xi   \, 
 \frac{1}{q^2_{3} q^2_{4} q_{5}^2 } \, \int d^2\bq_a\, d^2\bq_b\,  
 (\eta_5+[5\bq_b]) \, (\eta_1 + \xi [1\bq_a]   + \bx [1 \bq_b] )\nt 
&\qquad \times \delta^{(2)}(\bq_a \tilde q_5 + \ket{2}\eta_2 + \ket{3}\eta_3)
\, \delta^{(2)}(\bq_a \tilde q_4 + \bq_b \tilde q_3 + \ket{4}\eta_4) \,,
\end{align}
and similarly for $\cA_4$.
To determine $\cA_1$ it is convenient to set, e.g., $\eta_{1,2,3}=0$ in \p{5.1}, so that $\delta^{(2)}(Q) = 2 \vev{45} \eta_4\eta_5$. After some simple algebra we find 
\begin{align}\label{52}
  \cA_{1} 
&=  \frac{ 1}{\vev{15}}\int_0^1 \, \frac{d\xi}{(\xi p_1 + p_4 +p_5)^2 } 
= \frac{1}{\vev{15} }\, \, \frac{\log (s_{23}/s_{45})}{s_{23}-s_{45}}  \,.
\end{align}
The second contribution to the anomaly $\cA_4$  is obtained by permuting legs $1\leftrightarrow4\,, \ 2\leftrightarrow3$.

According to \p{e513} the lhs of Eq.~\p{WIboxtail} can be rewritten in terms of the collinearity operator $F_{123}^\a$. The complete superconformal Ward identity takes the form 
\begin{align}\label{e529}
F_{123}^\a \, \cI = \frac{\la_1^\a}{\vev{15} }\, \, \frac{a_1(p)}{s_{23}-s_{45}} +
\frac{\la_4^\a}{\vev{45} }\, \, \frac{a_4(p)}{s_{23}- s_{15}}\,,
\end{align}
where $a_1$ and $a_4$ are pure weight-one functions  
\begin{align}
a_1(p) =\log (s_{23}/s_{45}) \; ,\qquad 
a_4(p) = \log (s_{23}/ s_{15}) \,. \label{a1a4}
\end{align}
It is easy to check that $\cA_1,\,  \cA_4$ satisfy the consistency relation \p{consist5}.

The anomaly functions \p{e529}, \p{a1a4} have discontinuities at, e.g., $s_{45}=0$, etc. They correspond to the so-called `holomorphic anomaly' \cite{Cachazo:2004by}.  When applied to a unitarity cut of the integral \p{422}, the collinearity operator produces contact terms. 
 In App.~\ref{appE} we  show that the holomorphic anomaly of the cuts of $\cI$ reproduces exactly the discontinuities of the anomaly functions.

Now we want  to find $\cI$ by solving the differential equation \p{e529}.
It is convenient to introduce the helicity-free function 
\begin{align} \label{Cf}
f(p) = [14] \vev{45} [23] \, \cI = [14] \int \frac{d^4 \ell}{i\pi^2}\frac{\bra{1}q_2 \tilde q_3\ket{4}}{q^2_{1} q^2_{2} q^2_{3} q^2_{4}}\,.
\end{align}
This is a one-loop Feynman integral \p{422}  normalized to have unit leading singularity. Consequently $f(p)$ should be a pure function, i.e. it should be a $\mathbb{Q}$-linear combination of iterated integrals.  In particular $f(p)$ is dimensionless. 

We identify the relevant kinematic variables of the problem as the three Mandelstam invariants $s_{14}, s_{15}, s_{45}$. The  dimensionless function $f(p)$ depends only on their ratios. We can choose $f=f(x_1,x_2)$ with \cite{Chicherin:2018ubl}
\begin{align}\label{e4.22}
x_1 = -1 -\frac{s_{14}}{s_{15}} \;,\qquad 
x_2 = -1 -\frac{s_{14}}{s_{45}}\,.
\end{align}
Then we project  the spinor equation \p{e529} on $\la_1^\a$ and $\la_4^\a$, replace $\cI$ by the pure function \p{Cf} and obtain the differential equation
\begin{align} \label{DEx1x2}
d f(x_1,x_2) = a_1(x_1,x_2) d \log x_1 + a_4(x_1,x_2) d \log x_2 \,.
\end{align} 
The pure functions $a_1$ and $a_4$ \p{a1a4} take the following form in the new variables,
\begin{align}
a_1(x_1,x_2) = \log \frac{1+x_1}{1-x_1 x_2} \;,\qquad
a_4(x_1,x_2) = \log \frac{1+x_2}{1-x_1 x_2}\,. \label{a1a4x1x2}
\end{align} 
The consistency relation \p{consist5} for the Ward identity \p{e529} is translated into the integrability condition for the inhomogeneous differential equation \p{DEx1x2}
\begin{align}
d^2 f =0 \ \Rightarrow \ x_2 \pa_{x_2} \, a_1  = x_1 \pa_{x_1} \, a_4 \,. \label{integrab}
\end{align} 
Using \p{a1a4x1x2} one immediately verifies \p{integrab}.

Equation \p{DEx1x2} is easy to solve. Integrating its rhs  along a path connecting, e.g., the points $(-1,-1)$ and $(x_1,x_2)$, we find $f(x_1,x_2)$ up to an arbitrary constant. The latter is fixed by a natural boundary condition. Indeed, the Feynman integral $\cI$ should be finite at $[14] \to 0$, i.e. $s_{14}\to0$. Then the function $f(p)$ \p{Cf} has to vanish in this limit,    $f(-1,-1) = 0$. We also observe that $a_1(x_1,-1) = a_4(-1,x_2) = 0$. Consequently,   
\begin{align}
f(x_1,x_2) = \int^{x_1}_{-1} \frac{d t}{t} a_1(t,x_2) = \int^{x_2}_{-1} \frac{d t}{t} a_2(x_1,t) \,.
\end{align}
Doing the integration we find the explicit expression for the
Feynman integral \p{Cf}, 
\begin{align}\label{e428}
f(x_1,x_2) ={\rm Li}_2(x_1 x_2) - {\rm Li}_2(-x_1) - {\rm Li}_2(-x_2) + {\rm Li}_2(1)\,, 
\end{align}
which coincides with the finite part of the one-mass box integral.

\subsection{Five-leg double box 
} \label{sec_dblbox}

\begin{figure}
\begin{center}
\includegraphics[width = 7cm]{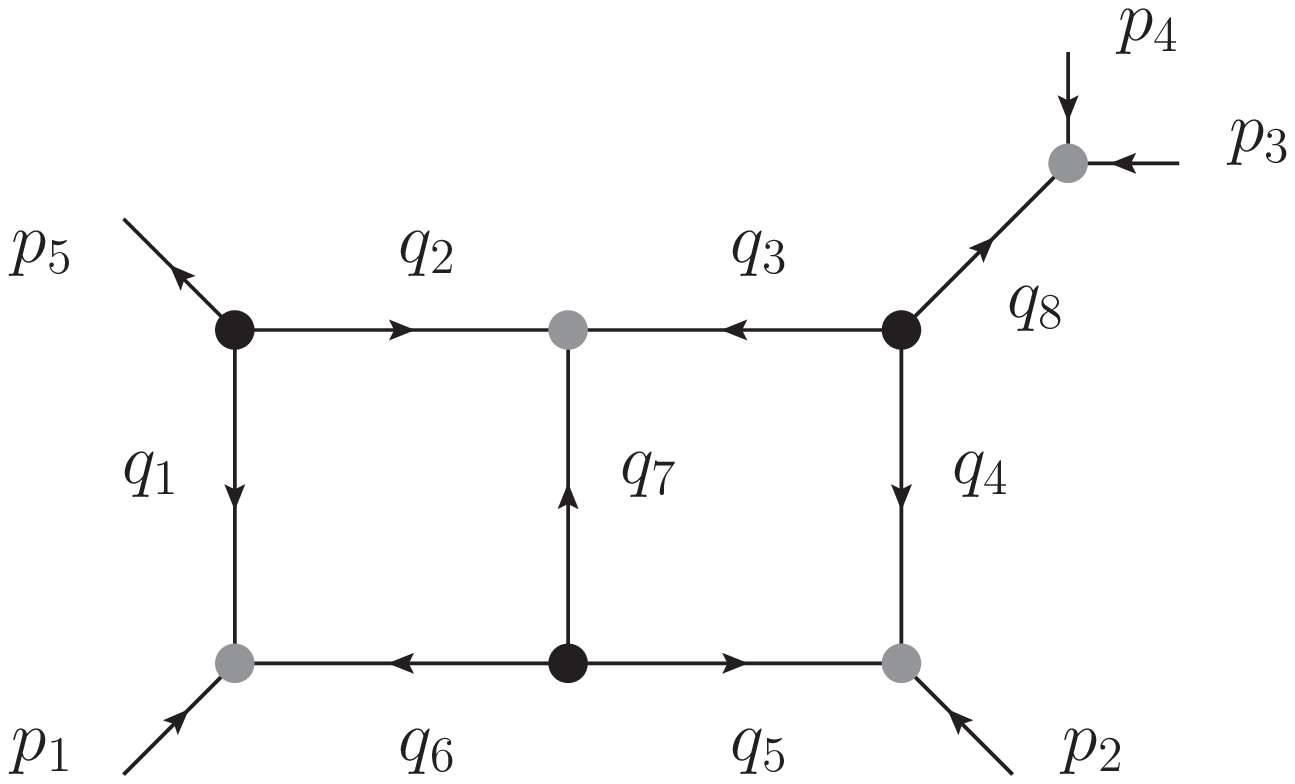} \quad
\includegraphics[width = 6.4cm]{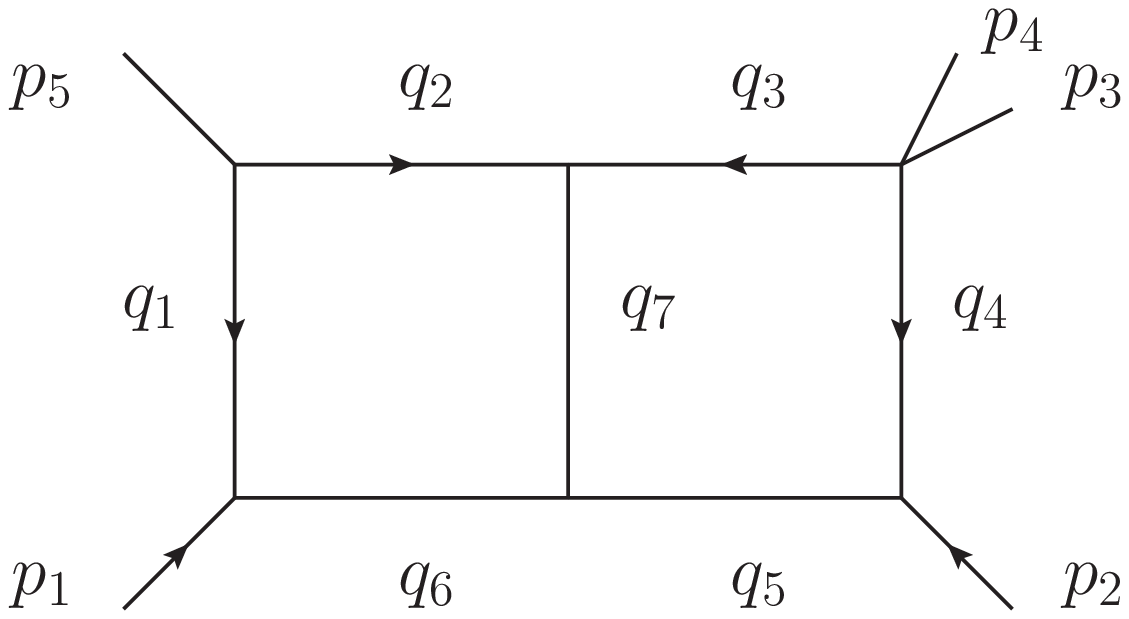}
\end{center}
\caption{Left: Two-loop ${\rm NMHV}_5$ on-shell supergraph of the double-box topology contributing to the amplitude $\vev{\bar\Phi(1)\bar\Phi(2)\bar\Phi(3)\bar\Phi(4)\Psi(5)}$. 
Right: The corresponding bosonic integral $\cal I$, Eq.~\p{Cdblbox}, of the double-box topology with one massive leg.} \label{dblbox}
\end{figure}

A more involved example of a five-leg NMHV amplitude is the two-loop supergraph on the left of Fig.~\ref{dblbox}. It has the topology of a double box with a tail (massive leg) attached. Its expression can be obtained from the Feynman rules in Sect.~\ref{sectgraphs}. The generic form of the supergraph compatible with supersymmetry is  (recall  \p{5leg} and \p{414})  
\begin{align}
A_{4,1} = \vev{\bar\Phi(1)  \bar\Phi(2)   \bar\Phi(3)\bar\Phi(4) \Psi(5) } = \delta^{(4)}(P) \, \delta^{(2)}(Q) \, \Xi_{345} \,  \cI\,.
\end{align}
To find the bosonic Feynman integral $\cI$ from the supergraph we extract its component $\eta_1 \eta_2 \eta_3^0 \eta_4^0 \eta_5$ corresponding to the component amplitude $\vev{\psi_- \psi_- \bar\phi \bar\phi \psi_+}$. In this way we obtain
\begin{align}
\cI = - \frac{1}{\vev{12}[34]} \int \frac{d^4 \ell_1\, d^4 \ell_2}{(i\pi^2)^2} \, \frac{\vev{1|q_1 \tilde q_2 q_3 \tilde q_4|2}}{\prod_{i=1}^7 q_i^2  }\,, \label{Cdblbox}
\end{align}
which is a double box with a numerator and a massive leg, see the rhs of Fig.~\ref{dblbox}. In principle, all master integrals for the one-mass double-box topology are known \cite{Gehrmann:2000zt}, and the integral \p{Cdblbox} can be evaluated by expanding it in this basis. Here we wish to derive the explicit expression for this Feynman integral directly from superconformal symmetry.

The superconformal Ward identity for the supergraph in Fig.~\ref{dblbox} has an anomaly originating from the antichiral legs 1 and 2 (as before, the vertex function with on-shell antichiral legs 3 and 4 forming the tail does not have an anomaly),
\begin{align}\label{WI12}
S^\a  A_{4,1} = \delta^{(4)}(P) \, \delta^{(2)}(Q)\,   \left( \la_{1}^{\a}\cA_{1}  + \la_{2}^{\a}\cA_{2}  \right) .
\end{align}
The anomaly contributions $\cA_{1}$ and $\cA_2$ are obtained by replacing the antichiral vertex functions adjacent to legs $1$ and $2$ by their anomalies \p{e411}.

After the Grassmann integrations the anomaly contribution from leg 1 is given by a one-parameter integral of the scalar three-mass triangle ${\rm Tri}$ defined  in \p{3mass}, 
\begin{align}
&\cA_1 = - \frac{1}{\vev{15}} 
\int^1_0 d\xi\; {\rm Tri}(s_{34},\xi s_{12}, \bar\xi s_{15})\,.\label{A1tri}
\end{align}
The explicit expression for ${\rm Tri}$   involves square roots of the Mandelstam invariants. They can be rationalized by a rational change of variables $\zeta=\zeta(\xi;s_{12},s_{15},s_{34})$ that facilitates the one-fold integration in \p{A1tri}. Instead, we prefer to introduce a Feynman parameter representation for 
${\rm Tri}$ in Eq. \p{A1tri},
\begin{align}
&\cA_1 = - \frac{1}{\vev{15}} \int^1_0 d\xi \int 
\frac{d\alpha_1 d\alpha_2 d\alpha_3 \, \delta(\alpha_1 + \alpha_2 + \alpha_3 -1)}{\alpha_1 \alpha_2 s_{34} + \alpha_2\alpha_3 \xi s_{12} + \alpha_1 \alpha_3 \bar\xi s_{15}}\,, \label{xialph}
\end{align}
where the Feynman parameters are integrated over the domain  $\alpha_1,\alpha_2,\alpha_3 \geq 0$. Then we change the order of integrations in \p{xialph}, integrating first over $\xi$ and then over $\a_i$. In this way no square roots arise at any step of the calculation. The iterated integral is linear reducible and is given by Goncharov polylogarithms,
\begin{align}
&\cA_1 = \frac{a_1(p)}{\vev{15}(s_{25}-s_{34})}\,,
\label{A1a1}
\end{align}
where $a_1(p)$ is a weight-3 pure function. We refrain from showing its explicit expression.

Similarly, after the Grassmann integrations the anomaly contribution from leg 2 is given by 
a $\xi$-parameter  integral of the one-mass box Feynman integral `${\rm box}$'  with a numerator defined by \p{boxnum},
\begin{align}
&\cA_2 =  -\frac{1}{\vev{15}} \int^1_0 \frac{d\xi}{\xi s_{34}+ \bar\xi s_{15}} 
{\rm box}(\xi p_2,p_1,p_5)\,.  \label{A2box}
\end{align}
Substituting the explicit expression \p{boxexpl} for `${\rm box}$' in \p{A2box}, we obtain the   anomaly function 
\begin{align}
\cA_2 & =  -\frac{1}{\vev{25}} \int^1_0 \frac{d\xi}{\xi s_{34}+ \bar\xi s_{15}} 
\biggl[ {\rm Li}_2\left( 1 - \frac{\xi s_{34} + \bar\xi s_{15}}{\xi s_{12}} \right) 
+ {\rm Li}_2\left( 1 - \frac{\xi s_{34} + \bar\xi s_{15}}{s_{15}} \right) \nt
&  +\frac{1}{2}\log^2 \left(\frac{\xi s_{12}}{s_{15}}\right) + \frac{\pi^2}{6}\Bigr] = 
 \frac{a_2(p)}{\vev{25}(s_{15} - s_{34})}\,, \label{A2a2}
\end{align}
where $a_2(p)$ is a weight-3 pure function resulting from the $\xi$ integration.

The kinematics of the double-box Feynman integral  is specified by two dimensionless ratios of the three Mandelstam invariants $s_{12},\,s_{15},\,s_{34}$, e.g. 
\begin{align} \label{xy}
x = \frac{s_{12}}{s_{34}} \;\;,\;\; 
y = \frac{s_{15}}{s_{34}}\,.
\end{align}
The pure functions $a_1 = a_1(x,y) $ and $a_2 = a_2(x,y) $ in Eqs.~\p{A1a1} and \p{A2a2} are given by iterated integrals corresponding to the following six-letter alphabet 
\begin{align}
\{ x , y , 1 - x - y , 1 - x ,  1 - y , x + y \}\,. \label{2dHPL}
\end{align}

Taking into account \p{twco} we rewrite the superconformal Ward identity \p{WI12}
in terms of the collinearity operator,
\begin{align}\label{e550}
F^\a_{345} \, {\cal I} = 
\frac{\la^{\a}_1}{\vev{15}} \frac{a_{1}(p)}{(s_{25}-s_{34})} + \frac{\la^{\a}_2}{\vev{25}}\frac{a_{2}(p)}{(s_{15}-s_{34})}\,.
\end{align}
We can further simplify this equation by properly normalizing the integral ${\cal I}$. We introduce the helicity-free dimensionless function  
\begin{align}   
f(x,y) = s_{12}[34]\, \cI = 
[21]\int \frac{d^4 \ell_1\, d^4 \ell_2}{(i\pi^2)^2} \, \frac{\vev{1|q_1 \tilde q_2 q_3 \tilde q_4|2}}{\prod_{i=1}^7 q_i^2  } \,.\label{fdblbox}
\end{align}
A simple calculation reveals that it has a unit leading singularity. 

Switching to the  variables \p{xy} in the Ward identity \p{e550} we find that it takes the following total differential form 
\begin{align}
d f(x,y) = a_1(x,y)\, d \log \left(\frac{x+y}{y}\right) + a_2(x,y)\,  d \log \left(\frac{1-y}{1-x-y}\right) \,.  \label{DEdblbox}
\end{align}
This differential equation can be easily integrated after choosing the appropriate boundary condition.
If we are looking for the symbol ${\cal S}(f)$ of the pure integral \p{fdblbox} then \p{DEdblbox} already provides the solution,
\begin{align}
{\cal S}(f) = {\cal S}(a_1) \otimes \log \left(\frac{x+y}{y}\right) + {\cal S}(a_2) \otimes \log \left(\frac{1-y}{1-x-y}\right)\,. \label{Sf}
\end{align}

Working at the symbol level we can easily demonstrate the power of the consistency relations \p{consist5}. The commutativity of the collinearity operators is equivalent to $d^2 f = 0$. Thus the differential equation \p{DEdblbox} implies 
\begin{align}
d {\cal S}(a_1) \wedge d \log \left(\frac{x+y}{y}\right) + d {\cal S}(a_2) \wedge d \log \left(\frac{1-y}{1-x-y}\right)= 0\,. \label{consist_symb}
\end{align}
The symbols of the anomalies $a_i$ are of weight three. The allowed first entries of the symbols are $x$ and $y$ since the Feynman integral could have nonzero discontinuities only in these letters of the alphabet \p{2dHPL}. There are 18 weight-three integrable symbols with the first entires $x$ and $y$. Substituting the weight-three ansatz of integrable symbols for $\cS(a_1)$ and $\cS(a_2)$, which contains $2\times 18$ free coefficients, in Eq.~\p{consist_symb} we fix $\cS(a_1)$ and $\cS(a_2)$ up to 7 free coefficients. Then \p{Sf} provides the symbol of the integral \p{fdblbox} up to 7 free coefficients. 
Thus without any detailed supergraphs calculations of the anomaly we could put severe constraints on the integrals.

In conclusion, let us also mention that using the same method,  in \cite{Chicherin:2018ubl} we calculated a much more nontrivial five-particle  two-loop non-planar hexa-box integral.

\subsection{Interpretation of the anomaly as a collinear limit of a six-leg amplitude}

Here we interpret the anomaly \p{5.1} according to the schematic relation `one leg more, one loop less' in \p{equation_superconformal_ward_identity_intro}.   Let us cut the graph in Fig.~\ref{fig1} through the lines $q_1$ and $q_2$, remove the lower corner and create a six-leg tree amplitude. The second graph on the rhs of Fig.~\ref{fig7} illustrates the procedure.  This tree amplitude of Grassmann degree 3 has two new chiral legs $\Psi(p_{1'},\eta_{1'})$ (attached to the vertex $\bq_a$) and $\Psi(p_{1''},\eta_{1''})$ (attached to the vertex $\bq_b$):
\begin{align}\label{e4.21}
A_{3,3}^{\rm tree} &= \vev{\Psi(p_{1''},\eta_{1''}) \Psi(p_{1'},\eta_{1'})  \bar\Phi(p_2,\eta_2) \bar\Phi(p_3,\eta_3)\bar\Phi(p_4,\eta_4) \Psi(p_5,\eta_5) }_{\rm tree}\nt &= \frac{1}{q^2_{3} q^2_{4} q_{5}^2 } \, \int d^2\bq_a\, d^2\bq_b\,  
 (\eta_5+[5\bq_b]) \,  (\eta_{1''}+[1''\bq_b]) \,  (\eta_{1'}+[1'\bq_a]) \nt 
&\qquad \times \delta^{(2)}(\bq_a \tilde q_5 + \ket{2}\eta_2 + \ket{3}\eta_3)
  \delta^{(2)}(\bq_a \tilde q_4 + \bq_b \tilde q_3 + \ket{4}\eta_4) \,.
\end{align}

To obtain the five-leg anomaly \p{5.1} of Grassmann degree 2, we need to make the following steps: (i) take the collinear limit $p_{1'}\sim p_{1''} \sim p_1$ in a way asymmetric between $\la$ and $\tl$:  $\la_{1'}=\la_{1''}=\la_1$ and $\tl_{1'}=\xi \tl_1$, $\tl_{1''}=\bx \tl_1$, so that $p_1'+p_1''=p_1$; (ii) integrate over $\xi$; (iii) eliminate one of the new odd variables by making the change $\eta_1= \eta_{1'}+\eta_{1''}\,, \ \eta= \eta_{1'}-\eta_{1''}$ and then integrating out $\eta$:
\begin{align}\label{}
 \int_0^1 d\xi   \, \int d\eta\, (\frac1{2}(\eta_1-\eta) + \xi  [1\bq_b]) \, (\frac1{2}(\eta_1+\eta) + \bx [1\bq_a])   \,    (\bullet) = \int_0^1 d\xi   \, (\eta_1 + \xi [1\bq_a]   + \bx [1 \bq_b] ) (\bullet) \,.
\end{align}
Note that this procedure does not predict the overall spinor factor $\la^\a_1 $ of the anomaly. The anomaly associated with the other antichiral leg is treated likewise. We conclude by rewriting \p{equation_superconformal_ward_identity_intro} in the more explicit form
\begin{align}\label{}
S^\a A_{4,1}^{\rm one\, loop} = \la^\a_1 \int_0^1 d\xi   \, \int d(\eta_{1'}-\eta_{1''})\, \lim_{1',1'' \to 1}A_{3,3}^{\rm tree}(1', 1'') + ({\rm leg}\,1 \leftrightarrow {\rm leg}\, 4)\,.
\end{align}

Due to the local origin of the anomaly, we can apply the same argument to the two chiral corners (grey blobs) in the two-loop graph in Fig.~\ref{dblbox}. In fact, all the further examples in the following sections admit the same interpretation.

The integral $ \int_0^1 d\xi   \, \int d\eta$ resembles the integration over the extra point on the super-Wilson loop contour in Refs.~\cite{Bullimore:2011kg,CaronHuot:2011kk}. The main difference is that there the collinear limit,  in which the new point approaches a segment  of the old contour,  creates a pole and one extracts the residue. Here the limit is smooth; the collinear singularity, which breaks the superconformal symmetry, appeared in the vertex adjacent to leg 1, which we have removed before defining the six-leg tree amplitude \p{e4.21}.

\section{$\overline{\rm MHV}$-like six-leg amplitudes}\label{6leg}

In this section we consider six-particle on-shell supergraphs of the $\rm N^{2}MHV=\rm\overline{MHV}$ type. Like the supergraphs of the previous Section, they are specified by a single bosonic function ${\cal I}(p)$. However they carry a different R-charge, so the structure of the superconformal Ward identities changes.

\subsection{General properties and superconformal Ward identity} \label{6leg_gen}

We study the  $\cN = 1$ six-particle on-shell amplitude \p{43} with $(m,n) = (6,0)$,
\begin{align}\label{131}
A_{6,0} = \vev{\bar\Phi(p_1,\eta_1)\ldots\bar\Phi(p_6,\eta_6)}= \delta^{(4)}(P) \delta^{(2)}(Q)\,  \cA_{6,0} (p,\eta)\,.
\end{align}
 According to Tables \ref{e16} and ~\ref{smallN} the reduced amplitude $\cA_{6,0}$ carries zero helicity at all points and the maximal possible R-charge in the six-particle case $R_{\cal A} = 2$. Thus the amplitude is of the $\rm N^{2}MHV$ type and it coincides with the $\rm\overline{MHV}$ amplitude. At six points there is a unique Poincar\'e supersymmetry invariant of Grassmann degree 2 (up to a bosonic  factor). Indeed  using $Q$ and $\bar Q$ transformations we can eliminate four of the six odd variables $\eta$. We construct the invariant as a product of two $\Xi_{ijk}$ \p{XIijk}. Thus the reduced amplitude is specified by a single bosonic function ${\cal I}(p)$,  
\begin{align}\label{131'}
\cA_{6,0} (p,\eta)= \Xi_{123} \Xi_{456}\, \cI(p)\,.
\end{align}

Then we study the superconformal properties of the amplitude. 
First we act with the $S$-generator on \p{131} and push it through the super-momentum conservation delta functions \p{SbarA}. Then we take into account the anticommutation relation \p{twco} and reduce the action of $S$ to collinearity operators,
\begin{align}\label{SA=XiFXiF}
S^\a A_{6,0} =  \delta^{(4)}(P) \delta^{(2)}(Q)\,  (\Xi_{456} F^\a_{123}- \Xi_{123} F^\a_{456})\, \cI(p)\,.
\end{align}
On the other hand, we can evaluate the anomaly of $A_{6,0}$ using the anomaly of the vertex function \p{e411}. 
The generator $S$ lowers the Grassmann degree by one unit. After stripping off the supercharge conservation $\delta^{(2)}(Q)$ we find that the anomaly has Grassmann degree one, and consequently we can expand it in a basis of $\Xi$-invariants (consisting of two invariants in the six point case),
\begin{align}\label{SA=Xi-Xi}
S^\a A_{6,0} = \delta^{(4)}(P) \delta^{(2)}(Q)\, \sum_{i = 1}^6 \la_i^\a \left(\Xi_{456} \cA_i(p) - \Xi_{123} \cA_i'(p) \right)\,.
\end{align}
Here the bosonic functions $\cA_i$ and $\cA_i'$ are given by Feynman integrals of one loop order less than the supergraph $A_{6,0}$. Comparing Eqs.~\p{SA=XiFXiF} and \p{SA=Xi-Xi} we obtain a pair of spinorial differential equations for the bosonic Feynman integral ${\cI}(p)$,   
\begin{align}\label{DEN2MHV}
F^\a_{123} \, \cI(p) = \sum_{i = 1}^6 \la_i^\a \, \cA_i(p) \, ,\qquad
F^\a_{456} \, \cI(p) = \sum_{i = 1}^6 \la_i^\a \, \cA'_i(p) \,.
\end{align}
In the following subsections we provide one- and two-loop examples of the Ward identities \p{DEN2MHV} and solve them.

The $\bar S$-symmetry of $A_{6,0}$ is not anomalous since there are no chiral legs and so all vertex functions constituting $A_{6,0}$ are invariant under $\bar S$. This can also be seen from the superconformal algebra perspective. We take into account that $\bar S$ commutes with the total supercharge and the total momentum conservation, Eq.~\p{SbarA}, 
\begin{align}
\bar S^{\da} A_{6,0} = \delta^{(4)}(P) \delta^{(2)}(Q)\, \bar S_{\da} \left[ \Xi_{123} \Xi_{456}\, \cI \right]\,.
\end{align}
Then we use $Q$ and $\bar Q$ transformations to eliminate four odd variables, $\eta_2,\eta_3,\eta_5,\eta_6 \to 0$. Consequently, $\Xi_{123} \Xi_{456} \sim \eta_1 \eta_4$ and the $\bar S$ generator \p{42} simplifies to $\bar S_{\da} = \eta_1 \frac{\pa}{\pa \tl_1^\da} + \eta_4 \frac{\pa}{\pa \tl_4^\da}$.
So we have $\bar S_{\da} \left[ \Xi_{123} \Xi_{456}\, \cI \right] =0 $. 

The absence of an $\bar S$ anomaly for the $\rm N^{2}MHV$ amplitude $A_{6,0}$ is equivalent to the absence of an $S$-anomaly for the conjugate amplitude $A_{0,6}$, which is of the $\rm MHV$ type. The latter is evidently not anomalous since the reduced amplitude $\cA_{0,6}$ has zero R-charge and it vanishes upon odd differentiations, see Eq.~\p{42}.

\subsection{Six-leg one-loop amplitude}

\begin{figure}
\begin{center}
\includegraphics[height=4cm]{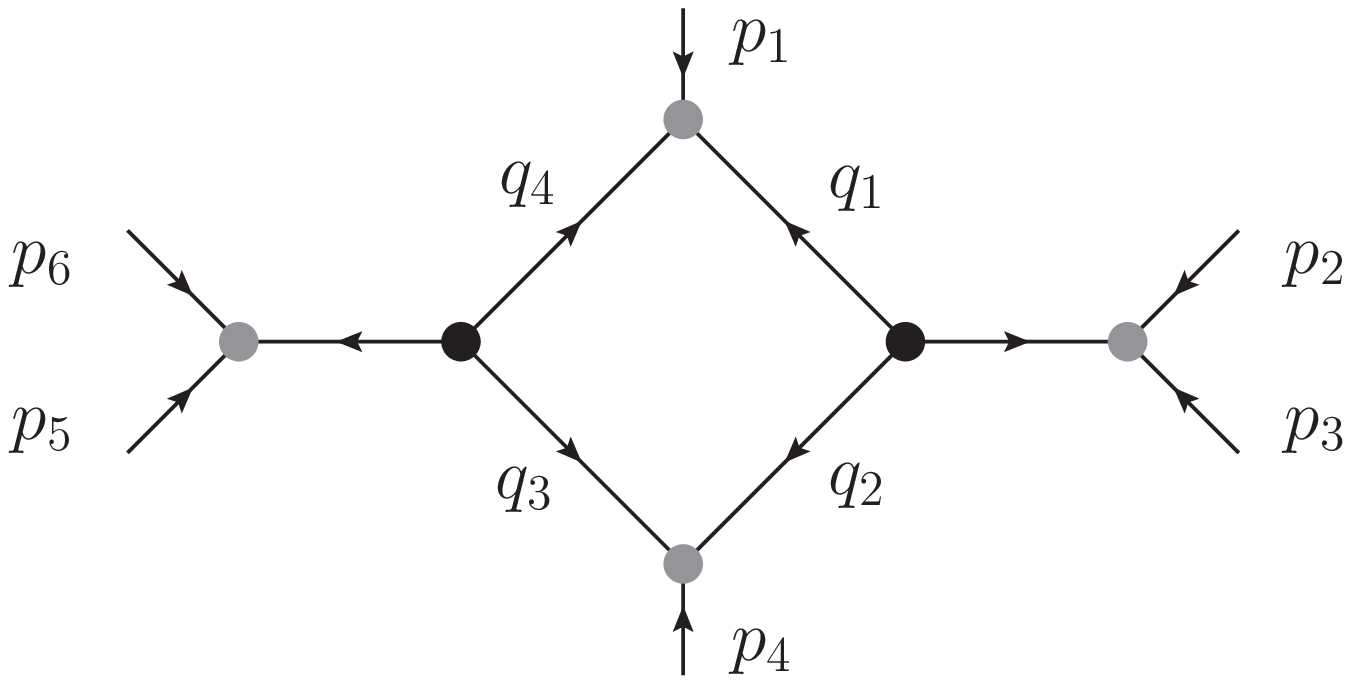} \qquad\includegraphics[height=4cm]{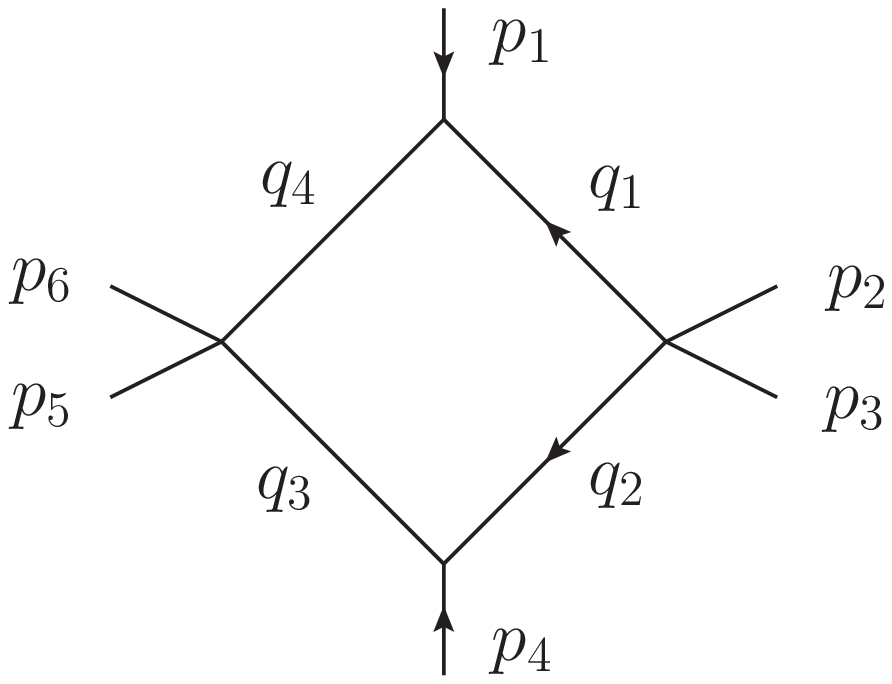}
\end{center}
\caption{LHS: One-loop six-leg supergraph of the $\overline{\rm MHV}$ type. RHS: The corresponding Feynman integral representing the bosonic factor $\cI(p)$.}
\label{fig2meboxsusy}
\end{figure}

We consider the one-loop six-particle amplitude supergraph $A_{6,0}$ on the lhs of Fig.~\ref{fig2meboxsusy}.
The corresponding expression can be easily found by splitting $A_{6,0}$ into vertex functions and using the Feynman rules of Sect.~\ref{sectgraphs}. The supergraph has to be of the form given by Eqs.~\p{131} and \p{131'}. In order to identify the single bosonic function $\cI(p)$ specifying the supergraph we extract its component $\eta_1\eta_2\eta_3\eta_4 \eta_5^0 \eta_6^0$ which corresponds to the amplitude $\vev{\psi_- \psi_- \psi_- \psi_-\bar\phi \bar\phi}$. In this way we find
\begin{align}\label{e448}
\cI(p) = \frac1{s_{123}[23][56]}\int  \frac{d^4\ell}{i\pi^2} \frac{\bra{1} q_1 \tilde q_2 \ket{4}}{q_1^2 q_2^2 q_3^2 q^2_4} \,,
\end{align}
which is represented by the two-mass easy box with a magic numerator on the rhs of Fig.~\ref{fig2meboxsusy}. Of course, this one-loop integral is well known but here we want to calculate it using only the superconformal symmetry. It serves as a toy example, preceding a highly nontrivial nonplanar two-loop integral elaborated in Sect.~\ref{s5.3}. 

The anomalous Ward identities for this supergraph are of the form \p{DEN2MHV}. In order to find the functions $\cA_i$ and $\cA_i'$ we need to replace the vertex function involving the $i$-th leg by its $S$-anomaly expression \p{e411}. 
The contribution from leg 1
is given by
\begin{align}
&\frac{\la^\a_1}{s_{23}s_{56}} \int_0^1 d\xi \frac{1}{q^2_{2} q^2_{3}} \int d^2\bq_a d^2\bq_b\, (\eta_1 + \xi [1 \bq_a] + \bar\xi[1\bq_b])\, \delta^2(\bq_a \tilde q_{2} + \bq_b \tilde q_{3} +\eta_4 \la_{4})\nt
&\times \delta^2(\eta_2 \la_{2} +\eta_3 \la_{3} -\bq_a p_{23})\, \delta^2(\eta_5 \la_{5} +\eta_6 \la_{6} -\bq_b p_{56})\,,
\end{align}
where we lifted the loop integration with the help of the delta function in \p{e411}. Setting $\eta_{1,2,3}=0$ or $\eta_{4,5,6}=0$, we obtain the contributions to the rhs of \p{SA=Xi-Xi} corresponding to the invariants $\Xi_{456}$  or $\Xi_{123}$, respectively,
\begin{align}\label{}
&\cA_{1} = \frac{1}{s_{123}[56]}\frac{\bra{4}p_{56}|1]}{s_{123} s_{156} - s_{23} s_{56}}
\left( \frac{s_{56}}{s_{156} - s_{56}} \log\left(\frac{s_{56}}{s_{156}}\right) - \frac{s_{123}}{s_{123} - s_{23}} \log\left(\frac{s_{23}}{s_{123}}\right) \right)\,, \nt
&\cA'_{1} = \frac{1}{s_{123}[23]} \frac{ \log\left(\frac{s_{56}}{s_{156}}\right) }{s_{156} - s_{56}}\,. \label{A1A1'}
\end{align}

The anomaly contribution of leg 4 is obtained by a reflection of the graph, $(123) \leftrightarrow (456)$. Legs $2,3$ and $5,6$ do not contribute to the anomaly according to Sect.~\ref{se3.3}. Thus the anomalous Ward identity takes the following form   
\begin{align}\label{F123F456}
F_{123} \,\cI(p) =  \ket{1} \cA_{1}(p) + \ket{4} \cA_{4}(p)\,, \qquad 
F_{456} \,\cI(p) =  \ket{1} \cA'_{1}(p) + \ket{4} \cA'_{4}(p) \,,
\end{align}
where $\cA_{1}$ and $\cA_1'$ are given by \p{A1A1'}, and $\cA_{4}$ and $\cA'_{4}$ are obtained by  permutation,   
\begin{align}
\cA_4 = \cA_1'|_{(123) \leftrightarrow (456)} \;,\qquad
\cA_4' = \cA_1|_{(123) \leftrightarrow (456)}\,.
\end{align}


Now we would like to solve the Ward identity \p{F123F456}.
It is convenient to introduce a dimensionless helicity-free function $f(p)$ by changing the normalization of $\cI$ \p{e448},
\begin{align}\label{e448'}
f(p) = s_{123}[23][14][56] \cI(p) = [14] \int  \frac{d^4\ell}{i\pi^2} \frac{\bra{1} q_1 \tilde q_2 \ket{4}}{q_1^2 q_2^2 q_3^2 q^2_4}\,.
\end{align}
The leading singularity analysis implies that the Feynman integral $f(p)$ is a pure function. 

The kinematics of the problem is specified by four variables, and $f(p)$ depends only three dimensionless variables. We choose  
\begin{align}
z_1 = \frac{s_{23}}{s_{123}}\,,\quad z_2= \frac{s_{56}}{s_{123}}\,, \quad z_3 = \frac{s_{156}}{s_{123}}\,,
\end{align}
so that $f=f(z_1,z_2,z_3)$. Projecting Eqs.~\p{F123F456} with the spinors $\bra{1}$ and $\bra{4}$ we reduce this system of spinorial differential equations to four scalar differential equations. However only three of them turn out to be independent. They can be conveniently written in the total differential form 
\begin{align}
d f(z_1,z_2,z_3) =& \log\left(z_1 \right) d \log \frac{(z_3 - z_1 z_2)}{(z_3- z_1)(1-z_{1})} 
+ \log\left(z_2 \right) d \log \frac{(z_3 - z_1 z_2)}{(z_3- z_2)(1-z_2)} \nt
&- \log \left( z_3 \right) d\log \frac{(z_3 - z_1 z_2)}{(z_3- z_1)(z_3-z_2)} \,. \label{df2me}
\end{align}
In order to integrate the differential equation \p{df2me} we need an appropriate boundary condition. The Feynman integral is not singular at $s_{14} \to 0$, moreover $f(p)$ has to vanish in this limit because of the normalization factor $[14] \to 0$, see Eq.~\p{e448'}. Hence, $f$ has to vanish on the surface $z_3 = z_1 +z_2 -1$ and we evaluate $f$ integrating the rhs of \p{df2me} 
along the $z_3$ axis from the boundary point $z_{3,0} = z_1 + z_2 -1$,
\begin{align}
f(z_1,z_2,z_3) = \int^{z_3}_{z_{3,0}} d z_3' \, df(z_1,z_2,z_3')\,.
\end{align}
Explicitly the solution is 
\begin{align}\label{e456}
f =& -{\rm Li}_2 \left(1-\frac{z_1 z_2}{z_3}\right) + {\rm Li}_2 \left( 1-z_1\right) + {\rm Li}_2 \left( 1-\frac{z_1}{z_3}\right) \nt
& + {\rm Li}_2 \left( 1-z_2\right) + {\rm Li}_2 \left( 1-\frac{z_2}{z_3}\right) +\frac12 \log^2 z_3 \,.
\end{align}
This pure weight-2 function is the finite part of the two-mass-easy scalar box integral.

\subsection{Six-leg two-loop nonplanar amplitude}\label{s5.3}

\begin{figure}
\begin{center}
\includegraphics[height=4cm]{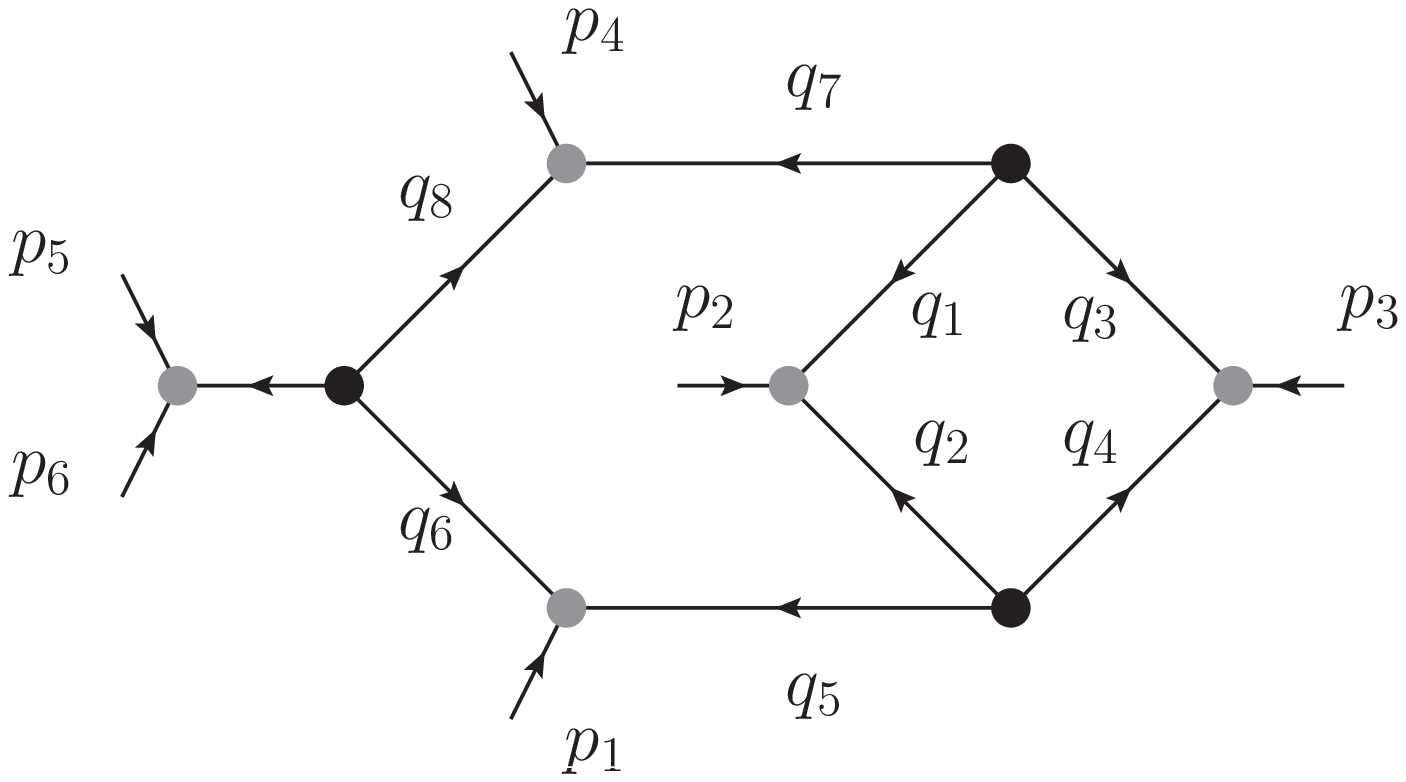} 
\quad
\includegraphics[height=4cm]{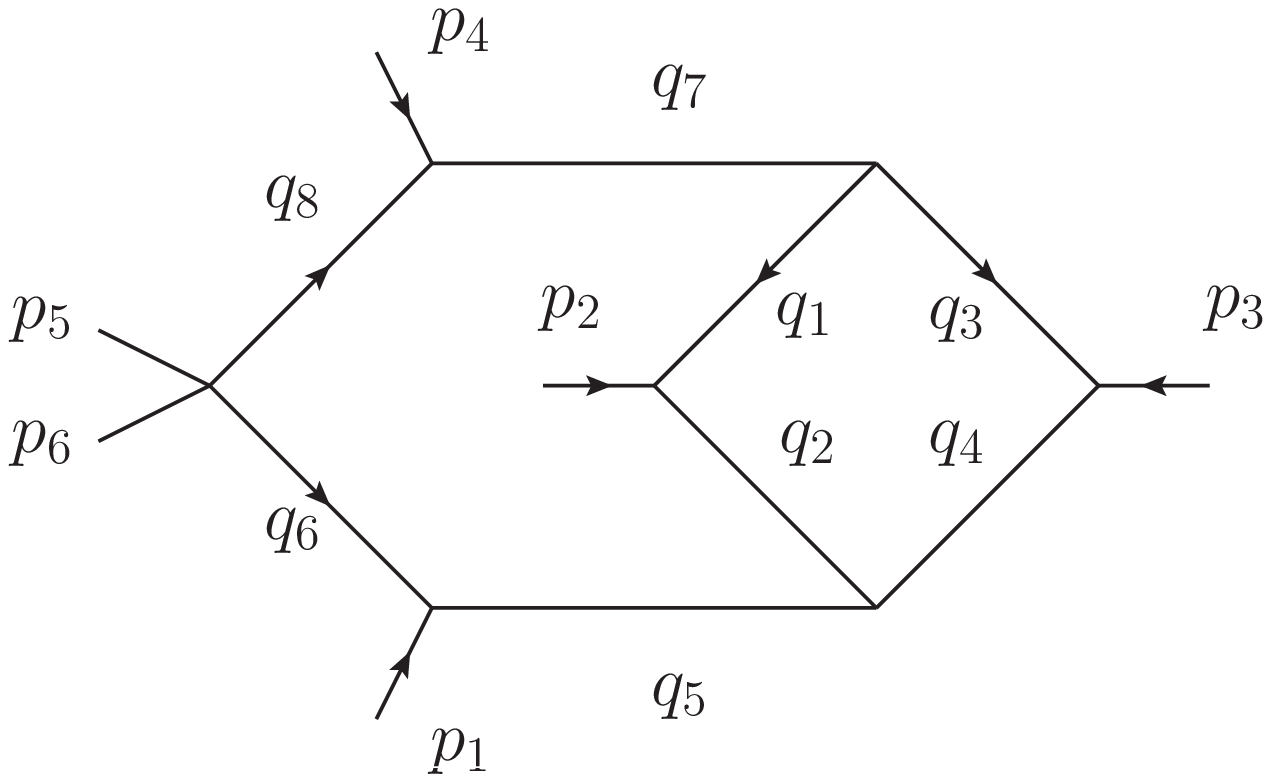}
\end{center}
\caption{Left: Two-loop nonplanar six-leg supergraph of the $\overline{\rm MHV}$ type. Right: The corresponding Feynman integral representing the bosonic function $\cI(p)$.}\label{figHiggs4j}
\end{figure}

Now we consider a much more nontrivial example of an $A_{6,0}$ supergraph, the non-planar two-loop graph of hexa-box topology on the lhs of Fig.~\ref{figHiggs4j}. The general structure of this type of amplitudes and of their superconformal properties was already discussed in Sect.~\ref{6leg_gen}, but 
now we prefer to use $\Xi$-invariants with different labels in order to take into account the discrete symmetries of the supergraph. So we have (cf. Eq.~\p{131'})
\begin{align} \label{AHiggs}
\cA_{6,0}(p,\eta) = \Xi_{156}\Xi_{456}\, \cI(p)\,.
\end{align}
We find $\cI(p)$ by evaluating the $\eta_1\eta_2\eta_3\eta_4 \eta_5^0 \eta_6^0$ component of the supergraph. By changing the normalization of $\cI(p)$ we define the helicity-free dimensionless function 
\begin{align}
f(p) =  \vev{23}[14][23][56]^2 \cI(p)= [14][23] \int \frac{d^4 \ell_1 d^4 \ell_2}{(i\pi^2)^2}  \frac{\vev{1|q_6 \tilde q_8|4}\vev{2|q_1 \tilde q_3|3}}{q_1^2 q_2^2 q_3^2 q_4^2 q_5^2 q_6^2 q_7^2 q_8^2} \,. \label{HiggsInt}
\end{align}
This Feynman integral is depicted on the rhs of Fig.~\ref{figHiggs4j}. The function $f(p)$ is invariant under the permutations $1 \leftrightarrows 4$ and $2 \leftrightarrows 3$ of the external legs. Integrals of this topology contribute to the Higgs production plus two jets.   

In order to obtain the $S$-symmetry Ward identities   we
act with the generator $S$ on \p{AHiggs}. The result is the twistor collinearity operators $F_{ijk}$ acting on $\cI(p)$. 
Thus we obtain a pair of spinorial differential equations     
\begin{align}\label{WIHiggs}
F^\a_{156} \, \cI(p) = \sum_{i = 1}^4 \la_i^\a \, \cA_i(p) \, ,\qquad
F^\a_{456} \, \cI(p) = \sum_{i = 1}^4 \la_i^\a \, \cA'_i(p) \,,
\end{align}
where $\cA_i$ and $\cA_i'$ are bosonic functions expressed through one-loop Feynman integrals. Legs 5 and 6, which form the massive tail, do not contribute to the anomaly. 

The rhs of the Ward identity \p{WIHiggs} comes from the supergraph calculation of the anomaly of $A_{6,0}$ using the $S$-anomaly of the vertex functions \p{e411}. In order to match it with the $S$-variation of Eq.~\p{AHiggs} we expand the result in the basis of  $\Xi_{156}$ and $\Xi_{456}$,  
\begin{align}\label{}
S^\a A_{6,0} = \delta^{(4)}(P) \delta^{(2)}(Q)\, \sum_{i = 1}^4 \la_i^\a \left(\Xi_{456} \cA_i(p) - \Xi_{156} \cA_i'(p) \right)\,. \label{Xi456-Xi156}
\end{align}
The delta function in \p{e411} freezes one of the loop integrations in the supergraph. The anomaly of leg 1 results in a one-mass box topology, and working out the Grassmann integrations we find     
\begin{align}
\cA_1(p) = \int^1_0 d\xi \frac{\xi \bra{4}p_{23}|1]}{\vev{23}[56](\xi s_{56}+\bar\xi s_{156})(\xi s_{456}+\bar\xi s_{23})} {\rm Box}(p_2,\xi p_1,p_3) \,,
\end{align}
with the one-mass `${\rm Box}$' integral defined in \p{boxnum1}. We then split the integrand into simple fractions leading to the  decomposition
\begin{align}
\cA_1(p) = \frac{\bra{4} p_{23}|1]}{[56]s_{23}(s_{156}s_{456}-s_{23}s_{56})} \left[  \frac{s_{156}}{s_{156}-s_{56}} a_1(p) + \frac{s_{23}}{s_{456}-s_{23}} \hat a_1(p) \right] \label{A1p}
\end{align}
in terms of two pure weight-3 functions (see Eq.~\p{F5}), 
\begin{align}
a_1(p) = & - \int^1_0 d\xi \frac{s_{156}-s_{56}}{\xi s_{56}+\bar\xi s_{156}}  \bigg[ 
{\rm Li}_2\left( 1-\frac{\bar\xi s_{23} +\xi s_{456}}{\xi s_{12}}\right) \nt 
& \quad + {\rm Li}_2\left( 1-\frac{\bar\xi s_{23} +\xi s_{456}}{\xi s_{13}}\right) + \frac12 \log^2\left( \frac{s_{12}}{s_{13}}\right) + \frac{\pi^2}{6}
\biggr] \,, \nt
\hat a_1(p) = & \int^1_0 d\xi \frac{s_{456}-s_{23}}{\xi s_{456}+\bar\xi s_{23}}  \bigg[ 
{\rm Li}_2\left( 1-\frac{\bar\xi s_{23} +\xi s_{456}}{\xi s_{12}}\right) \nt 
& \quad + {\rm Li}_2\left( 1-\frac{\bar\xi s_{23} +\xi s_{456}}{\xi s_{13}}\right) + \frac12 \log^2\left( \frac{s_{12}}{s_{13}}\right) + \frac{\pi^2}{6}
\biggr]\,.
\end{align}
Similarly, the contribution of leg 4 is given by the one-mass  ${\rm Box}(p_2,\xi p_4,p_3)$ \p{boxnum1}, 
\begin{align}
\cA_4(p) = \frac{a_4(p)}{[56]s_{23}(s_{456}-s_{56})}\,, \label{A4p}
\end{align}
with the pure weight-3 function 
$a_4(p) = a_1(p)|_{1\leftrightarrows 4}\,$.
The topology of the Feynman graph in Fig.~\ref{figHiggs4j} suggests that the anomaly contribution of leg 2 is given by a hexagon  integral with one massive corner. In reality, doing the Grassmann integrations we find that one of the propagators is canceled by the numerator and we obtain the pentagon integral \p{pentagon_int}, 
\begin{align}
\cA_2(p) = \frac{1}{\vev{23}[56]}\int^1_0 d\xi \, {\rm Pent}(p_4,\xi p_2 , p_3, p_1 +\bar\xi p_2)\,.
\end{align}
Substituting the explicit expression for the pentagon integral \p{pentint} we obtain
\begin{align}
\cA_2(p) = -\frac{a_2(p)}{\vev{23}[56][4|\tilde p_{56}p_2|3]}\,, \label{A2p}
\end{align}
where the pure weight-3 function $a_2$ is defined as follows
\begin{align}
a_2(p) = & \int^1_0
\frac{d\xi}{\frac{[4|\tilde p_{56}p_1|3]}{[4|\tilde p_{56}p_2|3]}+\bar\xi} 
\biggl[ 
{\rm Li}_2\left( 1 - \frac{\xi\bar\xi s_{12} s_{24}}{(\xi s_{13} + \bar\xi s_{456})(\bar\xi s_{34} + \xi s_{156})} \right) \nt
& \qquad + \log \left( \frac{\xi\bar\xi s_{12} s_{24}}{(\xi s_{13} + \bar\xi s_{456})(\bar\xi s_{34} + \xi s_{156})} \right)
\log \left( \frac{\xi s_{23} s_{56}}{s_{456}(\bar\xi s_{34} + \xi s_{156})} \right) \nt
& \qquad + {\rm Li}_2\left( 1 - \frac{\xi s_{23} s_{56}}{s_{456}(\bar\xi s_{34} + \xi s_{156})} \right)   -\frac{\pi^2}{6}
\biggr]\,. \label{a2p}
\end{align}
The anomalous contribution of leg $3$ is obtained from   \p{A2p} by exchanging legs $2$ and $3$: $\cA_3(p) = \cA_{2}(p)|_{2 \leftrightarrows 3}\,$, i.e. 
\begin{align}
\cA_3(p) = \frac{a_3(p)}{\vev{23}[56][4|\tilde p_{56}p_3|2]}
\end{align}
and $a_3(p) = a_2(p)|_{2 \leftrightarrows 3}\,$.
Thus we have found all the anomalous contributions $\cA_i$ accompanying the invariant $\Xi_{456}$ in Eq.~\p{Xi456-Xi156}. The anomalous contributions $\cA_i'$ corresponding to the invariant $\Xi_{156}$ (and the matching pure functions $a_1',\hat a'_1,a'_4,a'_2,a'_3$) are obtained from the expressions for $\cA_i$, Eqs.~\p{A1p}, \p{A4p}, \p{A2p} by exchanging points $1$ and $4$,
\begin{align}
\cA_i'(p) = \cA_i(p)|_{1 \leftrightarrows 4}\quad,\quad 
a_i'(p) = a_i(p)|_{1 \leftrightarrows 4}\,.
\end{align}
Let us note that only $a_1,\hat a_1,a_2$ are independent pure functions, the remaining ones are obtained from them by means of the discrete symmetries.

\subsubsection*{Solving the two-loop Ward identity in the spinor parametrization}

The six-particle scattering is not completely specified by the six Mandelstam invariants
\begin{align}
s_{12} \;,\; s_{23} \;,\; s_{34} \;,\; s_{456} \;,\; s_{56} \; , \;s_{156}\,,\label{Mandel}
\end{align}
since the amplitude can have a parity odd part. This is the reason why we introduce the following five dimensionless complex variables defined in terms of the helicity spinors  
\begin{align}
z_1 = \frac{s_{23}}{s_{156}} \;,\; 
z_2 = \frac{s_{23}}{s_{456}} \;,\; 
z_3 = \frac{\bra{4}p_3|1]}{\bra{4}p_{56}|1]} \;,\; 
z_4 = \frac{\bra{1}p_3|4]}{\bra{1}p_{56}|4]} \;,\;
z_5 = \frac{[12][34]}{[14][23]}\,. \label{zvar}
\end{align}  
The two-loop dimensionless integral \p{HiggsInt} is a function of them, $f=f(\{z_i\})$. The pure functions $a_i,a_i',\hat a_1,\hat a'_1$ describing the anomalies are also 5-variable functions of \p{zvar}.   

One of the merits of the spinor variables \p{zvar} is that they partially rationalize the alphabet which describes the set of iterated integrals emerging in the problem. Since parity-odd contributions are allowed in the six-particle scattering,  the pseudoscalar $\ep(p_1,p_2,p_3,p_4)$ naturally appears. In the parametrization by the Mandelstam invariants \p{Mandel} it is given by the square root of the Gram determinant, $\sqrt{\det ||s_{ij}||
_{i,j=1}^4 }$, but it is rational in the complex variables \p{zvar}. 
However the square roots are not completely resolved by \p{zvar}.
Doing the  integration in Eq.~\p{a2p} we encounter the square root
\begin{align*}
\sqrt{4s_{34}s_{456}(s_{12}s_{23}+s_{156}s_{23}-s_{23}s_{34})+(s_{12}s_{156}-s_{12}s_{23}+s_{23}s_{34}-s_{156}s_{456}+s_{34}s_{456})^2}
\end{align*}
which cannot be rationalized by any choice of the spinor parametrization.   
  
We want to reformulate the Ward identities \p{WIHiggs} in the new variables \p{zvar}. We project them with the spinors $\la_1$ and $\la_4$ and find four DEs which can be conveniently assembled together using the differential in the first four complex variables $\tilde d= \sum_{i = 1}^4 \pa_{z_i} d z_i$,
\begin{align}
& \tilde d f= \hat a_1 \,\tilde d \log(z-z_4) + \hat a_1' \,\tilde d \log(1+z+z_3) + a_1 \,\tilde d \log\left( \frac{z_1(1+z+z_3)}{z z_1 + z^2 z_1 + z z_3 - z_4 -z z_4 - z_3 z_4} \right) \nt
&+ a_1' \,\tilde d \log\left( \frac{z_2(z-z_4)}{z z_2 + z^2 z_2 + z z_3 - z_4 -z z_4 - z_3 z_4} \right) + a_2 \,\tilde d \log\left( \frac{z_2(z-z_4)}{1-z_2-z z_2 +z_4}\right) \nt
&+ a_3\, \tilde d \log\left( \frac{z_2(z-z_4)}{z z_2 - z_4}\right) + a'_2 \,\tilde d \log\left( \frac{z_1(1+z+z_3)}{1+ z z_1 +z_3}\right) + a'_3\, \tilde d \log\left( \frac{z_1(1+z+z_3)}{z_1 + z z_1 + z_3}\right). \label{DEHiggs}
\end{align}
Eq. \p{DEHiggs} is completely equivalent to the superconformal Ward identities.

We note that the integral \p{HiggsInt} vanishes at $p_1 = p_4 =0$ that corresponds to $z_1 = z_2 = 1$ and arbitrary $z_3, z_4, z_5$ (see \p{zvar}), 
\begin{align}
f(z_1 = 1 , z_2=1,z_3,z_4,z_5) =0\,.
\end{align}
Thus, in order to evaluate the integral $f$ at a generic kinematical point $\{z_{1*},z_{2*},z_3,z_4,z_5\}$ we need to integrate the DE \p{DEHiggs} along a path $\gamma$ lying in the plane $(z_1,z_2)$ and connecting points $(1,1)$ and $(z_{1*},z_{2*})$, so that  $z_3,z_4,z_5$ are fixed. For instance, we can choose the path $\gamma$ to be a straight line,
\begin{align}
z_1(t) = 1+(z_{1*}-1)t  \quad,\quad z_2(t) = 1+(z_{2*}-1)t \quad,\qquad \text{at} \quad 0\leq t \leq 1  \,.
\end{align}
The solution of the DE takes the following form
\begin{align}
f(p) & =  \int\limits_\gamma 
\biggl[ a_1 \, d \log\left( \frac{z_1}{z z_1 + z^2 z_1 + z z_3 - z_4 -z z_4 - z_3 z_4} \right) \nt
& + a_1' \, d \log\left( \frac{z_2}{z z_2 + z^2 z_2 + z z_3 - z_4 -z z_4 - z_3 z_4} \right) + a_2 \, d \log\left( \frac{z_2}{1-z_2-z z_2 +z_4}\right) \nt
&+ a_3\, d \log\left( \frac{z_2}{z z_2 - z_4}\right) + a'_2 \, d \log\left( \frac{z_1}{1+ z z_1 +z_3}\right) +  a'_3\, d \log\left( \frac{z_1}{z_1 + z z_1 + z_3}\right) \biggr]\,. \label{solutionH}
\end{align}
Thus, recalling that the anomalies $a_i$, $a_i'$ are given by one-fold integrals and substituting them in \p{solutionH}, we obtain a two-fold integral representation for $f(p)$ \p{HiggsInt}. We numerically compared this representation with the numeric integration of the Feynman integral in the Euclidean region and found a nice agreement.
 
It would be interesting to carry out the analytic integration in \p{solutionH} 
in terms of  hyperlogarithm functions and to study the associated alphabet.

\section{${\rm NMHV}$-like six-leg amplitude}\label{6nmhv}

In this section we discuss a genuine ${\rm NMHV}$-like amplitude with six legs. As explained in Sect.~\ref{s311} (see Eq.~\p{NMHV}), it is described by two independent bosonic functions $\cI_i(p)$. Consequently, the superconformal Ward identities become more complicated. The example we wish to show here has an enhanced  symmetry (dual conformal and cyclic), which greatly facilitates the solving of  the   Ward identities.

\subsection{General properties. One-loop amplitude}

We consider a supergraph with six alternating antichiral/chiral legs (see  \p{43})
\begin{align}\label{413}
A_{3,3} &=\vev{\bar\Phi(1)  \Psi(2) \bar\Phi(3)  \Psi(4)  \bar\Phi(5) \Psi(6) }   =\delta^{(4)}(P) \, \delta^{(2)}(Q)\, \, \cA_{3,3}(p,\eta)\,.
\end{align}
The R-weight counting shows that the reduced amplitude $\cA(p,\eta)$ is of   the NMHV type. 
 A convenient choice of basis (different but equivalent to that in \p{NMHV})  is 
\begin{align}\label{0414}
\cA_{3,3}(p,\eta) = \frac{1}{2\bra{1}p_{23}|4]}\left( \cI_{1}\, \Xi_{234} - \cI_{2}\, \Xi_{456} \right)\,.
\end{align}
The bosonic coefficient functions $\cI_1$ and $\cI_2$ are identified by comparing with the components $\eta_1 \eta_2 \eta_3$ and 
$\eta_1 \eta_5 \eta_6$:  
\begin{align}\label{C1C2}
\cI_1 = \vev{\psi_- \phi \psi_- \psi_+ \bar\phi \psi_+}\,, \qquad\qquad
\cI_2 = -\vev{\psi_- \psi_+ \bar\phi \psi_+ \psi_- \phi}  \,.  
\end{align}

\begin{figure}
\begin{center} 
\includegraphics[width = 8cm]{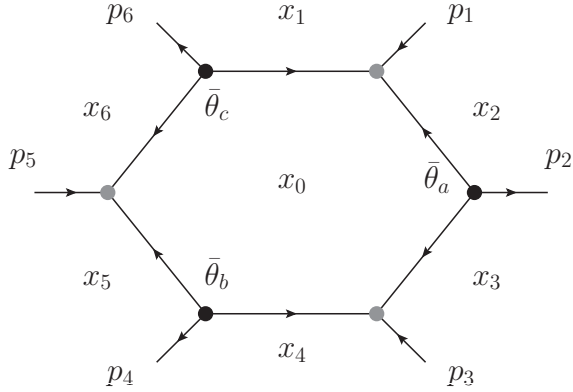}
\end{center}
\caption{One-loop supergraph for the amplitude \p{413}. } \label{fig5}
\end{figure}

In the following we consider the one-loop  supergraph in Fig.~\ref{fig5}.
The component $\cI_1$ is represented by a hexagon Feynman integral made from Yukawa vertices,   
\begin{align}\label{e64}
\cI_1 = &\int \frac{d^4 y_0}{\pi^2}\frac{1}{\prod_{j=1}^6 y_{0j}^2}  \,\bra{1} y_{20} y_{03} \ket{3}\, [4| y_{50} y_{06} |6] \,,
\end{align}
and similarly for $\cI_2$.
Here we have used the dual space notation $p_i=y_i-y_{i+1}$, with $y_0$ denoting the loop momentum.

The superamplitude \p{0414}  can be rewritten in a cyclic symmetric form using a redundant basis of $\Xi$ invariants: 
\begin{align}\label{1213}
&\cA_{3,3} =  R_{123}\, f_{123}(u_1,u_2,u_3) + {\rm cycle} (i \to i+2)  \,.
\end{align}
Here we introduced the $\cN=1$ analog of the $\cN=4$ dual supersymmetry  $R$ invariants \cite{Drummond:2008vq}
\begin{align}\label{1212}
R_{123} = \frac{\Xi_{123} }{\bra{4}p_{23}|1]\bra{6}p_{45}|3]}\,. 
\end{align}
Its coefficient  is a function of the dual conformal cross-ratios 
\begin{align}\label{e55}
u_1 = \frac{y^2_{24} y^2_{51}}{y^2_{25} y^2_{41}}\;\;,\;\;
u_2  = \frac{y^2_{35} y^2_{62}}{y^2_{36} y^2_{52}} \;\;,\;\;
u_3 = \frac{y^2_{46} y^2_{13}}{y^2_{63} y^2_{41}} \,.
\end{align}
The explicit expression of the integral \p{e64} (see \cite{ArkaniHamed:2010gh}) yields
\begin{align}\label{1214}
f_{123} = \frac1{2} \log u_1 \log u_3\,.
\end{align}

\subsection{Superconformal Ward identities and their solution}\label{s63}

According to Sect.~\ref{sect5NMHV}   the action of $S^{\a}$ on this amplitude should produce an anomalous Ward identity  of the form 
\begin{align}\label{}
S^{\a} A_{3,3}^{\rm 1-loop} = \delta^{(4)}(P) \delta^{(2)}(Q) \sum_{i=1,3,5} \la^\a_{i}\cA_{i}\,,
\end{align}
where $\cA_{i}(p)$ are bosonic functions. The anomaly originates from the antichiral vertices  adjacent to legs 1,3,5 in Fig.~\ref{fig5}. 
To compute $\cA_{i}(p)$ we replace the antichiral vertices by the anomaly contact term \p{e411}.
A short calculation then gives
\begin{align}\label{A1}
\cA_{1} =  -\frac{\bra{1}p_{23}|4]}{\vev{12} \vev{61} p_{123}^2 p_{156}^2} \, \frac{\log u_1}{u_1-1}\,,
\end{align}
while $\cA_{3}$ and $\cA_{5}$ are obtained from \p{A1} by cyclic shifts of the labels.

Let us now act with $S$ on the cyclic symmetric amplitude \p{1213}. The R-invariants \p{1212} are easily seen to be inert  and we obtain the anomalous Ward identity
\begin{align}\label{1230}
\frac{1 }{\bra{4}p_{23}|1]\bra{6}p_{45}|3]} \, F^\a_{123} \, f_{123}(u_i)+ {\rm cycle} (i \to i+2) = \sum_{i=1,3,5} \la^\a_{i}\cA_{i}\,.
\end{align} 
After some elementary spinor algebra we get the differential equation 
\begin{align}\label{6.23}
&\frac{p^{-1}_{23}|4] [23]}{\bra{6}p_{45}|3] p^2_{234}}\left(u_1 \pa_1 f_{123} -\frac1{2} \log u_3  \right)  + {\rm cycle} (i \to i+2)\nt
&+ \frac{p^{-1}_{34}|5] \bra{5} p^{-1}_{345}|3][12] + p^{-1}_{61} |5] \bra{5} p^{-1}_{561}|1][23]}{\bra{4}p_{23}|1]\bra{6}p_{45}|3]} \, \Big( u_2 \pa_2 f_{123}\Big)  + {\rm cycle} (i \to i+2) = 0\,.
\end{align}
We notice that each spinor structure in  \p{6.23} is dual conformally covariant,  but transforming with different (matrix) weights.  We conclude that their coefficients  must vanish,
\begin{align}\label{}
u_1 \pa_1 f_{123} -\frac1{2} \log u_3 = u_2 \pa_2 f_{123} = 0 \quad \mbox{and permutations}\,.
\end{align}
These equations have the general solution 
\begin{align}\label{6.25}
f_{123}=  \frac1{2} \log u_1 \log u_3 + \phi(u_3)\,.
\end{align}
The remaining freedom can be fixed by substituting \p{6.25} in the amplitude \p{1213} and requiring the absence of spurious singularities due to the denominator in \p{1212}. This implies $ \phi(u_3)= { \rm const}$, which drops out due to an identity for the R-invariants. In this way we recover the result \p{1214}.

To summarize, we have shown how to solve the $S$-supersymmetry Ward identities for a superamplitude described by more than one bosonic function. Our task was greatly simplified by the extra symmetries of the problem.

\section{Conclusions and outlook}

In this paper we made the first steps of a systematic study of the implications of superconformal symmetry of massless scattering amplitudes.  We consider its manifestations on the finite hard part of the scattering process. The superconformal  Ward identities  in momentum space are rather powerful  first-order inhomogeneous differential equations. They contain an anomaly due to collinear contact singularities of the loop integrand. The anomaly is given by an integral with one loop less than the original graph, so it is much easier to evaluate. 

 Working in an $\cN=1$ model of   massless supersymmetric matter, we derived and solved the  Ward identities for various scattering processes. We focused mostly on MHV-like superamplitudes with up to six external particles, at one and two loops. They are described by a single bosonic function of the momenta. We showed how to solve the first-order differential equations following from the anomalous superconformal symmetry. Together with physically motivated boundary conditions, they uniquely fix the answer.  Our most interesting example is a previously unknown five-point non-planar hexa-box integral
with an off-shell leg.  It gives first indications on the function space needed for Higgs plus two jet production at next-to-next-to leading order.

In this paper we consider only one example of an  NMHV-like six-particle one-loop superamplitude. The principle complication of non-MHV-like amplitudes is that they involve more than one bosonic function. Consequently, the superconformal Ward identities turn into a system of coupled differential equations which are harder to solve. The example presented here benefits from a powerful extra symmetry (cyclic and dual conformal), so that we can easily solve the system. In the generic case of non-MHV-like amplitudes we still need to develop a strategy for integrating such complicated systems. This will give us access to a richer choice of loop integral topologies.

Another line of future development is the systematic study of the function space of our differential equations. In all the cases studied in this paper, interestingly,  the solutions were given by uniform weight functions. Will this be a general feature of loop integrals obtained from superamplitudes? If so, how exactly does the (anomalous) superconformal symmetry fix the class of functions? 

The $\cN=1$ Wess-Zumino model of supersymmetric matter that we explored here has some limitations as to the variety of the  accessible integral topologies. The cubic interactions in superspace (or equivalently, the Yukawa and quartic vertices for the component fields) do not allow one-loop subdiagrams with an odd number of edges, e.g., pentagons etc.
A natural generalization of our work is to include $\cN=1$ gauge fields, which would allow us to discuss more general topologies.
Let us anticipate two obvious complications in doing so. 
Compared to the matter sector, where one can deal with individual graphs, in a gauge theory one has to consider sums of diagrams to achieve gauge invariance.
The other, more serious problem is that most of the diagrams with external gluon legs suffer from infrared divergences.

In a broader perspective, we wish to address the issue of broken superconformal symmetry in the presence of UV and/or IR divergences. 
This involves two steps. The first is to choose a suitable infrared finite part, called `remainder' or 'ratio' function \cite{Drummond:2008vq,Alday:2010ku,Henn:2011by} in the $\mathcal{N}=4$ literature, or hard function \cite{Collins:1989gx} in the QCD literature. Such a function captures the non-trivial new information not beyond the divergent terms \cite{Weinzierl:2011uz}. The second step is to understand the conformal properties of the latter.  We find this problem very interesting for future studies.

\section*{Acknowledgements}
This research received funding in part from the PRISMA Cluster of Excellence at Mainz University, and from the European Research Council (ERC) under the European Union's Horizon 2020 research and innovation programme (grant agreement No 725110), {\it{Novel structures in scattering amplitudes}}.

\appendix

\section{Spinor conventions}\label{Appendix_Spinor_Conventions}

We use the two-component spinor conventions of \cite{Galperin:1984av}. They include the definitions of the Levi-Civita tensors
\begin{align}\label{}
\ep_{12}=-\ep^{12}=
\ep_{\dot{1}\dot{2}}=-\ep^{\dot{1}\dot{2}}=1\, ,\qquad \ep^{\a\b} \ep_{\b\gamma}=\delta^\a_\gamma  
\end{align}
and of a four-vector as a two-by-two matrix,
\begin{align}\label{}
x_{\a\dot{\alpha}}=x^\mu(\sigma_\mu)_{\a\dot{\alpha}}\,, \qquad \tilde x^{\da\a} = x^\mu(\tilde\sigma_\mu)^{\da\a} = \ep^{\a\b} \ep^{\da\db} x_{\b\db}\,.
\end{align}
They satisfy the following identities (here $x \cdot y = x^\mu y_\mu$):
\begin{align}\label{}
& x_{\a\da} \tilde y^{\da\b} + y_{\a\da} \tx^{\da\b} = 2x \cdot y\, \delta_\a^\b\,, \nt
& x_{\a\da} \tx^{\da\b} = x^2 \delta_\a^\b\,, \nt
& x^2 = \frac1{2}  x_{\a\da} \tx^{\da\a}\,.
\end{align}
The space-time derivative is defined as $\pa_{\a\da} = \sigma^\mu_{\a\da} \pa_\mu$ and has the property
\begin{align}\label{A4}
\pa_{\a\da} \tx^{\db\b} = 2 \delta_\a^\b \delta_\da^\db\,, \qquad  \pa_{\a\da} x^2 = 2x_{\a\da} \,.  
\end{align}

A lightlike four-vector is written down in terms of a commuting chiral spinor $\la$ and its antichiral conjugate $\tl= \la^*$ as follows
\begin{align}\label{}
p_{\a\da} = \la_\a \tl_\da \ \Leftrightarrow \ p = \ket{\la} [\tl|\,.
\end{align}

For the Grassmann variables we use the conventions
\begin{align}
&\theta^2= \theta^{\a} \theta_{\a} \;\;,\;\;
\bar\theta^2= \bar\theta_{\da} \bar\theta^{\da} \;\;,\;\; \bq\tilde q\q= \bq_\da \tilde q^{\da\a} \q_\a  \;\;,\;\;
\vev{\la\q} = \la^\a \q_\a \;\;,\;\; [\tl\bq] = \tl_\da \bq^\da  \;\;,\;\;\notag\\
&\int d^2 \bar\theta \; \bar\theta^2 = 1 \;\;,\;\; 
\int d^2 \bar\theta\; \bar\theta^{\da} \bar\theta^{\db} = \frac{1}{2} \ep^{\da\db}  \;\;,\;\;
\int d^2 \theta \; \theta^2 = 1 \;\;,\;\; 
\int d^2 \theta\; \theta_{\a} \theta_{\b} = \frac{1}{2} \ep_{\a\b}\,.
\end{align}
We define the Grassmann delta functions by
\begin{align}\label{a6}
\delta^{(2)}(\q)=\q^2\,, \qquad \delta^{(2)}(\bq)=\bq^2\,.
\end{align}

\section{$\cN=1$ superconformal symmetry}\label{Appendix_Susy_Algebra}

The $\cN=1$ superconformal algebra $su(2,2|1)$ is given by the (anti)commutation relations
\begin{align} \notag
& \{{Q}_{\a}\,,\bar{{Q}}_{\da}\}  =    {P}_{\a \da}\,,
   \qquad \{{S}_{\a}\,,\bar{{S}}_{\da} \} =  {K}_{\a \da}\,,
\\ \notag
& {}[{P}_{\a \da}\,,{S}^{\b}] = \delta_{\a}^{\b} \bar{{Q}}_{\da}\,,
 \qquad [{K}_{\a \da}\,,{Q}^{\b}] = \delta_{\a}^{\b}
   \bar{{S}}_{\da}\,,
\\ \notag
& {}[{P}_{\a \da}\,,\bar{{S}}^{\db}]  =  \delta^{\db}_{\da} {Q}_{\a}\,,
\qquad [{K}_{\a \da}\,, \bar{{Q}}^{\db}]  =  \delta_{\da}^{\db} {S}_{\a}\,,
\\ \notag
& [{K}_{\a \da}\,,{P}^{\b \db}] = \delta_\a^\b \delta_\da^\db {D} +
M_{\a}{}^{\b}
 \delta_\da^\db + \bar{M}_{\da}{}^{\db} \delta_\a^\b\,,
\\ \notag
& \{{Q}^{\a}\,,{S}_\b\} =  M^{\a}{}_{\b}
   +  \tfrac{1}{2}\delta^{\a}_{\b}   (D -\tfrac{3}{2} R-H)\,,
\\ \label{b1}
& \{\bar{{Q}}_{\da}\,,\bar{{S}}^{\db}\} = \bar{M}_{\da}{}^{\db}     -  \tfrac{1}{2}\delta^{\da}_{\db}
({D}+\tfrac{3}{2} R+H)\,.
\end{align}
Here we see the generators of translations ($P$), conformal boosts ($K$), Lorentz transformations in the (anti)chiral spinor representation ($M,\bar M$) and dilatations ($D$), forming the conformal algebra $su(2,2)$. In fact, the whole superalgebra  \p{b1} is generated by the odd generators of Poincar\'e supersymmetry $Q, \bQ$  and special conformal supersymmetry $S, \bS$. They have dilatation weight and R-charge as follows:
\begin{align}\label{}
&[D, (Q,\bQ)]= \tfrac{1}{2}(Q,\bQ)\,, \qquad [D, (S,\bS)]= -\tfrac{1}{2}(S,\bS)\,,\nt
&[R, (Q,\bS)]=  (Q,\bS)\,, \qquad [R, (S,\bQ)]= - (S,\bQ)\,.
\end{align}
The bosonic generator  $H$ is a central charge counting the helicity of the on-shell states, see below; off shell it vanishes.  

\subsection{On-shell realization}

If we use both on-shell superfields $\Phi$ and $\bar\Phi$ \p{72}
to describe the external states, the relevant on-shell odd conformal supersymmetry generators are 
\begin{align}\label{b3}
& Q_\a = \sum_L  \la_\a \frac\pa{\pa\be} + \sum_R  \la_\a    \eta\,, \qquad \bar Q_\da = \sum_R \tl_\da   \frac\pa{\pa\eta}   + \sum_L \tl_\da \be  \,;\nt
&  S_\a = \sum_L \be \frac\pa{\pa \la^\a} + \sum_R  \frac{\pa^2}{\pa\eta\pa \la^\a} \,, \qquad \bar S_\da = \sum_R \eta  \frac\pa{\pa \tl^\da}   + \sum_L  \frac{\pa^2}{\pa\be\pa \tl^\da}\,,
\end{align}
where the sums go over the chiral (L) and antichiral (R) legs of the amplitude. As we have already explained, after the Fourier transform $\be\to\eta$ 
\p{FTgr}, we need only one type of odd variable $\eta$ to describe the external superstates of both types, $\bar\Phi(p,\eta)$ and $\Psi(p,\eta)$. Then the odd conformal supersymmetry generators become
\begin{align}\label{42}
& Q_\a = \sum_i  \la_{i\a} \eta_i  \,, \quad \bar Q_\da = \sum_i \tl_{i\da} \frac{\pa}{\pa\eta_i} \,,\quad
S_\a = \sum_i \frac{\pa^2}{\pa\eta_i\pa \la^\a_i} \,, \quad \bar S_\da =   \sum_i \eta_i\frac\pa{\pa \tl^\da_i}  \,. 
\end{align}

The on-shell odd generators satisfy the algebra \p{b1}, wherefrom we can read off the expressions for the dilatation, R-charge and helicity generators 
\begin{align}\label{b5}
&D = \frac1{2}\sum_{i=1}^n \left( \la^\a_i \frac{\pa}{\pa \la^\a_i } +  \tl^\da_i \frac{\pa}{\pa \tl^\da_i }  + 2\right) , \quad R =  \sum_{i=1}^n \eta_i \frac{\pa}{\pa \eta_i } 
,\nt
& H= \frac1{2}\sum_{i=1}^n \left( - \la^\a_i \frac{\pa}{\pa \la^\a_i } +  \tl^\da_i \frac{\pa}{\pa \tl^\da_i }+  \eta_i \frac{\pa}{\pa \eta_i } -2 \right)\,.
\end{align}

\subsection{Off-shell realization} 

The off-shell realization of the $\cN=1$ superconformal algebra depends on the type of superfields it acts upon. In this paper we are using (anti)chiral superfields, each type requiring the appropriate superspace basis \p{e2.2}.

The Poincar\'e supersymmetry generators are
\begin{align}\label{540}
&Q_\a = 
\left\{
\begin{array}{ll}
  L:&\frac\pa{\pa \q^\a}       \\
  & \\
  R:&  -i \bq^\da \frac\pa{x^{\da\a}_{R}}       
\end{array}
\right.
,\qquad 
\bar Q_\da =
\left\{
\begin{array}{ll}
  R:& - \frac\pa{\pa \bq^\da}      \\
  & \\
 L:& i \q^\a \frac\pa{x^{\da\a}_{L}}  
\end{array}
\right.  
,\qquad
P_{\a\da} =
\left\{
\begin{array}{ll}
  L:&i \frac\pa{x^{\da\a}_{L}}      \\
  & \\
  R:& i \frac\pa{x^{\da\a}_{R}}  
\end{array}
\right.   \ \ ,
\end{align}
where L and R stand for the chiral $(x_L,\q)$ and antichiral $(x_R,\bq)$ superspace bases. After Fourier transforming $x_L \to q$ or $x_R \to q$, we obtain the momentum space realization
\begin{align}\label{e111}
&Q_\a = 
\left\{
\begin{array}{ll}
  L:&\frac\pa{\pa \q^\a}       \\
  & \\
  R:&  - q_{\a\da}\bq^\da     
\end{array}
\right.
,\qquad
\bar Q_\da =
\left\{
\begin{array}{ll}
  R:& - \frac\pa{\pa \bq^\da}      \\
  & \\
 L:&  \q^\a q_{\a\da}    
\end{array}
\right.  
,\qquad
P_{\a\da} = q_{\a\da}  \ \  .
\end{align}

The superconformal generators are 
\begin{align}\label{549}
& S_\a = -\tfrac{1}{2}\q^2 \frac\pa{\pa \q^\a} - \tfrac{1}{2}(x_L)_\a^{\db}\q^\b \frac\pa{\pa x_L^{\db\b}}  - \tfrac{1}{2} \q_{\a}(d- \tfrac{3}{2}r)- \tfrac{1}{2}\q_\b\, m^\b_\a    +\tfrac{i}{2} (x_R)_{\a\da} \frac\pa{\pa \bq_{\da}} \ , \\
& \bar S^\da = -\tfrac{1}{2} \bq^2 \frac\pa{\pa \bq_{\da}} - \tfrac{1}{2} (x_R)^{\da\a}\bq^\db \frac\pa{\pa x_R^{\db\a}}  - \tfrac{1}{2}\bq^\da(d+\tfrac{3}{2}r)  - \tfrac{1}{2}\bq^\db\, \bar m^\da_\db  -\tfrac{i}{2} (x_L)^{\da\a} \frac\pa{\pa\q^\a} \ ,     \label{128}
\end{align}
where $d,r,m,\bar m$ are the matrix parts of the corresponding generators.
In each generator the first line applies to superfields of the same chirality, the second to the opposite chirality.     

The anticommutator $\{ S_\a, \bar S_\da\}$ defines the conformal boost generator $K_{\a\da}$, whose explicit form is not needed in this paper. The anticommutators $\{{Q}^{\a}\,,{S}_\b\} $ and $ \{\bar{{Q}}_{\da}\,,\bar{{S}}^{\db}\}$ define the generators of dilatation, R-charge,
\begin{align}\label{b10}
& D=\tfrac{1}{2} \q^\a \pd{\q^\a} + \tfrac{1}{2} \bq_\da \pd{\bq_\da} +\tfrac{1}{2} x^{\da\a}_L \pd{ x^{\da\a}_L} +\tfrac{1}{2} x^{\da\a}_R \pd{ x^{\da\a}_R} + d\,, \nt
& R= \q^\a \pd{\q^\a}-\bq_\da \pd{\bq_\da}+r\,, 
\end{align}
 and Lorentz transformations. The helicity $H$ does not appear off shell.

\subsection{Conformal supersymmetry properties of the (super)momentum conservation delta functions}\label{appeC}

Let us act on the amplitude \p{43} with the superconformal generator $S_\a$. We want to show  that the generator passes through the momentum and supercharge  delta functions,
\begin{align}\label{58}
S_\a A = \delta^{(4)}(P)\, \delta^{(2)}(Q)\, S_\a \cA\,.
\end{align}

According to the superconformal algebra \p{b1},  
\begin{align}\label{}
&[S_\a, \delta^{(4)}(P)] = \frac1{2} [S_\a, P^{\db\b}] \frac{\pa\delta^{(4)}(P)}{\pa P^{\db\b} } = \frac1{2}  \frac{\pa\delta^{(4)}(P)}{\pa P^{\da\a} } \bQ^\da\,,\nt
&[\bQ^\da, Q^2]= 2P^{\da\a} Q_\a\,,\nt
&[S^\a, Q^2] = 2 Q^\b [ M^{\a}{}_{\b}
   +  \tfrac{1}{2}\delta^{\a}_{\b}   (D -\tfrac{3}{2} R-H-2) ]\,.  \label{e32}
\end{align}
Putting all of this together, using the identity
\begin{align}\label{56}
P^{\b\da}\frac{\pa\delta^{(4)}(P)}{\pa P^{\a\da} } =-4 \delta_\a^\b\, \delta^{(4)}(P)\,,
\end{align}
and the $\bQ$ and Lorentz invariance of the reduced amplitude, $\bQ\cA=M \cA=0$, we obtain
\begin{align}\label{57}
S_\a A = 2 \delta^{(4)}(P) Q_\a\, (D-\tfrac{3}{2} R-H-6)\cA+ \delta^{(4)}(P) \delta^{(2)}(Q)\, S_\a \cA\,.
\end{align}
The reduced amplitude satisfies the following  condition on its total dilatation weight, helicity and R charge:
\begin{align}\label{e47}
(D-\tfrac{3}{2} R-H-6)\cA=0\,.
\end{align} 
This follows from the invariance of the whole amplitude, $(D-\tfrac{3}{2} R-H)A=0$, taking into account  the dimension and R-charge of  $\delta^{(4)}(P)\, \delta^{(2)}(Q)$. So, the first term on the rhs of \p{57} vanishes and we confirm \p{58}.

Let us now consider the $\bar{S}$ transformation \p{42} of the amplitude \p{43} with manifest $Q$ supersymmetry. In view of the commutation relations $\{Q,\bar S\}=0$ and $[P,\bar S] \sim Q$  (see \p{b1}), 
$\bar S$ goes through the two delta functions  and we get the antichiral analog of \p{58},
\begin{align}\label{SbarA}
\bar S_\da  A = \delta^4(P) \delta^{(2)}(Q)\, \bar S_\da \cA\,.
\end{align}

\section{ Vertex functions from the (anti)chiral vertices $\Phi^3$ and $\bar\Phi^3$}\label{Appendix_Derivation_Anomaly}

\begin{figure}
\begin{center}
\includegraphics[width = 4cm]{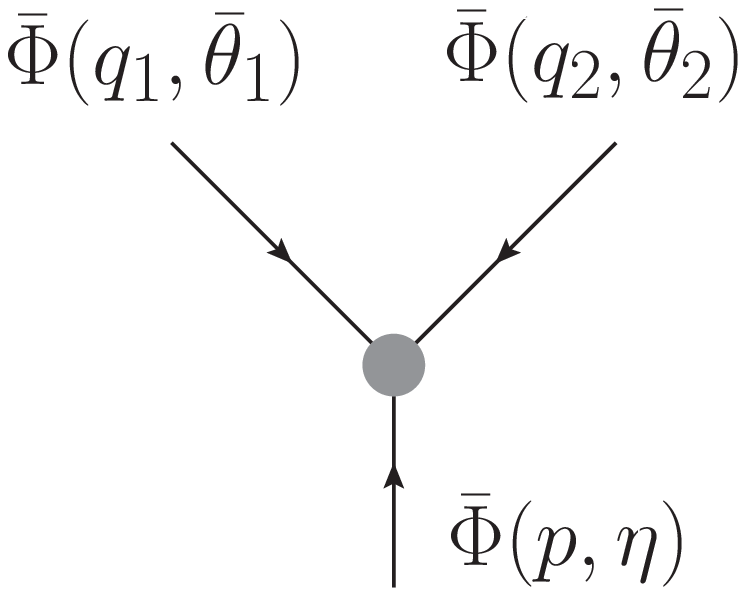}
\end{center}
\caption{Antichiral vertex function $\vev{\bar\Phi(q_1,\bq_1) \bar\Phi(q_2,\bq_2)|\bar\Phi(p,\eta)}_{\rm tree}$. } \label{fig01}
\end{figure}

The key elements in our treatment of the superconformal properties of $\cN=1$ matter amplitudes are the tree-level antichiral vertex function (super-form factor) in Fig.~\ref{fig01}, and its chiral counterpart. The antichiral vertex function is obtained by amputating one leg of the antichiral tree-level three-point function $\vev{\bar\Phi\bar\Phi\bar\Phi}$, and similarly for the chiral one. We first work out their manifestly supersymmetric expressions and then derive anomalous superconformal Ward identities for them.

\subsection{Supersymmetric  vertex functions}

Let us start with the antichiral vertex function. Using the propagator \p{e19}, the wave functions \p{534} and doing the Grassmann integration at the chiral 
interaction point (see the action \p{1.2}), we find
\begin{align}\label{219}
\bar\cF &\equiv \vev{\bar\Phi(q_1,\bq_1) \bar\Phi(q_2,\bq_2)|\bar\Phi(p,\eta)}_{\rm tree} \nt
&=  \delta^4(q_1+q_2+p)\,  \frac1{q^2_1 q^2_2} \, \int d^2\q_0\, e^{-(\bq_1 \tilde q_1 + \bq_2 \tilde q_2 + \eta \bra{\la})\q_0} \nt
&=  \delta^4(q_1+q_2+p)\,  \frac1{q^2_1 q^2_2} \, \delta^{(2)}(\bq_1 \tilde q_1 + \bq_2 \tilde q_2 + \eta \bra{\la}) \nt
& =  \frac1{q^2_1 q^2_2} \, \delta^4(P) \, \delta^{(2)}(Q)\,.
\end{align}
In the last line we have rewritten the Grassmann delta function (see \p{a6})  in terms of the  generator $Q$ from \p{e111} and \p{42}. In this form $Q$ supersymmetry is manifest, $\bQ$  supersymmetry follows from momentum conservation.

The same result could be obtained as follows. The general form of the vertex function compatible with translations $P$ and $\bar Q$ supersymmetry is
\begin{align}\label{e22}
\bar \cF  =   f(q_1,q_2,p)\, \delta^4(P) \, \delta^{(2)}(Q)\,,
\end{align}
with some function $f$ of the momenta. The latter can be determined  by comparing the Grassmann expansion of \p{e22} with some known component vertex, e.g.,
\begin{align}\label{}
 \bq_1^\da \bq_2^\db \eta^0\, \vev{\bar\psi_\da(q_1) \bar\psi_\db(q_2)|\bar\phi(p)}_{\rm tree} = \frac{\q_1 q_1 \tilde q_2 \q_2}{2q^2_1 q^2_2} \quad \Rightarrow \ f=\frac1{q^2_1 q^2_2} \,.
\end{align}

The chiral vertex function is obtained by conjugation, 
\begin{align}
\cF \equiv \vev{ \Phi(q_1,\q_1)  \Phi(q_2,\q_2)|\Phi(p,\be)}_{\rm tree} &=  \delta^4(q_1+q_2+p)\,  \frac1{4q^2_1 q^2_2} \, \delta^{(2)}(\q_1 \tilde q_1 + \q_2 \tilde q_2 +\be[\tl|) \nt
& =    \frac1{q^2_1 q^2_2}\, \delta^4(P) \, \delta^{(2)}(\bar Q)\,. \label{211}
\end{align}

\subsection{Superconformal anomaly of the antichiral vertex function}

In this subsection we want to show that the simple tree-level vertex functions  \p{219} and \p{211} have a superconformal anomaly. It originates from the  singularity in the propagator factor when both off-shell momenta become collinear with the on-shell one, $q_1\sim q_2 \sim p$. The anomaly takes the form of a contact term. This result is the key ingredient in our treatment of Feynman integrals in  the paper. 

\subsubsection{Naive superconformal invariance}

Let us consider the action of the generator $S_\a$
on the antichiral vertex function  \p{211}\,.  
Once $S_\a$ has gone past $Q^2$, it acts on the bosonic factor $\frac{\delta^{(4)}(P)}{q_1^2 q_2^2}$. The last term in the off-shell generator \p{549} relevant to the antichiral legs and the on-shell generator in \p{42} contain fermionic derivatives, so $S_\a\frac{\delta^{(4)}(P)}{q_1^2 q_2^2}=0$. Using the last of the relations \p{e32}   we get 
\begin{align}\label{224}
 S^\a\, \bar \cF =  2Q^\a (D -\tfrac{3}{2} R-H-2)  \, \frac{\delta^{(4)}(P)}{q_1^2 q_2^2}\,.
\end{align}
The vertex function satisfies the condition of invariance\footnote{Remember  that the generators \p{b10} contain the values $d,r$ of the dimension and R-charge, so that $DA=RA=0$;  the helicity of the on-shell leg is zero.} 
\begin{align}\label{}
0=(D -\tfrac{3}{2} R-H) \, \bar \cF = \delta^{(2)}(Q)\, (D -\tfrac{3}{2} R-H-2) \, \frac{\delta^{(4)}(P)}{q_1^2 q_2^2} \,,
\end{align}
hence  $S_\a \bar \cF=0$. This proves the (naive) superconformal invariance of the vertex function.

\subsubsection{The origin of the superconformal anomaly}\label{s232}

The above naive argument misses the hidden singularity in the propagator factor  in \p{224}.  We regularize it by introducing a Feynman parameter and a regulator of the analytic type:
\begin{align}\label{225}
\frac1{(q_1^2  + i 0)(q_2^2  + i 0)} = \lim_{\vep\to0}\, \int_0^1 \frac{d\xi}{((q_1 + \xi p)^2  + i 0 )^{2-\vep}}\,.
\end{align}
Then we examine the action of the dilatation operator in \p{224}. We have modified the dilatation weight of the regularized propagator factor \p{225}, therefore the action of $S_\a$ will give a non-vanishing result $\sim\vep$ 
\begin{align}\label{}
&S_\a\, \bar \cF 
= 2\delta^{(4)}(P)\,  Q_\a  \,  \int_0^1 \frac{ \vep\, d\xi}{((q_1 + \xi p)^2  + i 0)^{2-\vep}}  \,.
\end{align}
According to Ref.~\cite{Gelfand} the integrand is a singular distribution  with residue\footnote{This formula can be proven by Fourier transform.}
\begin{align}\label{}
 \lim_{\vep\to0}\, \vep\, ((q_1+\xi p)^2  + i 0)^{\vep-2} = i \pi^2 \delta^{(4)}(q_1+\xi p) \,.
\end{align}
Consequently, we find the superconformal anomaly (here $\bx = 1-\xi$)
\begin{align}\label{2.19}
S_\a \bar \cF & = \frac{i\pi^2}{2} \,  \int_0^1 d\xi\, Q_\a \, 
\delta^{(4)}(q_1+\xi p)\, \delta^{(4)}(q_2+\bx p)\nt& = \frac{i\pi^2}{2} \la_{\a}\,  \int_0^1 d\xi\, \Big(\eta + [\tl\bar\q_1]\xi + [\tl\bar\q_2] \bar\xi \Big) \, 
\delta^{(4)}(q_1+\xi p)\, \delta^{(4)}(q_2+\bx p)\nt
& = \frac{i\pi^2}{2} \la_{\a}\, \delta^{(4)}(P)\, \int_0^1 d\xi\, \Big(\eta + [\tl\bar\q_1]\xi + [\tl\bar\q_2] \bar\xi \Big) \, 
\delta^{(4)}(q_1+\xi p)\,.
\end{align}
We see that the anomaly is a contact term with support on the collinear configuration of the momenta $q_1\sim q_2 \sim p$.

The anomaly \p{2.19} is invariant under Poincar\'e supersymmetry,
\begin{align} \label{c11}
Q_{\b} S_\a \bar \cF =\bar Q_{\db} S_\a \bar \cF  = 0 \,. 
\end{align}
The first relation follows from the fact that $Q_\b$ together with $Q_\a$ in the first line in \p{2.19} form $Q^2$, which vanishes on the anomaly surface $q_1\sim q_2 \sim p$. The second relation can easily be shown using the generator $\bar Q_R$ from \p{e111} and the bosonic delta functions.  

The same argument shows the {absence of dilatation and $\bar S$ anomalies} of the antichiral vertex $\bar \cF$ \p{211}. Indeed,  the generators $D$ or  $\bar S_\da$  go through the fermionic factor $\delta^{(2)}(Q)$. The latter vanishes on the anomaly surface $q_1\sim q_2 \sim p$. 

Using the same approach, we can derive the $\bar S$ anomaly of the chiral vertex \p{211} with the on-shell super-state $\Psi(p,\eta)$, 
\begin{align}\label{e312}
&\bar S^\da \, F   =\frac{i\pi^2}{2} \tl^{\da}\, \delta^{(4)}(P)\,  \int_0^1 d\xi  \, e^{\eta(\vev{\la\q_1}\xi + \vev{\la\q_2}\bar\xi}  \, \delta^{(4)}(q_1+\xi p)
\end{align} 
and argue the absence of dilatation and $S$ anomalies.

Concluding  this subsection we wish to compare the role of the collinear singularities in the breakdown of conformal supersymmetry considered here, and that of conformal symmetry studied in  \cite{Chicherin:2017bxc}. Although the two mechanisms are very similar, the fact that the conformal boost generator $K_\mu$ is a second-order operator in the momenta makes its treatment more difficult. The conformal supersymmetry generators $S$, $\bar S$ are first order in the bosonic variables, which greatly simplifies the  derivation of the anomaly in \p{2.19}.

\subsection{The chiral vertex function in position/momentum space}

In this subsection we give an alternative derivation of the anomaly of the basic object, the chiral vertex function $\vev{\Phi\Phi|\Phi}$. This time we keep the off-shell legs in position space. This has the effect of smearing the contact term in \p{e312} and making it easier to detect. The anomaly is revealed by inserting the Lagrangian, see \cite{Chicherin:2017bxc} for a similar approach to the conformal collinear anomaly. The same method was originally used in the study of the conformal properties of Wilson loops in \cite{Drummond:2007au}.

\subsubsection{Computation of the vertex function}

We start by computing the vertex function
\begin{align}\label{737}
\cF(x_1,\q_1; x_2,\q_2; p,\be) := \vev{\Phi(x_1,\q_1)\, \Phi(x_2,\q_2)|\Phi(p,\be)}_{\rm tree}\,,
\end{align}
obtained from the chiral vertex $\Phi^3$ by putting one of its legs on shell, in the mixed representation where the off-shell legs are in position space and the on-shell leg is in momentum space. Of course, it can be obtained by Fourier transforming $q_i \to x_i$ the expression \p{219}, but it is instructive to do the calculation directly. 

In position space we have the superpropagator 
\begin{align}
& \vev{\bar\Phi(x_{1R},\bq_1) \Phi(x_{2L},\q_2)} = \frac1{\hat x_{12}^2}\,, \qquad \hat x_{12}^{\da\a} = x^{\da\a}_{1R} - x^{\da\a}_{2L} - 2 i \bq_1^\da \q_2^\a \,.
  \label{533}
\end{align}
 The mixed off/on-shell propagator (wave function) that we need is
\begin{align}
&\vev{\bar\Phi(x_{R},\bq) \Phi(p,\be)} =e^{ipx_{R} + [\tl\bq]\be}  \,. \label{534'} 
\end{align}

Using these Feynman rules   we get
\begin{align}\label{735}
\cF(x_1,\q_1; x_2,\q_2; p,\be) = \int d^{D}x_{0R} d^2\bq_0\, \frac{   e^{ipx_{0 R} + [\tl\bq_0]\be}}{(x_{10_R} - 2i\q_1 \bq_0)^2\, (x_{20_R} - 2i\q_2 \bq_0)^2}\,.
\end{align}
Note that the interaction vertex for three chiral superfields is in fact antichiral. We are using a dimensional regulator $D=4-2\ep$ in the measure because the Fourier integral, without the Grassmann shifts,  diverges if $p^2=0$. Introducing Schwinger parameters and doing the integration over $x_0$, we find
\begin{align}\label{2.3}
\cF&= \pi^{\frac{D}{2}}\int  d^2\bq_0\, e^{[\tl\bq_0]\be}\, \int_0^1 d\xi\, \int_0^\infty d\rho\, \rho^{-1+\ep}\nt
&\times \exp\big[-\frac{p^2}{4\rho}
 -\xi\bx (x_{12} - 2i\q_{12} \bq_0)^2 \rho + ip(x_1 - 2i \q_1\bq_0 )\xi  + ip(x_2 - 2i\q_2\bq_0 )\bx \; \big]\,.
\end{align}
Here, for the time being, we consider the spinor $\tl_\da$ as unrelated to $p^2\neq0$.

The integral over $\rho$ is potentially divergent after identifying $p=\ket{\la}[\tl|$,
\begin{align}\label{2.4}
\cF(p^2=0) &= \pi^{\frac{D}{2}} \Gamma(\ep) \int  d^2\bq_0\,  \int_0^1 d\xi\, \big(\xi\bx (x_{12} - 2i\q_{12} \bq_0)^2  \big)^{-\ep}\nt
&\times \exp\big( \,[\tl\bq_0 ]\be+ip(x_1 - 2i\q_1\bq_0 )\xi  + ip(x_2 - 2i\q_2\bq_0 )\bx \; \big)\,,
\end{align}
but the odd integral over $\bq_0$ makes it finite. To get the necessary factor $\bq^2_0$, we need at least one power of $\bq_0$ from the expansion of $(x_{12} - 2i\q_{12} \bq_0)^{-2\ep}$. In the limit $\ep\to0$ we obtain
\begin{align}\label{ch3Expl}
\cF &= \frac{2\pi^2}{x^2_{12}} \int  d^2\bq_0\,   \big(i \bra{\q_{12}} x_{12} |\bq_0]  +   \q^2_{12} \bq^2_0 \big) \nt
&\qquad\qquad \times \int_0^1 d\xi\, \exp\big(\, [\tl\bq_0]\be+ip(x_1 - 2i\q_1\bq_0 )\xi  + ip(x_2 - 2i\q_2\bq_0 )\bx \; \big)\nt
 &= \frac{\pi^2}{x^2_{12}} \int_0^1 d\xi\, e^{ip x_1 \xi  + i p x_2 \bar\xi} \Bigl[ 2\theta_{12}^2 + i \bra{\q_{12}} x_{12} |\tl] \Theta  \Bigr] \,,
\end{align}
where we use the shorthand notation
\begin{align} \label{Xi}
\Theta \equiv \be + \vev{\la  \q_1}\xi + \vev{\la  \q_2} \bar\xi\,, \qquad   Q_\a \Theta=0\,.
\end{align}

\subsubsection{Superconformal anomaly from Lagrangian insertion}

The anomaly originates from the regularized measure $d^{D}x_{0R}$ in \p{735}.
In the generator $\bar S$ from \p{128} we see the term $\bar S^\da \sim \bq^\da d$, where $d$ is the conformal weight. When this generator acts on the integral  \p{735}, we find different sources of weight factors $ \bq^\da_0 d$. If the measure were not regularized, all such factor would cancel exactly and the integral would be $\bar S$ invariant. The regulator creates a mismatch of the superconformal weights $\sim \ep \bq_0$, to be inserted in the integral \p{2.4}. In other words, we have effectively inserted the antichiral Lagrangian $\ep \bq_0 L(x_0,\bq_0)$ into the vertex function, as a probe for a possible anomaly: 
\begin{align}\label{743}
\bar S^\da \cF &=\lim_{\ep\to0} i\pi^{\frac{D}{2}} \Gamma(\ep)\int  d^2\bq_0\,   \ep\bq^\da_0 \int_0^1 d\xi\,  \big(\xi\bx (x_{12}-2i\q_{12} \bq_0)^2  \big)^{-\ep}\nt
&\qquad\qquad \times \exp\bigl( [\tl\bq_0]\be+ip(x_1 -2i \q_1\bq_0 )\xi  + ip(x_2 - 2i \q_2\bq_0 )\bx \; \bigr)\nt
&=i\pi^2 \int  d^2\bq_0\, \bq^\da_0\,   \int_0^1 d\xi\,  \exp\bigl( [\tl\bq_0 ]\be+ip(x_1 - 2i \q_1\bq_0 )\xi  + ip(x_2 - 2i \q_2\bq_0 ) \bar\xi\, \bigr)\nt
&= \frac{i\pi^2}{2} \tilde \la^\da\, \int_0^1 d\xi\, [\be + \vev{\la  \q_1}\xi + \vev{\la  \q_2} \bar\xi]\,  e^{ i p x_1 \xi  + i p x_2 \bar\xi} 
\equiv \cA^\da_{\bar S}\,.
\end{align}
This is the anomalous superconformal Ward identity \p{e312}, Fourier transformed to position space. The anomaly originates from the pole in the integral \p{2.4}. It is not compensated by the  odd integration anymore because of the inserted $\bq^\da_0$.

The other generator $S^\a$ in \p{549} contains no weight factor in the antichiral realization, therefore it is not anomalous. The anticommutator $\{S,\bar S\} = K$ then generates the conformal anomaly. We note that the anomaly \p{743} contains no component $\sim \q^0_1 \q^2_2 \be$, in accord with the absence of an anomaly in the auxiliary form factor $\vev{\phi(x_1) F(x_2)|\phi(p)}$.

\subsubsection{Conformal boost anomaly}

The conformal generator $K^\mu=\ldots + x^\mu d$  contains a term which measures the conformal weight. The  conformal anomaly of the form factor  \p{737} is obtained by inserting $x^\mu_{0\,R} L(x_0,\bq_0)$. Adapting the argument from Ref.~\cite{Chicherin:2017bxc} to the integral \p{2.3}, we find 
\begin{align}\label{231}
&\int d^{D}x_{0R} d^2\bq_0\,  i x^{\a\da}_{0\, R}\,   \vev{\Phi(1)\Phi(2) L(0)|\Phi(p,\be)}\nt
&=\pi^{\frac{D}{2}}\int_0^1 d\xi\, e^{ip(x_1\xi+x_2\bx)}  \int   d^2\bq_0 \, e^{[\tl\bq_0] \Theta}\int_0^\infty d\rho\, \rho^{-1+\ep} \, 
\exp\bigl(-\xi\bx (x_{12}-2i\q_{12} \bq_0)^2 \rho   \bigr)  \nt
& \times\left(-\frac{p}{2\rho} +i  (x_1 -2i\q_1\bq_0 )  \xi  + i(x_2 -2i\q_2\bq_0 )   \bx + O(\rho)   \right)^{\da\a}\, .
\end{align}
The  $\rho$ integral has a pole at $\rho\to0$ with residue
\begin{align}\label{}
&\ep \int_0^\infty d\rho\, (\ldots)^{\da\a} =    i (x_1 - 2i\q_1\bq_0 )^{\da\a}  \xi  + i(x_2 - 2i\q_2\bq_0 )^{\da\a}  \bx  +\frac1{2}  \xi\bx (x_{12}-2i\q_{12} \bq_0)^2 \, \la^\a\tl^\da \nt
& \rightarrow \ 2(\q_1^\a\xi+ \q_2^\a \bx)\bq_0^\da  - \xi\bx ( i \bra{\q_{12}} x_{12} |\bq_0] + \q_{12}^2 \bq_0^2) \, \la^\a\tl^\da\,.
\end{align}
In the second line we have dropped the terms $\sim (\bq_0)^0$ because the exponential can supply at most one power of $\bq_0$. Doing the integral $ \int   d^2\bq_0$ and integrating $\xi$ by parts yields the conformal anomaly 
\begin{align}\label{Kanom}
\cA^{\a\da}_K &= \pi^2 \tl^\da\int_0^1 d\xi\, e^{ip(x_1\xi+x_2\bx)} 
\Bigl[ - \tfrac{i}{2} \xi\bar\xi \vev{\la\q_1}\vev{\la\q_2} \tl_{\db} x_{12}^{\db\a} + \bar\xi \q^{\a}_{1} \vev{\la \q_{2}} \notag\\
&- \xi\vev{\la \q_{1}} \q^{\a}_{2}  + \bigl( \q_1^\a\bar\xi+ \q_2^\a \xi - \tfrac{i}{2}\xi\bar\xi \vev{\la\q_{12}} \tl_{\db}x^{\db\a}_{12} \bigr)\be  \Bigr]\,,
\end{align}
where the $\theta_1^2$ and $\theta_2^2$ components are  absent. This implies that the component  vertex  functions involving the auxiliary field $F$ are anomaly free. 
So, the conformal anomaly occurs only in the Yukawa vertex with an on-shell scalar state $\phi(p)$ or an on-shell fermion state $\psi_{+}(p)$ (see Eq.~\p{72}). This anomaly was first revealed in Ref.~\cite{Chicherin:2017bxc}.

The  algebra \p{b1} implies  the following identities for the conformal anomaly:
\begin{align}\label{}
&K^{\a\da} \cF = \cA^{\a\da}_K \ \stackrel{Q_\b}{\longrightarrow} \ [Q_\b, K^{\a\da}] \cF= \delta^\a_\b \bar S^\da \cF   \ \Rightarrow\  Q_\b   \cA^{\a\da}_K=  \delta^\a_\b \cA^\da_{\bar S}\,,\nt
&K^{\a\da} \cF = \cA^{\a\da}_K \ \stackrel{\bar Q_\db}{\longrightarrow} \ [\bar Q_\db, K^{\a\da}] \cF= \delta^\da_\db S^\a \cF=0   \ \Rightarrow\  \bar Q_\db  \cA^{\a\da}_K=  0\,, \nt
&\bar S^\da \cF = \cA^\da_{\bar S}  \ \stackrel{S^\a}{\longrightarrow} \ \{S^\a, \bar S^\da\} \cF = K^{\a\da} \cF \ \Rightarrow\  S^\a \cA^\da_{\bar S} = \cA^{\a\da}_K \label{idK3}\,.
\end{align}
We have checked that they hold indeed.

\section{One-loop integrals}\label{Appendix1LoopIntegrals}

In this Appendix we collect $D=4$ one-loop Feynman integrals which arise in the calculation of superconformal anomalies of two-loop graphs in Sects.~\ref{sec_dblbox}, \ref{s5.3}. All of these integrals are explicitly UV- and IR-finite.
\begin{figure}
\begin{center}
\includegraphics[width = 6cm]{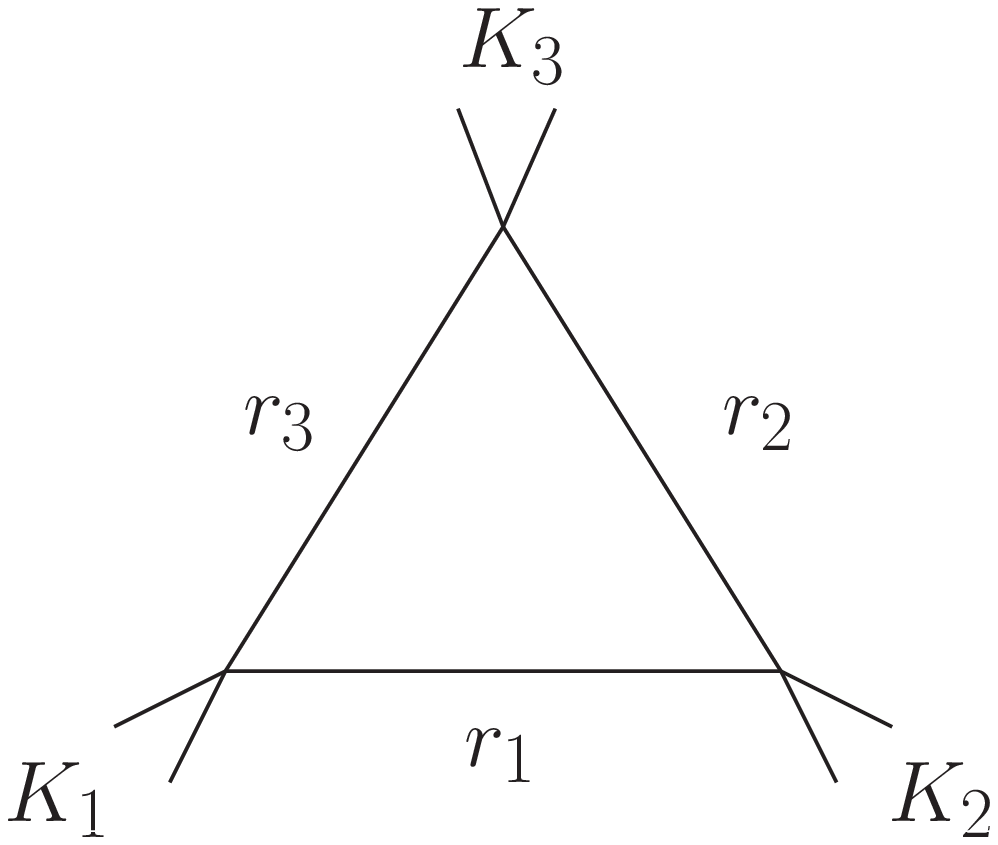}\qquad\quad \includegraphics[width = 6cm]{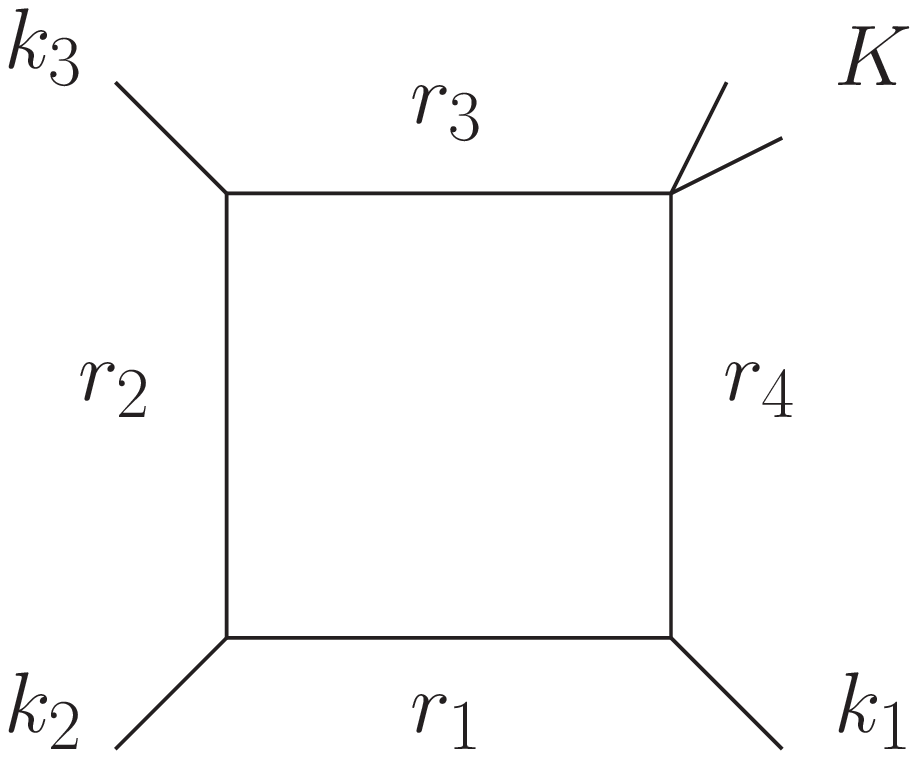}
\end{center}
\caption{Left: Three-mass scalar triangle with $K_i^2 \neq 0$, $i=1,2,3$. Right: One-mass box  with numerator $[1|\tilde K r_3 \tilde r_2 \ket{2}$ defined in Eq.~\p{boxnum}. Legs $k_1, k_2, k_3$ are massless, i.e. $k_i = \ket{i}[i|$, and $K=-k_1-k_2-k_3$ is massive.} \label{box1mnum}
\end{figure}

\begin{figure}
\begin{center}
\includegraphics[width = 6cm]{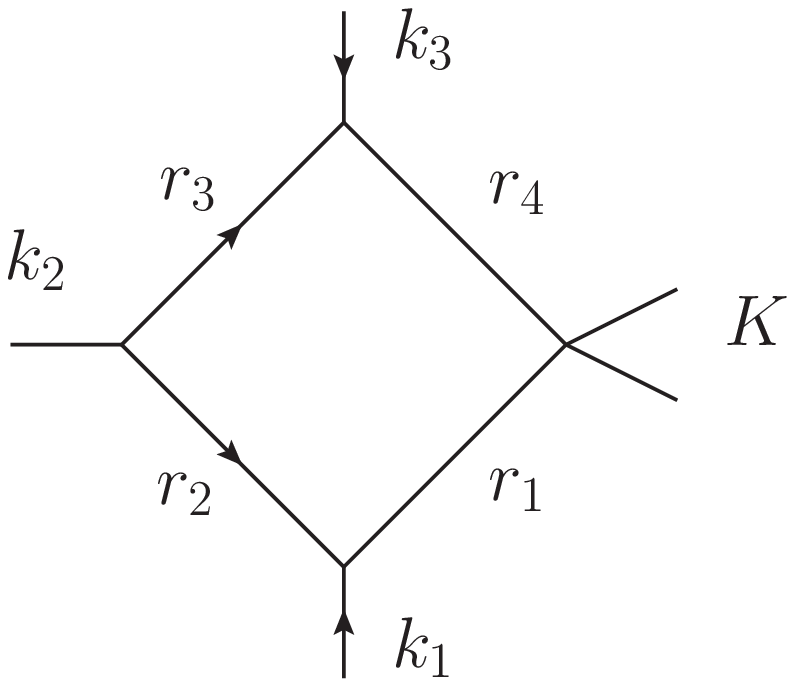}\qquad\quad \includegraphics[width = 5.5cm]{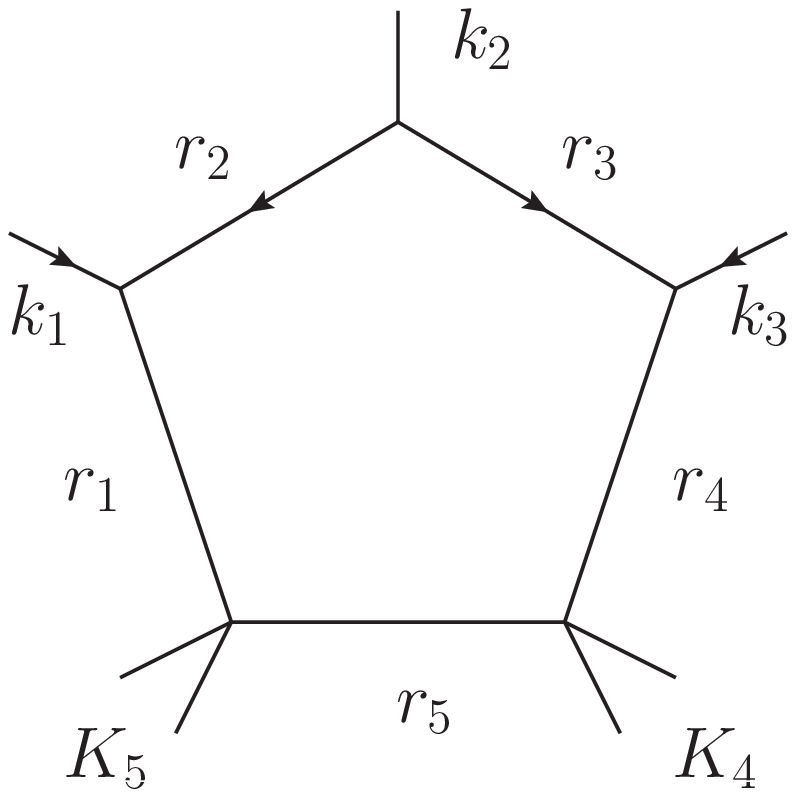}
\end{center}
\caption{Left: One-mass box  with numerator $\vev{1| r_2 \tilde r_3 |3}$ defined in Eq.~\p{boxnum1}. Legs $k_1, k_2, k_3$ are massless and $K=-k_1-k_2-k_3$ is massive. Right: Pentagon with numerator $\vev{1| r_2 \tilde r_3 |3}$ defined in Eq.~\p{pentagon_int}. Legs $k_1, k_2, k_3$ are massless and legs $K_4, K_5$ are massive. 
} \label{boxpentnum}
\end{figure}

The scalar three-mass triangle is defined as usual,
\begin{align}
{\rm Tri}(K_1^2,K_2^2,K_3^2) = \int  \frac{d^4 \ell}{i\pi^2}\frac{1}{r_1^2 r_2^2 r_3^2}\, \label{3mass}
\end{align}
with off-shell $K_i^2 \neq 0$, $i=1,2,3$, and $K_1+K_2+K_3 = 0$. The off-shell momenta $r_1,r_2,r_3$ flowing along the box edges are defined on the lhs of Fig.~\ref{box1mnum}; $\ell$ is the loop momentum.
The triangle integral is obviously IR finite since all of its corners are massive. This integral has been evaluated e.g. in \cite{Usyukina:1992jd}.

The one-mass box (rhs of Fig.~\ref{box1mnum}) with a numerator defined as follows
\begin{align}
{\rm box}(k_1,k_2,k_3) = \int \frac{d^4 \ell}{i\pi^2} \frac{[1|\tilde K r_3 \tilde r_2 \ket{2}}{r_1^2 r_2^2 r_3^2 r_4^2} \label{boxnum}
\end{align}
for massless $k_1^2 = k_2^2 = k_3^2 =0$, i.e. $k_i = \ket{i}[i|$, and massive $K=-k_1-k_2-k_3$, i.e. $K^2 \neq 0$. 
Due to the numerator this Feynman integral is IR finite in $D=4$ dimensions. A simple calculation of the maximal cut of the box leads to the following expression 
\begin{align}
{\rm box}(k_1,k_2,k_3) = -\frac{\vev{23}}{\vev{13}}\left[  {\rm Li}_2\left( 1 - \frac{K^2}{t_{12}} \right) 
+ {\rm Li}_2\left( 1 - \frac{K^2}{t_{23}} \right)  +\frac{1}{2}\log^2 \left(\frac{t_{12}}{t_{23}}\right) + \frac{\pi^2}{6}\right] \,,\label{boxexpl}
\end{align}
where $t_{ij}=(k_i+k_j)^2$ are the two-particle Mandelstam invariants.

The one-mass box (lhs of Fig.~\ref{boxpentnum}) with a numerator defined as follows
\begin{align}
{\rm Box}(k_1,k_2,k_3) = \int \frac{d^4 \ell}{i\pi^2} \frac{\vev{1| r_2 \tilde r_3 |3}}{r_1^2 r_2^2 r_3^2 r_4^2} \label{boxnum1}
\end{align}
for massless $k_1, k_2, k_3$ and massive $K=-k_1-k_2-k_3$. This integral coincides with the finite part of the one-mass scalar box,
\begin{align}\label{F5}
{\rm Box}(k_1,k_2,k_3) = \frac{1}{[13]} \left[ {\rm Li}_2 \left( 1 - \frac{K^2}{t_{12}}\right) + {\rm Li}_2 \left( 1 - \frac{K^2}{t_{23}}\right) + \frac{1}{2} \log^2 \left( \frac{t_{12}}{t_{23}}\right)+ \frac{\pi^2}{6} \right]\,.
\end{align}

The pentagon  with two massive corners and a numerator (rhs of Fig.~\ref{boxpentnum})
\begin{align}
{\rm Pent}(k_1,k_2,k_3,K_4,K_5) = \int \frac{d^4 \ell}{i\pi^2} \frac{\vev{1| r_2 \tilde r_3 |3}}{r_1^2 r_2^2 r_3^2 r_4^2 r_5^2} \, \label{pentagon_int}
\end{align}
where $k_1+k_2+k_3+K_4+K_5 =0$.
This integral evaluates to the following expression
\begin{align}
{\rm Pent}(k_1,k_2,k_3,K_4,K_5) =& 
\frac{1}{[1|K_5 \tilde K_4|3]} \biggl[ - \frac{\pi^2}{6} + \log\left(\frac{t_{12} K_4^2}{ t_{34} t_{45} }\right)\log\left(\frac{t_{23} K_5^2}{ t_{15} t_{45}}\right) \nt 
&+ 
{\rm Li}_2 \left( 1- \frac{t_{12} K_4^2}{ t_{34}t_{45}} \right)  + {\rm Li}_2 \left( 1- \frac{t_{23} K_5^2}{t_{15} t_{45}} \right) \biggr]\,.
\label{pentint}
\end{align}

\section{Collinearity at the integrand level and holomorphic anomaly} \label{appE}

In this appendix we explain the relationship between the $S$-supersymmetry anomaly of loop integrals studied in this paper and  the so-called `holomorphic anomaly' of Ref.~\cite{Cachazo:2004by}. We show that both are manifestations of the same  phenomenon of collinear singularities of the integrand and the associated breakdown of the naive twistor collinearity. 

Let us revisit the one-mass box integral with a numerator \p{422} (see the rhs of Fig.~\ref{fig1}),
\begin{align}\label{1601}
\cI(p) = \frac{1}{\vev{45} [23]} 
\int \frac{d^4 \ell}{\pi^2}\frac{\bra{1}q_2 \tilde q_3\ket{4}}{q^2_{1} q^2_{2} q^2_{3} q^2_{4}}\,.
\end{align}
 In Sect.~\ref{sec_1loop} we showed that the collinearity operator  $F_{123}$ acting on $\cI(p)$ is anomalous, Eq.~\p{e529}. Now we want to demonstrate it without invoking the superconformal symmetry. Instead of the integral itself we can study its various cuts, where we encounter the holomorphic anomaly.
 
\begin{figure}
\begin{center}
\includegraphics[width = 6cm]{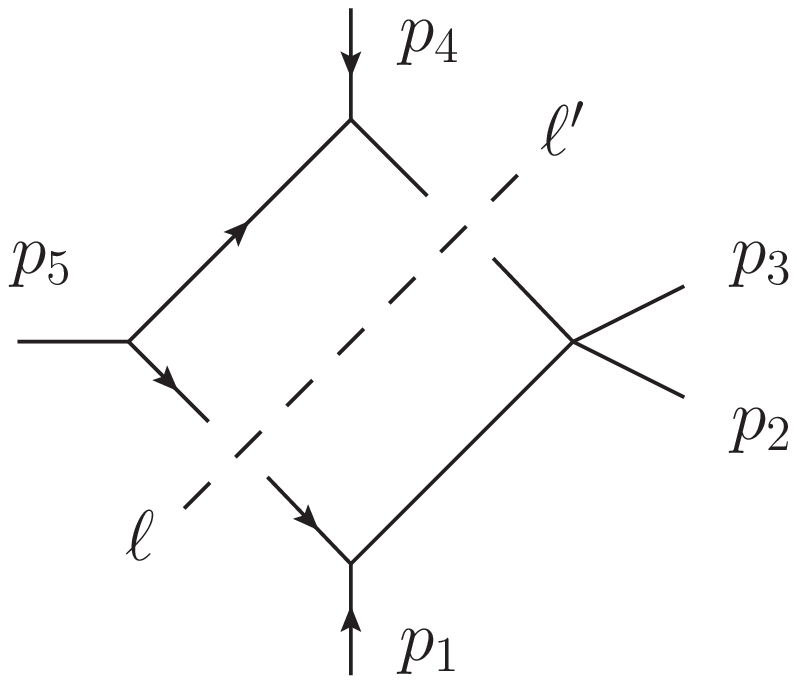} \qquad
\includegraphics[width = 6cm]{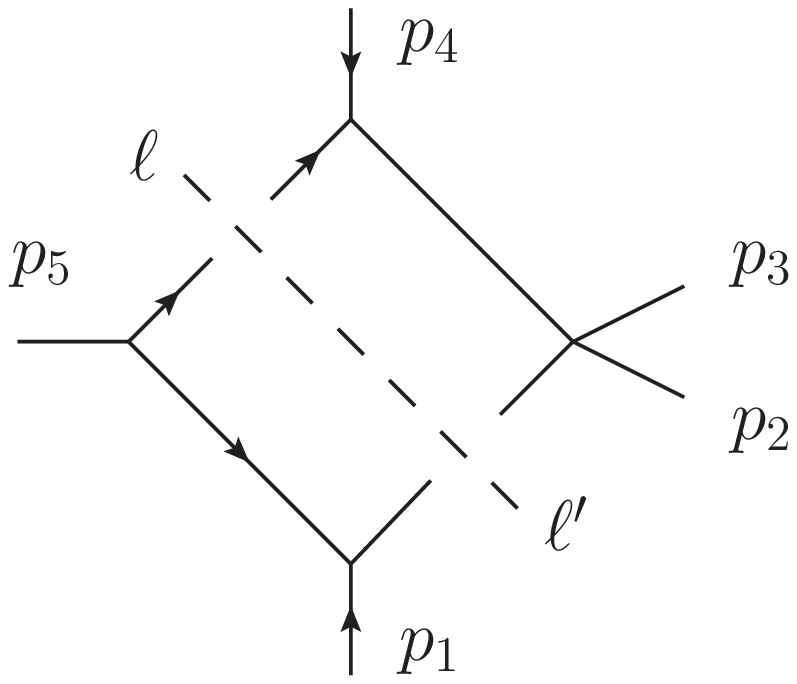}
\end{center}
\caption{Dicsontinuities of the one-mass box integral \p{1601} in $s_{45}$ (left) and in $s_{15}$ (right).} \label{boxx}
\end{figure}

First we cut the integral through the momenta $q_2 \equiv \ell$, $q_4 \equiv \ell'$
as on the lhs of Fig.~\ref{boxx}, 
\begin{align}\label{1632}
{\rm Disc}_{s_{45}}\, \cI &= \int d^4 \ell\, \delta^{(+)}(\ell^2)\, \delta^{(+)}((\ell+p_{45})^2)  \, \frac{1}{[23][1|\tilde \ell\ket{5}} \,.
\end{align}
Naively,   the collinearity operator $F^\a_{123}$ annihilates the integrand. 
In reality, due to the pole at $\tilde \mu \equiv \tilde l \ket{5} \sim \tl_1$  the operator  generates  the contact term (`holomorphic anomaly')
\begin{align}\label{hola}
\pd{\la^\a_1}\, \frac{1 }{[\tilde \mu 1]} = 2\pi i\, \ket{1}_\a \, \delta(\mu,1)\delta(\tilde\mu,\tilde 1)\,,
\end{align}
which is the covariant version of the well-known complex relation $\pa/\pa z (1/\bar z) \sim \delta^2(z,\bar z)$.
Following \cite{Cachazo:2004dr,Cachazo:2004kj}, we integrate over the future light cone of $\ell$ in terms of $\la,\tl$, using the parametrization for null vectors $\ell = t\la \tl$:
\begin{align}\label{e.5}
&\int d^4 \ell\, \delta^{(+)}(\ell^2) \, (\ldots) = \int_0^\infty dt \, t \int d\Omega \, (\ldots) \,, \qquad  d\Omega = \vev{\la d\la}\, [\tl d\tl]\,.
\end{align}
So, the integral in \p{1632}  becomes
\begin{align}\label{E5}
 \int_0^\infty dt \, t \int d\Omega  \,  \delta^{(+)}(t\bra{\la}p_{45}|\tl])  \, \frac{1}{[23][1\tl]\, t\, \vev{\la 5}} =     \int d\Omega  \,    \frac{1}{ [23] [1\tl]\vev{\la 5}\bra{\la}p_{45}|\tl]}\,.
\end{align}
Now we act with $F_{123}$ and apply the relation \p{hola}. The delta function removes the integral $ \int d\Omega$ and identifies $\ell=p_1$. 
With this we obtain the anomalous relation 
\begin{align}\label{1638}
F_{123}\, {\rm Disc}_{s_{45}}\, \cI =-2\pi i\, \frac{\ket{1} }{\vev{15}\,   (s_{23}-s_{45})  }\,,
\end{align}
whose rhs is precisely the $s_{45}$ cut of the anomaly $a_1$ in \p{e529}, \p{a1a4}.

Next we cut the momenta $q_3 \equiv \ell$, $q_1 \equiv \ell'$
as on the rhs of Fig.~\ref{boxx}.  Using  \p{e.5} we get 
\begin{align}\label{e.9}
{\rm Disc}_{s_{15}}\, \cI &   =   \int d\Omega  \,    \frac{\vev{15}}{\vev{45}[23][4\tl]\vev{\la 5}\bra{\la}p_{15}|\tl] }\,.
\end{align}
Then we act with  $F^\a_{123}$. Only leg 1 is involved  and we find (naively!)
\begin{align}\label{e.11}
(F_{123})_\a {\rm Disc}_{s_{15}}\, \cI & = -\frac1{\vev{45}  }\int d\Omega  \,   \frac1{ [4\tl] } \pd{\la^\a}\, \left[\frac{1 }{\bra{\la}p_{15}|\tl] }\right]= 0\,.
\end{align}
In reality, when integrating by parts the total derivative in the last relation, we generate  the contact term (`holomorphic anomaly')
\begin{align}\label{}
\pd{\la^\a}\, \frac{1 }{[4\tl]} =-2\pi i\, \ket{4}_\a \, \delta(\la,4)\delta(\tl,\tilde 4)\,.
\end{align} 
With this we obtain the anomalous relation 
\begin{align}\label{1638}
F_{123}\, {\rm Disc}_{s_{15}}\, \cI =-2\pi i \frac{\ket{4} }{\vev{45}\,   (s_{23}-s_{15})  }\,,
\end{align}
which is precisely the $s_{15}$ cut of the anomaly $a_4$ in \p{e529}, \p{a1a4}.

The examples above illustrate the relationship between the superconformal anomalies of the $\cN=1$ amplitudes   in this paper and  of the $\cN=4$ amplitude in \cite{Bargheer:2009qu,Korchemsky:2009hm}. The reason is a singularity when an external on-shell momentum becomes collinear with a loop momentum. In \cite{Bargheer:2009qu,Korchemsky:2009hm} this happens when two helicity spinors become collinear. Our new mechanism described in App.~\ref{s232} involves three collinear momenta. The former phenomenon can only tell us what happens on an unitarity cut of a loop integral, where all the momenta are already on shell. The latter concerns a whole region in the loop momentum space. This is why  we can derive Ward identities for the loop integral itself, not just its discontinuity. 

Here we reveal an important feature of the anomaly mechanism. In \p{E5} the naive twistor collinearity  is obvious, while in \p{e.11} we first have to integrate by parts. A similar phenomenon is behind the naive property of the loop integrals  in this paper. We derive the twistor collinearity Ward identity \p{e513} from the $S$-variation of a supergraph. Naively, one can argue that the super-integral must be $S$-invariant because it originates from an invariant Lagrangian. However, if we forget the supersymmetric origin of a specific integral, e.g., $\cI$ in \p{422}, we cannot immediately tell why it should  satisfy the naive Ward identity $F_{123}\, \cI=0$ (cf. \p{e529}). Like in \p{e.11}, we would have to show that the collinearity operator acting on the integrand produces a   total derivative. At present we do not know a simple criterion whether a given loop integral might be naively twistor-collinear. It would be interesting to find out which property of the loop integrand is responsible for this phenomenon.

\bibliographystyle{JHEP}
\bibliography{refs}

\providecommand{\href}[2]{#2}\begingroup\raggedright\begin{thebibliography}{10}

\bibitem{Braun:2003rp}
V.~M. Braun, G.~P. Korchemsky, and D.~Mueller, {\it {The Uses of conformal
  symmetry in QCD}},  {\em Prog. Part. Nucl. Phys.} {\bf 51} (2003) 311--398,
  [\href{http://xxx.lanl.gov/abs/hep-ph/0306057}{{\tt hep-ph/0306057}}].

\bibitem{Bendavid:2018nar}
J.~R. Andersen et~al., {\it {Les Houches 2017: Physics at TeV Colliders
  Standard Model Working Group Report}},  in {\em {10th Les Houches Workshop on
  Physics at TeV Colliders (PhysTeV 2017) Les Houches, France, June 5-23,
  2017}}, 2018.
\newblock \href{http://xxx.lanl.gov/abs/1803.0797}{{\tt arXiv:1803.0797}}.

\bibitem{Chicherin:2018ubl}
D.~Chicherin, J.~M. Henn, and E.~Sokatchev, {\it {Amplitudes from
  superconformal Ward identities}},  {\em Phys. Rev. Lett.} {\bf 121} (2018),
  no.~2 021602, [\href{http://xxx.lanl.gov/abs/1804.0357}{{\tt
  arXiv:1804.0357}}].

\bibitem{Collins:1989gx}
J.~C. Collins, D.~E. Soper, and G.~F. Sterman, {\it {Factorization of Hard
  Processes in QCD}},  {\em Adv. Ser. Direct. High Energy Phys.} {\bf 5} (1989)
  1--91, [\href{http://xxx.lanl.gov/abs/hep-ph/0409313}{{\tt hep-ph/0409313}}].

\bibitem{Catani:1998bh}
S.~Catani, {\it {The Singular behavior of QCD amplitudes at two loop order}},
  {\em Phys. Lett.} {\bf B427} (1998) 161--171,
  [\href{http://xxx.lanl.gov/abs/hep-ph/9802439}{{\tt hep-ph/9802439}}].

\bibitem{Aybat:2006mz}
S.~M. Aybat, L.~J. Dixon, and G.~F. Sterman, {\it {The Two-loop soft anomalous
  dimension matrix and resummation at next-to-next-to leading pole}},  {\em
  Phys. Rev.} {\bf D74} (2006) 074004,
  [\href{http://xxx.lanl.gov/abs/hep-ph/0607309}{{\tt hep-ph/0607309}}].

\bibitem{Dixon:2008gr}
L.~J. Dixon, L.~Magnea, and G.~F. Sterman, {\it {Universal structure of
  subleading infrared poles in gauge theory amplitudes}},  {\em JHEP} {\bf 08}
  (2008) 022, [\href{http://xxx.lanl.gov/abs/0805.3515}{{\tt
  arXiv:0805.3515}}].

\bibitem{Becher:2009cu}
T.~Becher and M.~Neubert, {\it {Infrared singularities of scattering amplitudes
  in perturbative QCD}},  {\em Phys. Rev. Lett.} {\bf 102} (2009) 162001,
  [\href{http://xxx.lanl.gov/abs/0901.0722}{{\tt arXiv:0901.0722}}]. [Erratum:
  Phys. Rev. Lett.111,no.19,199905(2013)].

\bibitem{Almelid:2015jia}
{\O}.~Almelid, C.~Duhr, and E.~Gardi, {\it {Three-loop corrections to the soft
  anomalous dimension in multileg scattering}},  {\em Phys. Rev. Lett.} {\bf
  117} (2016), no.~17 172002, [\href{http://xxx.lanl.gov/abs/1507.0004}{{\tt
  arXiv:1507.0004}}].

\bibitem{Wess:1973kz}
J.~Wess and B.~Zumino, {\it {A Lagrangian Model Invariant Under Supergauge
  Transformations}},  {\em Phys. Lett.} {\bf 49B} (1974) 52.

\bibitem{Ferrara:1974fv}
S.~Ferrara, J.~Iliopoulos, and B.~Zumino, {\it {Supergauge Invariance and the
  Gell-Mann - Low Eigenvalue}},  {\em Nucl. Phys.} {\bf B77} (1974) 413.

\bibitem{Chicherin:2017bxc}
D.~Chicherin and E.~Sokatchev, {\it {Conformal anomaly of generalized form
  factors and finite loop integrals}},  {\em JHEP} {\bf 04} (2018) 082,
  [\href{http://xxx.lanl.gov/abs/1709.0351}{{\tt arXiv:1709.0351}}].

\bibitem{Bargheer:2009qu}
T.~Bargheer, N.~Beisert, W.~Galleas, F.~Loebbert, and T.~McLoughlin, {\it
  {Exacting N=4 Superconformal Symmetry}},  {\em JHEP} {\bf 11} (2009) 056,
  [\href{http://xxx.lanl.gov/abs/0905.3738}{{\tt arXiv:0905.3738}}].

\bibitem{Korchemsky:2009hm}
G.~P. Korchemsky and E.~Sokatchev, {\it {Symmetries and analytic properties of
  scattering amplitudes in N=4 SYM theory}},  {\em Nucl. Phys.} {\bf B832}
  (2010) 1--51, [\href{http://xxx.lanl.gov/abs/0906.1737}{{\tt
  arXiv:0906.1737}}].

\bibitem{Beisert:2010gn}
N.~Beisert, J.~Henn, T.~McLoughlin, and J.~Plefka, {\it {One-Loop
  Superconformal and Yangian Symmetries of Scattering Amplitudes in N=4 Super
  Yang-Mills}},  {\em JHEP} {\bf 04} (2010) 085,
  [\href{http://xxx.lanl.gov/abs/1002.1733}{{\tt arXiv:1002.1733}}].

\bibitem{Cachazo:2004by}
F.~Cachazo, P.~Svrcek, and E.~Witten, {\it {Gauge theory amplitudes in twistor
  space and holomorphic anomaly}},  {\em JHEP} {\bf 10} (2004) 077,
  [\href{http://xxx.lanl.gov/abs/hep-th/0409245}{{\tt hep-th/0409245}}].

\bibitem{Bidder:2004tx}
S.~J. Bidder, N.~E.~J. Bjerrum-Bohr, L.~J. Dixon, and D.~C. Dunbar, {\it {N=1
  supersymmetric one-loop amplitudes and the holomorphic anomaly of unitarity
  cuts}},  {\em Phys. Lett.} {\bf B606} (2005) 189--201,
  [\href{http://xxx.lanl.gov/abs/hep-th/0410296}{{\tt hep-th/0410296}}].

\bibitem{CaronHuot:2011ky}
S.~Caron-Huot, {\it {Superconformal symmetry and two-loop amplitudes in planar
  N=4 super Yang-Mills}},  {\em JHEP} {\bf 12} (2011) 066,
  [\href{http://xxx.lanl.gov/abs/1105.5606}{{\tt arXiv:1105.5606}}].

\bibitem{Bullimore:2011kg}
M.~Bullimore and D.~Skinner, {\it {Descent Equations for Superamplitudes}},
  \href{http://xxx.lanl.gov/abs/1112.1056}{{\tt arXiv:1112.1056}}.

\bibitem{Dixon:2011pw}
L.~J. Dixon, J.~M. Drummond, and J.~M. Henn, {\it {Bootstrapping the three-loop
  hexagon}},  {\em JHEP} {\bf 11} (2011) 023,
  [\href{http://xxx.lanl.gov/abs/1108.4461}{{\tt arXiv:1108.4461}}].

\bibitem{Caron-Huot:2016owq}
S.~Caron-Huot, L.~J. Dixon, A.~McLeod, and M.~von Hippel, {\it {Bootstrapping a
  Five-Loop Amplitude Using Steinmann Relations}},  {\em Phys. Rev. Lett.} {\bf
  117} (2016), no.~24 241601, [\href{http://xxx.lanl.gov/abs/1609.0066}{{\tt
  arXiv:1609.0066}}].

\bibitem{Drummond:2008vq}
J.~M. Drummond, J.~Henn, G.~P. Korchemsky, and E.~Sokatchev, {\it {Dual
  superconformal symmetry of scattering amplitudes in N=4 super-Yang-Mills
  theory}},  {\em Nucl. Phys.} {\bf B828} (2010) 317--374,
  [\href{http://xxx.lanl.gov/abs/0807.1095}{{\tt arXiv:0807.1095}}].

\bibitem{Goncharov:2010jf}
A.~B. Goncharov, M.~Spradlin, C.~Vergu, and A.~Volovich, {\it {Classical
  Polylogarithms for Amplitudes and Wilson Loops}},  {\em Phys. Rev. Lett.}
  {\bf 105} (2010) 151605, [\href{http://xxx.lanl.gov/abs/1006.5703}{{\tt
  arXiv:1006.5703}}].

\bibitem{Duhr:2011zq}
C.~Duhr, H.~Gangl, and J.~R. Rhodes, {\it {From polygons and symbols to
  polylogarithmic functions}},  {\em JHEP} {\bf 10} (2012) 075,
  [\href{http://xxx.lanl.gov/abs/1110.0458}{{\tt arXiv:1110.0458}}].

\bibitem{Witten:2003nn}
E.~Witten, {\it {Perturbative gauge theory as a string theory in twistor
  space}},  {\em Commun. Math. Phys.} {\bf 252} (2004) 189--258,
  [\href{http://xxx.lanl.gov/abs/hep-th/0312171}{{\tt hep-th/0312171}}].

\bibitem{ArkaniHamed:2010gh}
N.~Arkani-Hamed, J.~L. Bourjaily, F.~Cachazo, and J.~Trnka, {\it {Local
  Integrals for Planar Scattering Amplitudes}},  {\em JHEP} {\bf 06} (2012)
  125, [\href{http://xxx.lanl.gov/abs/1012.6032}{{\tt arXiv:1012.6032}}].

\bibitem{Chicherin:2017dob}
D.~Chicherin, J.~Henn, and V.~Mitev, {\it {Bootstrapping pentagon functions}},
  {\em JHEP} {\bf 05} (2018) 164,
  [\href{http://xxx.lanl.gov/abs/1712.0961}{{\tt arXiv:1712.0961}}].

\bibitem{Gehrmann:2018yef}
T.~Gehrmann, J.~M. Henn, and N.~A. Lo~Presti, {\it {Pentagon functions for
  massless planar scattering amplitudes}},  {\em JHEP} {\bf 10} (2018) 103,
  [\href{http://xxx.lanl.gov/abs/1807.0981}{{\tt arXiv:1807.0981}}].

\bibitem{Abreu:2018rcw}
S.~Abreu, B.~Page, and M.~Zeng, {\it {Differential equations from unitarity
  cuts: nonplanar hexa-box integrals}},
  \href{http://xxx.lanl.gov/abs/1807.1152}{{\tt arXiv:1807.1152}}.

\bibitem{Chicherin:2018mue}
D.~Chicherin, T.~Gehrmann, J.~M. Henn, N.~A. Lo~Presti, V.~Mitev, and
  P.~Wasser, {\it {Analytic result for the nonplanar hexa-box integrals}},
  \href{http://xxx.lanl.gov/abs/1809.0624}{{\tt arXiv:1809.0624}}.

\bibitem{Chawdhry:2018awn}
H.~A. Chawdhry, M.~A. Lim, and A.~Mitov, {\it {Two-loop five-point massless QCD
  amplitudes within the IBP approach}},
  \href{http://xxx.lanl.gov/abs/1805.0918}{{\tt arXiv:1805.0918}}.

\bibitem{Boehm:2018fpv}
J.~Böhm, A.~Georgoudis, K.~J. Larsen, H.~Schönemann, and Y.~Zhang, {\it
  {Complete integration-by-parts reductions of the non-planar hexagon-box via
  module intersections}},  {\em JHEP} {\bf 09} (2018) 024,
  [\href{http://xxx.lanl.gov/abs/1805.0187}{{\tt arXiv:1805.0187}}].

\bibitem{Badger:2017jhb}
S.~Badger, C.~Brønnum-Hansen, H.~B. Hartanto, and T.~Peraro, {\it {First look
  at two-loop five-gluon scattering in QCD}},  {\em Phys. Rev. Lett.} {\bf 120}
  (2018), no.~9 092001, [\href{http://xxx.lanl.gov/abs/1712.0222}{{\tt
  arXiv:1712.0222}}].

\bibitem{Abreu:2017hqn}
S.~Abreu, F.~Febres~Cordero, H.~Ita, B.~Page, and M.~Zeng, {\it {Planar
  Two-Loop Five-Gluon Amplitudes from Numerical Unitarity}},  {\em Phys. Rev.}
  {\bf D97} (2018), no.~11 116014,
  [\href{http://xxx.lanl.gov/abs/1712.0394}{{\tt arXiv:1712.0394}}].

\bibitem{Abreu:2018jgq}
S.~Abreu, F.~Febres~Cordero, H.~Ita, B.~Page, and V.~Sotnikov, {\it {Planar
  Two-Loop Five-Parton Amplitudes from Numerical Unitarity}},
  \href{http://xxx.lanl.gov/abs/1809.0906}{{\tt arXiv:1809.0906}}.

\bibitem{Papadopoulos:2015jft}
C.~G. Papadopoulos, D.~Tommasini, and C.~Wever, {\it {The Pentabox Master
  Integrals with the Simplified Differential Equations approach}},  {\em JHEP}
  {\bf 04} (2016) 078, [\href{http://xxx.lanl.gov/abs/1511.0940}{{\tt
  arXiv:1511.0940}}].

\bibitem{Wess:1992cp}
J.~Wess and J.~Bagger, {\em {Supersymmetry and supergravity}}.
\newblock Princeton University Press, Princeton, NJ, USA, 1992.

\bibitem{Drummond:2008cr}
J.~M. Drummond and J.~M. Henn, {\it {All tree-level amplitudes in N=4 SYM}},
  {\em JHEP} {\bf 04} (2009) 018,
  [\href{http://xxx.lanl.gov/abs/0808.2475}{{\tt arXiv:0808.2475}}].

\bibitem{Dixon:2011ng}
L.~J. Dixon, J.~M. Drummond, and J.~M. Henn, {\it {The one-loop six-dimensional
  hexagon integral and its relation to MHV amplitudes in N=4 SYM}},  {\em JHEP}
  {\bf 06} (2011) 100, [\href{http://xxx.lanl.gov/abs/1104.2787}{{\tt
  arXiv:1104.2787}}].

\bibitem{Gehrmann:2000zt}
T.~Gehrmann and E.~Remiddi, {\it {Two loop master integrals for $\gamma^*
  \longrightarrow 3$ jets: The Planar topologies}},  {\em Nucl. Phys.} {\bf
  B601} (2001) 248--286, [\href{http://xxx.lanl.gov/abs/hep-ph/0008287}{{\tt
  hep-ph/0008287}}].

\bibitem{CaronHuot:2011kk}
S.~Caron-Huot and S.~He, {\it {Jumpstarting the All-Loop S-Matrix of Planar N=4
  Super Yang-Mills}},  {\em JHEP} {\bf 07} (2012) 174,
  [\href{http://xxx.lanl.gov/abs/1112.1060}{{\tt arXiv:1112.1060}}].

\bibitem{Alday:2010ku}
L.~F. Alday, D.~Gaiotto, J.~Maldacena, A.~Sever, and P.~Vieira, {\it {An
  Operator Product Expansion for Polygonal null Wilson Loops}},  {\em JHEP}
  {\bf 04} (2011) 088, [\href{http://xxx.lanl.gov/abs/1006.2788}{{\tt
  arXiv:1006.2788}}].

\bibitem{Henn:2011by}
J.~M. Henn, S.~Moch, and S.~G. Naculich, {\it {Form factors and scattering
  amplitudes in N=4 SYM in dimensional and massive regularizations}},  {\em
  JHEP} {\bf 12} (2011) 024, [\href{http://xxx.lanl.gov/abs/1109.5057}{{\tt
  arXiv:1109.5057}}].

\bibitem{Weinzierl:2011uz}
S.~Weinzierl, {\it {Does one need the O($\epsilon$)- and O($\epsilon^2$)-terms
  of one-loop amplitudes in an NNLO calculation?}},  {\em Phys. Rev.} {\bf D84}
  (2011) 074007, [\href{http://xxx.lanl.gov/abs/1107.5131}{{\tt
  arXiv:1107.5131}}].

\bibitem{Galperin:1984av}
A.~Galperin, E.~Ivanov, S.~Kalitsyn, V.~Ogievetsky, and E.~Sokatchev, {\it
  {Unconstrained N=2 Matter, Yang-Mills and Supergravity Theories in Harmonic
  Superspace}},  {\em Class. Quant. Grav.} {\bf 1} (1984) 469--498. [Erratum:
  Class. Quant. Grav.2,127(1985)].

\bibitem{Gelfand}
I.~M. Gelfand and G.~E. Shilov, {\em {Generalized functions. Vol. 1, Properties
  and operations}}.
\newblock Academic Press, New York, NY, USA, 1964.

\bibitem{Drummond:2007au}
J.~M. Drummond, J.~Henn, G.~P. Korchemsky, and E.~Sokatchev, {\it {Conformal
  Ward identities for Wilson loops and a test of the duality with gluon
  amplitudes}},  {\em Nucl. Phys.} {\bf B826} (2010) 337--364,
  [\href{http://xxx.lanl.gov/abs/0712.1223}{{\tt arXiv:0712.1223}}].

\bibitem{Usyukina:1992jd}
N.~I. Usyukina and A.~I. Davydychev, {\it {An Approach to the evaluation of
  three and four point ladder diagrams}},  {\em Phys. Lett.} {\bf B298} (1993)
  363--370.

\bibitem{Cachazo:2004dr}
F.~Cachazo, {\it {Holomorphic anomaly of unitarity cuts and one-loop gauge
  theory amplitudes}},  \href{http://xxx.lanl.gov/abs/hep-th/0410077}{{\tt
  hep-th/0410077}}.

\bibitem{Cachazo:2004kj}
F.~Cachazo, P.~Svrcek, and E.~Witten, {\it {MHV vertices and tree amplitudes in
  gauge theory}},  {\em JHEP} {\bf 09} (2004) 006,
  [\href{http://xxx.lanl.gov/abs/hep-th/0403047}{{\tt hep-th/0403047}}].

\end{thebibliography}\endgroup

\end{document}